\documentclass[final,3p]{elsarticle}
\biboptions{numbers,sort&compress}

\usepackage{amssymb}
\usepackage[]{siunitx}
\usepackage{multirow}
\usepackage{wasysym}
\usepackage{xcolor}
\usepackage{subcaption}
\usepackage{graphicx}
\usepackage[utf8]{inputenc}
\usepackage{makecell}
\usepackage[page,toc,titletoc,title]{appendix}
\usepackage[subfigure]{tocloft}
\usepackage[section]{placeins} 

\usepackage[export]{adjustbox}

\usepackage{siunitx} 
\DeclareSIUnit\sq{\ensuremath{\Box}}                           

\usepackage[rawfloats=true]{floatrow}

\usepackage{xcolor}

\usepackage{hyperref}

\journal{Nucl. Instrum. Meth. A}

\begin{document}

\begin{frontmatter}



\title{Results for pixel and strip centimeter-scale AC-LGAD sensors with a 120 GeV proton beam}


\author[a]{Irene Dutta \corref{cor1}}\ead{idutta@fnal.gov}\cortext[cor1]{Corresponding author}
\author[a]{Christopher Madrid}
\author[b]{Ryan Heller}
\author[c]{Shirsendu Nanda}
\author[c]{Danush Shekar}
\author[d,e]{Claudio San Mart\'in}
\author[d,e]{Mat\'ias Barr\'ia}
\author[a]{Artur Apresyan}
\author[b,c]{Zhenyu Ye}
\author[d,e,f]{William K. Brooks}
\author[b]{Wei Chen}
\author[g]{Gabriele D'Amen}
\author[g]{Gabriele Giacomini}
\author[b]{Alessandro Tricoli}
\author[h]{Aram Hayrapetyan}
\author[i]{Hakseong Lee}
\author[j]{Ohannes Kamer Köseyan}
\author[a]{Sergey Los}
\author[k]{Koji Nakamura}
\author[l]{Sayuka Kita}
\author[l]{Tomoka Imamura}
\author[a]{Cristi\'an Pe\~na}
\author[a,m]{Si Xie}

\address[a]{Fermi National Accelerator Laboratory, PO Box 500, Batavia IL 60510-5011, USA}
\address[b]{Lawrence Berkeley National Laboratory, Berkeley, CA, 94720, USA}
\address[c]{University of Illinois at Chicago, Chicago, IL 60607, USA}

\address[d]{Departamento de F\'isica y Astronom\'ia, Universidad Técnica Federico Santa María, Valparaiso, Chile}
\address[e]{Centro Cient\'ifico Tecnol\'ogico de Valpara\'iso-CCTVal, Universidad T\'ecnica Federico Santa Mar\'ia, Casilla 110-V, Valpara\'iso, Chile}
\address[f]{Millennium Institute for Subatomic Physics at the High-Energy Frontier (SAPHIR) of ANID, Fern\'andez Concha 700, Santiago, Chile}
\address[g]{Brookhaven National Laboratory, Upton, 11973, NY, USA}
\address[h]{A.~Alikhanyan National Science Laboratory, Yerevan, Armenia}
\address[i]{Kyungpook National University, Daegu, South Korea}

\address[j]{Department of Physics and Astronomy, The University of Iowa, Iowa City, Iowa, USA}

\address[k]{High Energy Research Organization, Oho 1-1, Tsukuba, Ibaraki, 305-0801, Japan}
\address[l]{University of Tsukuba, 1-1-1 Tennodai, Tsukuba, Ibaraki, 305-8571, Japan}
\address[m]{California Institute of Technology, Pasadena, CA, USA}

\begin{abstract}
We present the results of an extensive evaluation of strip and pixel AC-LGAD sensors tested with a \SI{120}{\GeV} proton beam, focusing on the influence of design parameters on the sensor temporal and spatial resolutions. 
Results show that reducing the thickness of pixel sensors significantly enhances their time resolution, with 20-\si{\um}-thick sensors achieving around \SI{20}{\ps}. 
Uniform performance is attainable with optimized $n^{+}$ sheet resistance, making these sensors ideal for future timing detectors. 
Conversely, 20-\si{\um}-thick strip sensors exhibit higher jitter than similar pixel sensors, negatively impacting time resolution, despite reduced Landau fluctuations with respect to the 50-\si{\um}-thick versions. 
Additionally, it is observed that a low resistivity in strip sensors limits signal size and time resolution, whereas higher resistivity improves performance. 
This study highlights the importance of tuning the $n+$ sheet resistance and suggests that further improvements should target specific applications like the Electron-Ion Collider or other future collider experiments.
In addition, the detailed performance of four AC-LGADs sensor designs is reported as examples of possible candidates for specific detector applications.
These advancements position AC-LGADs as promising candidates for future 4D tracking systems, pending the development of specialized readout electronics.
\end{abstract}


\end{frontmatter}

\clearpage
\tableofcontents


\section{Introduction}\label{sec:intro}

Three-dimensional (3D) silicon-based tracking systems have played a crucial role in particle physics experiments (for e.g., the pixel detector of the CMS experiment\cite{Adam:2748381}), allowing physicists to trace the trajectories of charged particles with precision.  
However, as future colliders move to higher energy and increased luminosity, with increased particle occupancy, the need for 4D (spatial and temporal) tracking systems becomes extremely important to maintain the desired particle reconstruction efficiency~\cite{Berry:2807541,Cartiglia_2022}.
Additionally, the use of timing information is critical for particle identification (PID) and meeting future collider physics goals.
Tracking detectors capable of achieving 5–50 \si{\ps} timing resolution and 5–30 \si{\um} spatial resolution are needed for many proposed future colliders, including the FCC-ee \cite{Benedikt:2651299}, Muon colliders \cite{MuonColliderForum2022}, and the Electron–Ion Collider (EIC) \cite{AbdulKhalek:2021gbh}. Low-gain avalanche diode (LGAD) based timing layers for the HL-LHC upgrades, such as the MIP Timing Detector (MTD)~\cite{CMS:2667167} or the High Granularity Timing Detector (HGTD)~\cite{CERN-LHCC-2020-007}, are a first step towards 4D reconstruction of particles. These dedicated timing layers are necessary for enhancing track-to-vertex association in the high pileup environment of the HL-LHC but differ from the integrated 4D tracking detectors envisioned for future colliders, which aim to combine spatial and timing information in a single layer. The development of fast readout electronics and advanced timing reconstruction for these LGAD-based timing layers, along with their implementation in data reconstruction and analysis, will play a pivotal role in shaping and informing the design of future 4D tracking detectors. 

A major breakthrough in 4D tracking detector technology in recent years has been the development of silicon-based sensors such as AC-coupled Low Gain Avalanche Diodes (AC-LGAD) \cite{giacomini_fabrication_2021,8846722,RSD_NIM,firstAC}. 
These sensors have been demonstrated to be capable of achieving \SI{30}{\ps} time resolution and 5-30 \si{\um} spatial resolution in a single sensing layer~\cite{Apresyan:2020ipp,Heller_2022,Madrid:2022rqw,TORNAGO2021165319,OTT2023167541}.In this paper, we present the performance of centimeter-scale AC-LGAD devices including strip sensors with varying pitch (coarse and narrow), and pixel sensors of varying metal electrode size and geometry. 
These results are a continuation of studies following recent campaigns~\cite{Madrid:2022rqw} for developing centimeter-scale devices that, among many future applications, will be ideal for the TOF-PID system of the ePIC detector~\cite{ABDULKHALEK2022122447} at the EIC.
Strip sensors with narrow pitch provide good position reconstruction capabilities, however coarser pitch may offer better cost and performance for applications where channel count or electrical power density should be economized. 
Pixel sensors with a higher metal area fraction and smaller active thickness provide faster signal risetimes, leading to better time resolution capabilities. 
The sensors tested are produced by Brookhaven National Laboratory (BNL) and Hamamatsu Photonics (HPK). 
Section~\ref{sec:exp_methods} presents a description of the experimental setup at the Fermilab Test Beam Facility (FTBF)~\cite{FTBF}, the AC-LGAD sensors, and the used signal reconstruction techniques. 
Experimental results are presented in Section~\ref{sec:results}. 
These results include measurements of sensor signal properties, detection efficiency, and position and time resolution. 
Conclusions and outlook are presented in Section~\ref{sec:discussion}.

\section{Experimental methods}\label{sec:exp_methods}

\subsection{\textbf{Beam test experimental setup}}\label{sec:setup}
\begin{figure}[H]
    \centering
    \includegraphics[width=0.9\textwidth]{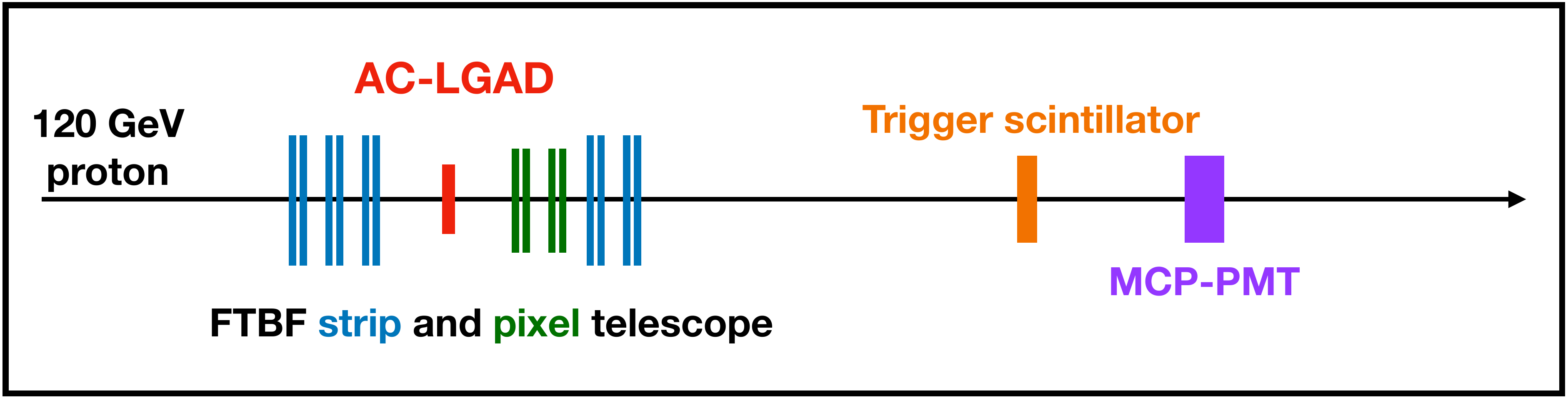}
    \caption{Diagram of the tracking system and the AC-LGAD under test along the beamline.}
    \label{fig:diagram_telescope2022}
\end{figure}

All measurements reported in this article are from a dataset that was collected at FTBF, using \SI{120}{\GeV} protons. The FTBF beam was tuned to provide a burst of approximately 50,000--100,000 protons every minute with each spill lasting \SI{4.2}{\s}. A schematic diagram of the setup is shown in Figure~\ref{fig:diagram_telescope2022}. The reference positions of the incident proton tracks are given by a silicon tracking telescope, consisting of four pixel layers and ten strip layers with a combined resolution of roughly \SI{5}{\um}. Good quality tracks are required to have two and six hits from the four-pixel and ten-strip detectors respectively, reduced $\chi^2 < 3.0$ and slopes less than $10^{-4}$ (with respect to the beam axis). The trigger signal is generated by a scintillator detector and is distributed to the tracker and the oscilloscope. The reference time of the proton hits is provided by a fast microchannel plate detector (Photek 240 MCP-PMT) with a precision of about \SI{10}{\pico \s}~\cite{RONZHIN2015288}. For all temporal and spatial resolution results mentioned in this article, the \SI{10}{\pico \s} MCP-PMT contribution and \SI{5}{\um} tracker contribution have been subtracted in quadrature respectively.

The AC-LGAD sensors were read out using the Fermilab 16-channel readout board~\cite{Heller:2021mht}. 
For comparing the performances, we also used the low noise single-channel UCSC board~\cite{Cartiglia201783} to read out certain pixel sensors. 
The sensor with the read-out board was mounted on an aluminum cooling block. 
An ethylene glycol-water mix solution was circulated through the block to help maintain a uniform and constant temperature of \SI{20}{\celsius}. 
Each readout channel on the board contains a 2-stage amplifier (Mini-Circuits GALI-66+ integrated circuit) chain. 
The amplifiers used a \SI{50}{\ohm} input impedance, \SI{1}{\GHz} bandwidth and an equivalent transimpedance of approximately \SI{4.3}{\kilo\ohm}. 
The AC-LGAD and MCP-PMT waveforms were recorded using an eight-channel Lecroy Waverunner 8208HD oscilloscope featuring a bandwidth of \SI{2}{\GHz} and a sampling rate of \SI{10}{\giga S / \s} per channel. 
Further details on our experimental setup are described in Refs.~\cite{Apresyan:2020ipp, Heller_2022, Madrid:2022rqw}.

\subsection{\textbf{AC-LGAD sensors}}\label{sec:sensors}

\begin{figure}[H]
    \centering
    \includegraphics[width=0.5\textwidth]{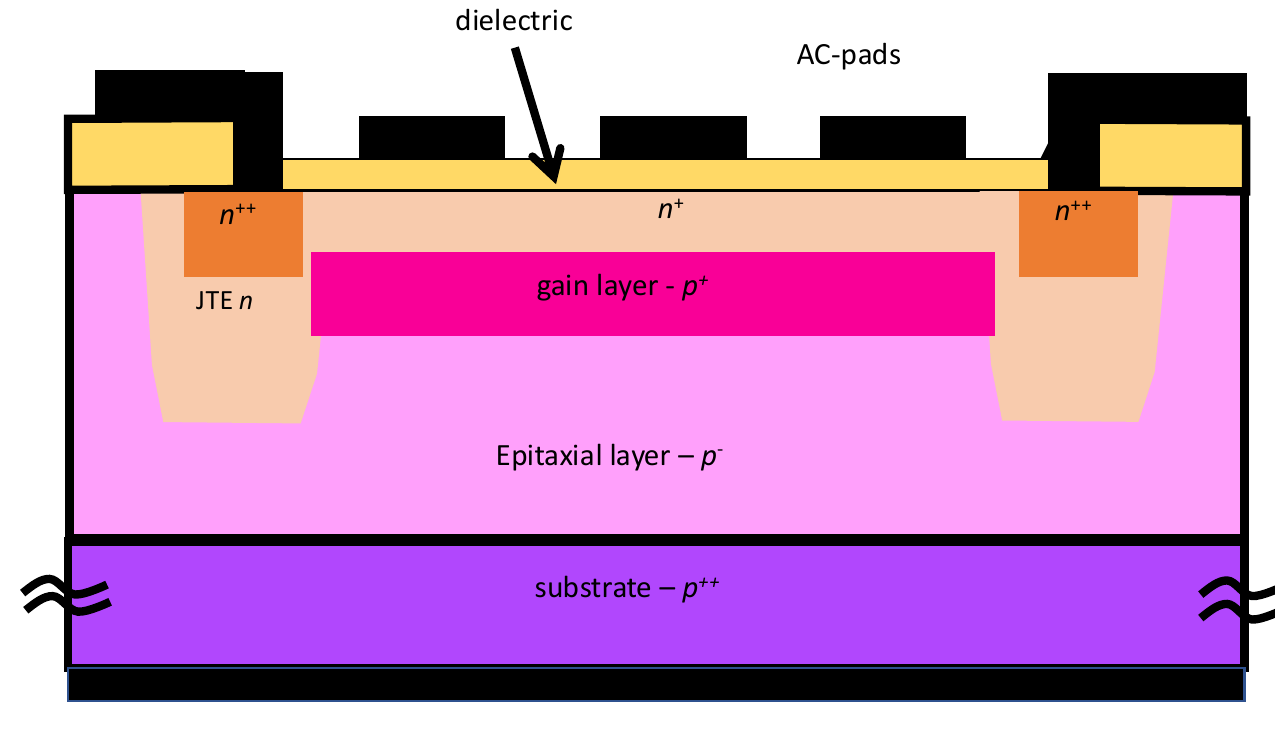}  
    \caption{A cross-sectional view of a typical AC-LGAD device.}
    \label{fig:sensor_xsec}
\end{figure}

A cross-sectional view of a typical AC-LGAD device is shown in Fig.~\ref{fig:sensor_xsec}. AC-LGAD sensors of varying geometrical features were tested in this campaign. A detailed description of HPK sensors and their fabrication parameters are described in~\cite{HPKsensorRef,KITA2023168009}. The HPK sensors achieve full depletion of the voltage gain layer at 10--25 V, depending on thickness.


The BNL sensors were manufactured on 4-inch p-type epitaxial wafers. The epitaxial layer is composed of high-resistivity silicon with varying thicknesses (20, 30, or 50 $\mu$m). The substrate is a low-resistivity wafer of 200-300 $\mu$m thickness with an unpatterned metal layer on the back, and provides mechanical support. Aluminum (approximately 0.7 $\mu$m thick) for the AC-coupled strips is patterned on top of a 150 nm layer of PECVD silicon oxide, which was deposited over a uniform low-dose, high-resistivity phosphorus implant. Strip sensor wafers were passivated with either polyimide or silicon dioxide, while pixel sensor wafers were passivated with a dielectric stack of PECVD nitride and oxide. The gain layer, consisting of a deep low-dose boron implant, is situated beneath the $n^{+}$ resistive layer. In the sensors studied in ~\cite{Madrid:2022rqw}, some signal amplitude non-uniformity was observed due to non-uniformity in the gain implant. However, this issue was resolved in the BNL sensor production campaign for the current study, where all sensors were found to have a uniform gain implant. The BNL sensors achieve full depletion of the voltage gain layer at 25 V.

A junction termination edge (JTE) was created at the boundary of the resistive $n^{+}$ layer surrounding the full device. The JTE consists of a deep phosphorus implantation contacted by a metal frame, which is grounded during sensor biasing. The termination region, including guard rings, is the same for all devices irrespective of the vendor. For BNL strip sensors, the full length of the metalized strip is exposed and available for wire-bonding, while HPK sensors have specific wire-bonding pads at both ends of each strip.

\begin{figure}[H]
    \centering
    \begin{subfigure}{0.49\textwidth}
     
        \includegraphics[height=2.5in]{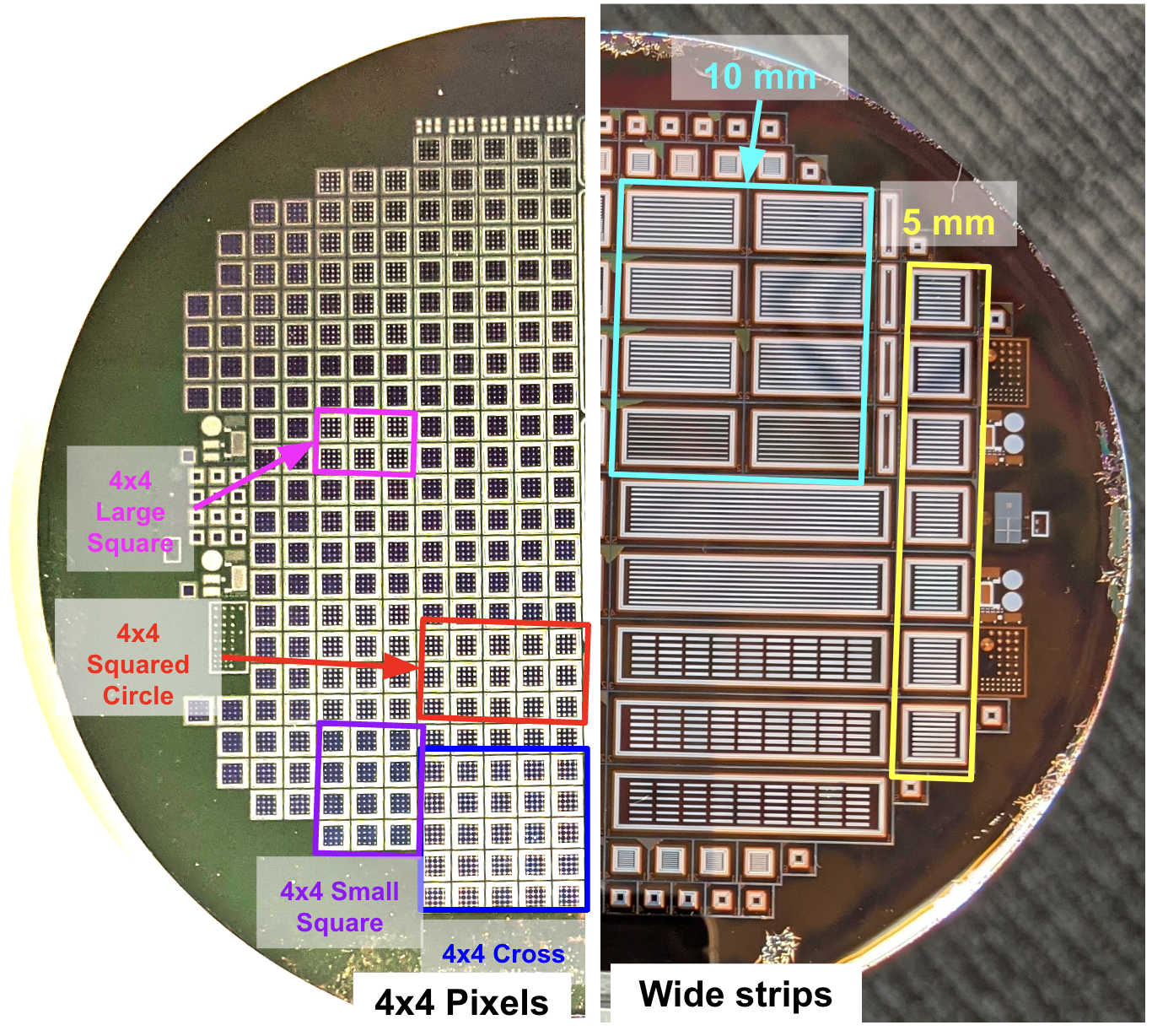}
        \caption{}
        \label{fig:Wafer1_BNL}
    \end{subfigure}
\hfill
    \begin{subfigure}{0.49\textwidth}
    
        \includegraphics[height=2.5in]{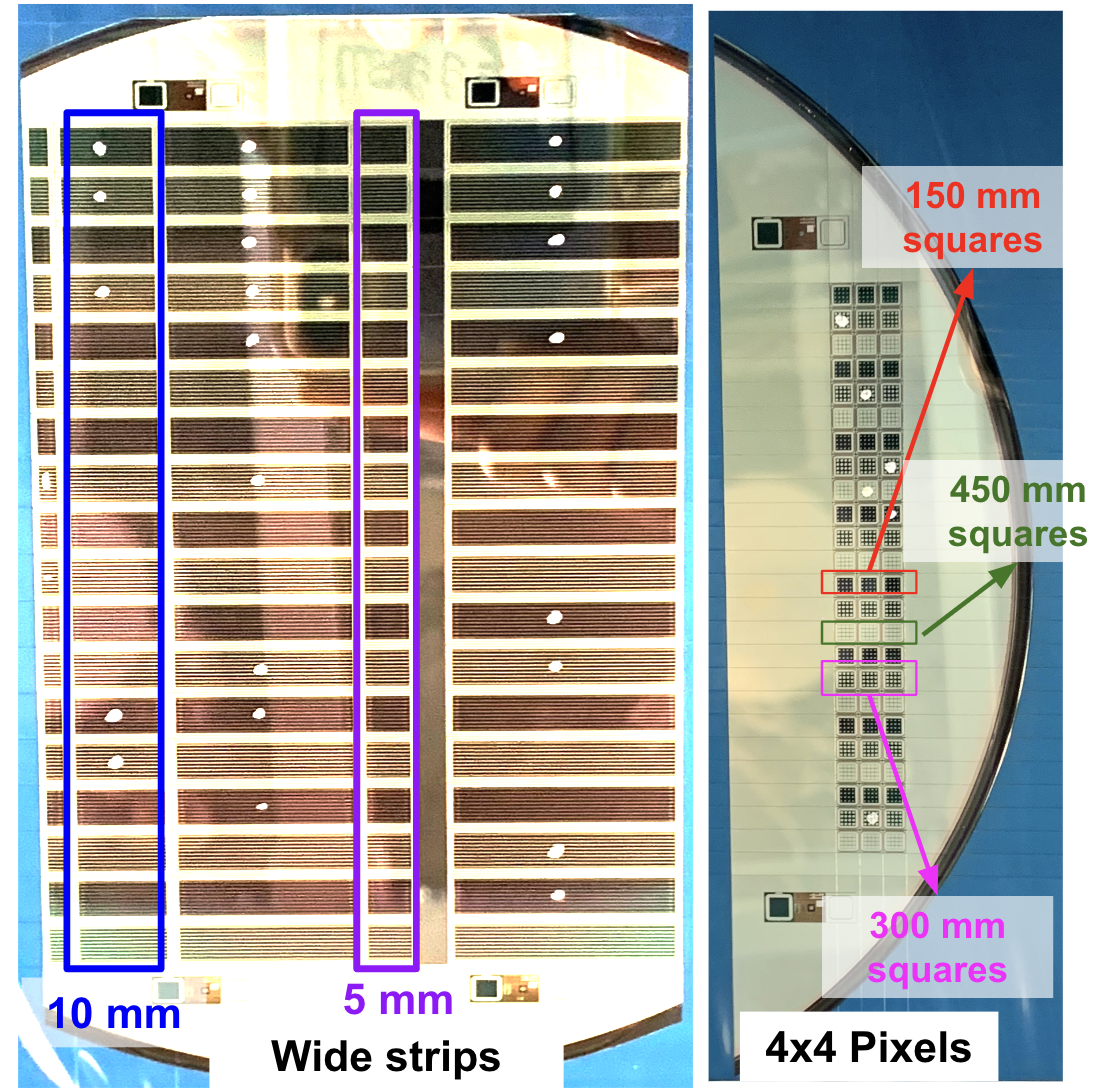}
        \caption{}
        \label{fig:hpk_wafer_may}
    \end{subfigure}
    \caption{Typical wafers from (a) BNL, and (b) HPK. The highlighted regions indicate AC-LGAD devices of varying length, pitch and metal geometry that were used for this study.}
    \label{fig:sensor_wafer}
\end{figure}

 Figure~\ref{fig:sensor_wafer} shows typical wafers from BNL and HPK that were used in this study.

\begin{figure}[H]
\centering
    \begin{subfigure}[b]{0.3\textwidth}
    \centering
        \includegraphics[width=\textwidth,height=1.7 in]{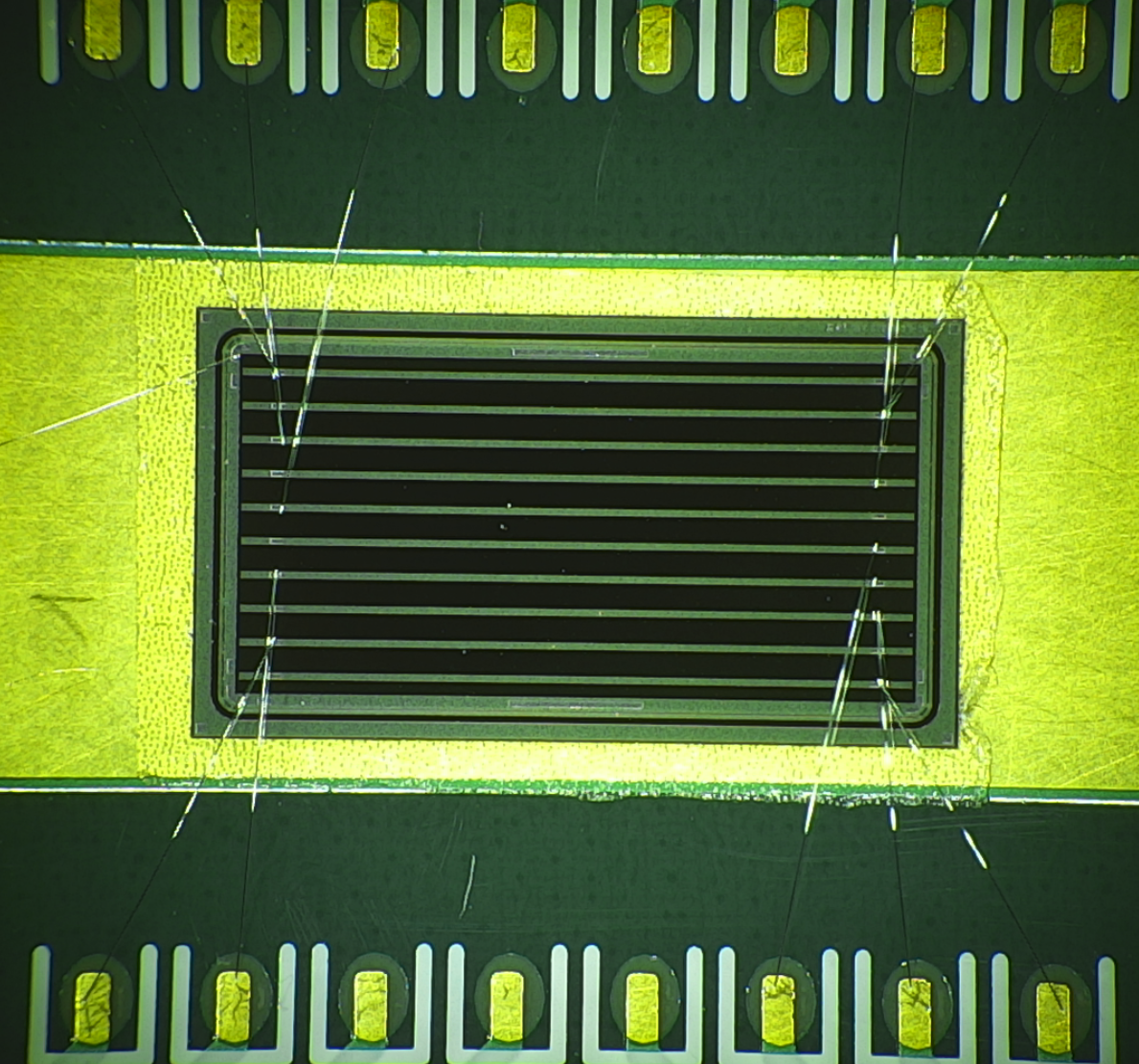}
        \caption{}
        \label{fig:hpk_1cmstrip}
    \end{subfigure}
    \begin{subfigure}[b]{0.3\textwidth}
    \centering
        \includegraphics[width=\textwidth,height=1.7 in]{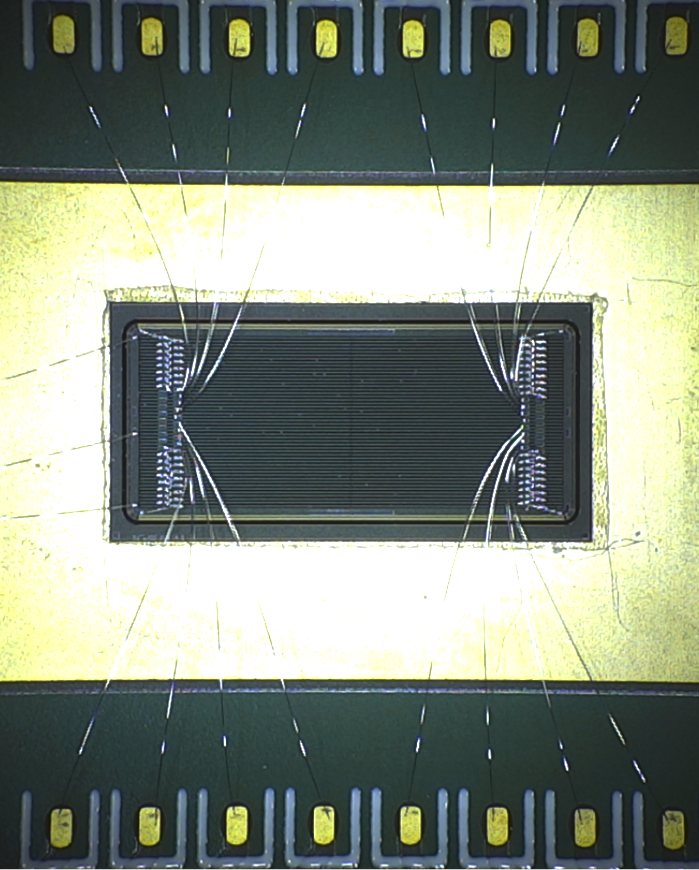}
        \caption{}
        \label{fig:hpk_narrowstrip}
    \end{subfigure}
    \hfill
    \begin{subfigure}[b]{0.49\textwidth}
        \includegraphics[width=0.6\textwidth,height=1.7 in,right]{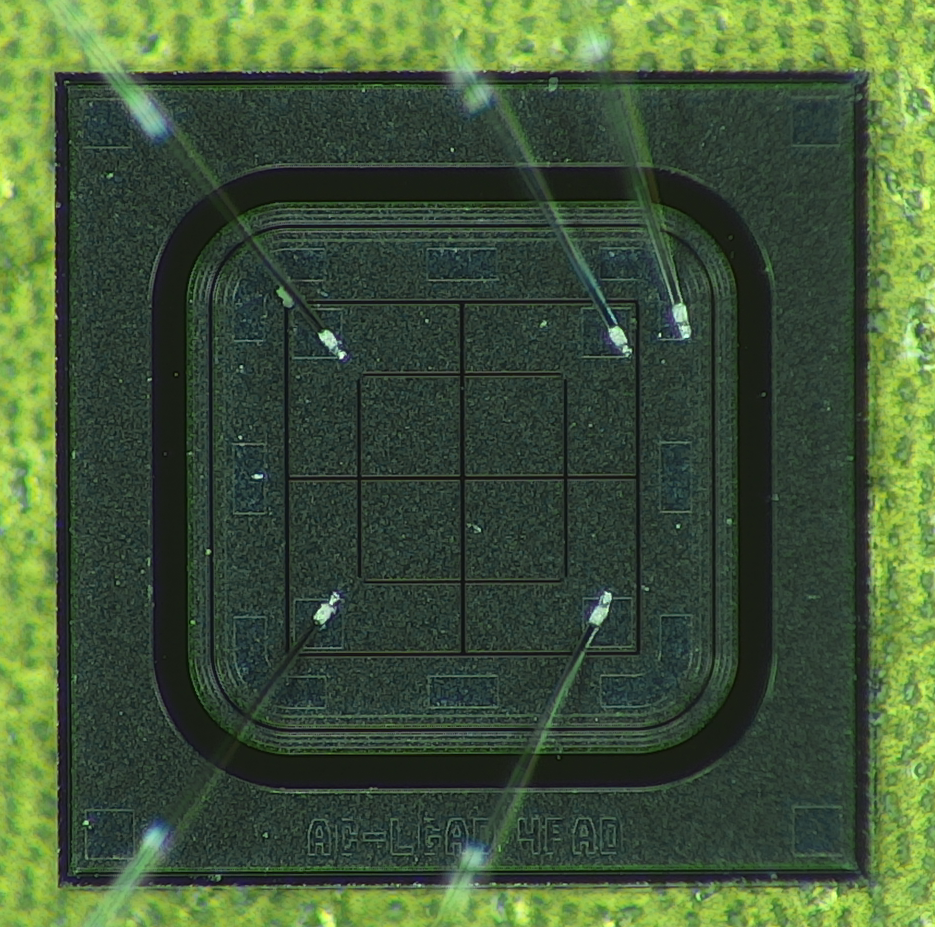}
        \caption{}
        \label{fig:hpk_2x2}
    \end{subfigure}
\hfill
    \begin{subfigure}{0.49\textwidth}
        \includegraphics[width=0.6\textwidth,height=1.7 in]{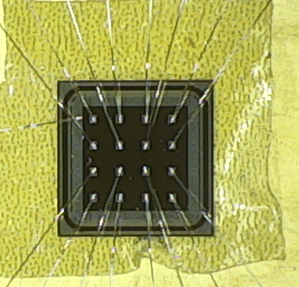}
        \caption{}
        \label{fig:hpk_4x4}
    \end{subfigure}
    \caption{ Photographs of some of the HPK AC-LGAD devices tested in this campaign: a) SH7, b) SHN2, c) PH1, and d) PH4.}
    \label{fig:hpk_sensor_pics}
\end{figure}

\begin{figure}[H]
\centering
    \begin{subfigure}[b]{0.3\textwidth}
    \centering
        \includegraphics[width=\textwidth,height=1.7 in]{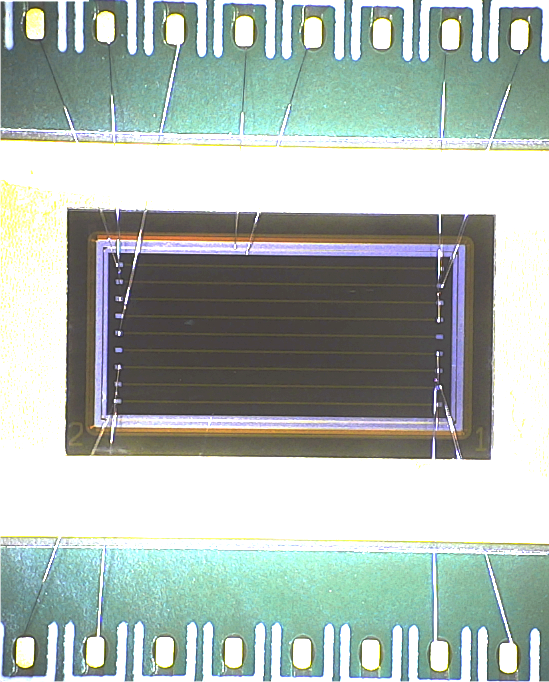}
        \caption{}
        \label{fig:bnl_1cmstrip}
    \end{subfigure}
    \hfill
    \begin{subfigure}[b]{0.33\textwidth}
    \centering
        \includegraphics[width=\textwidth,height=1.7 in]{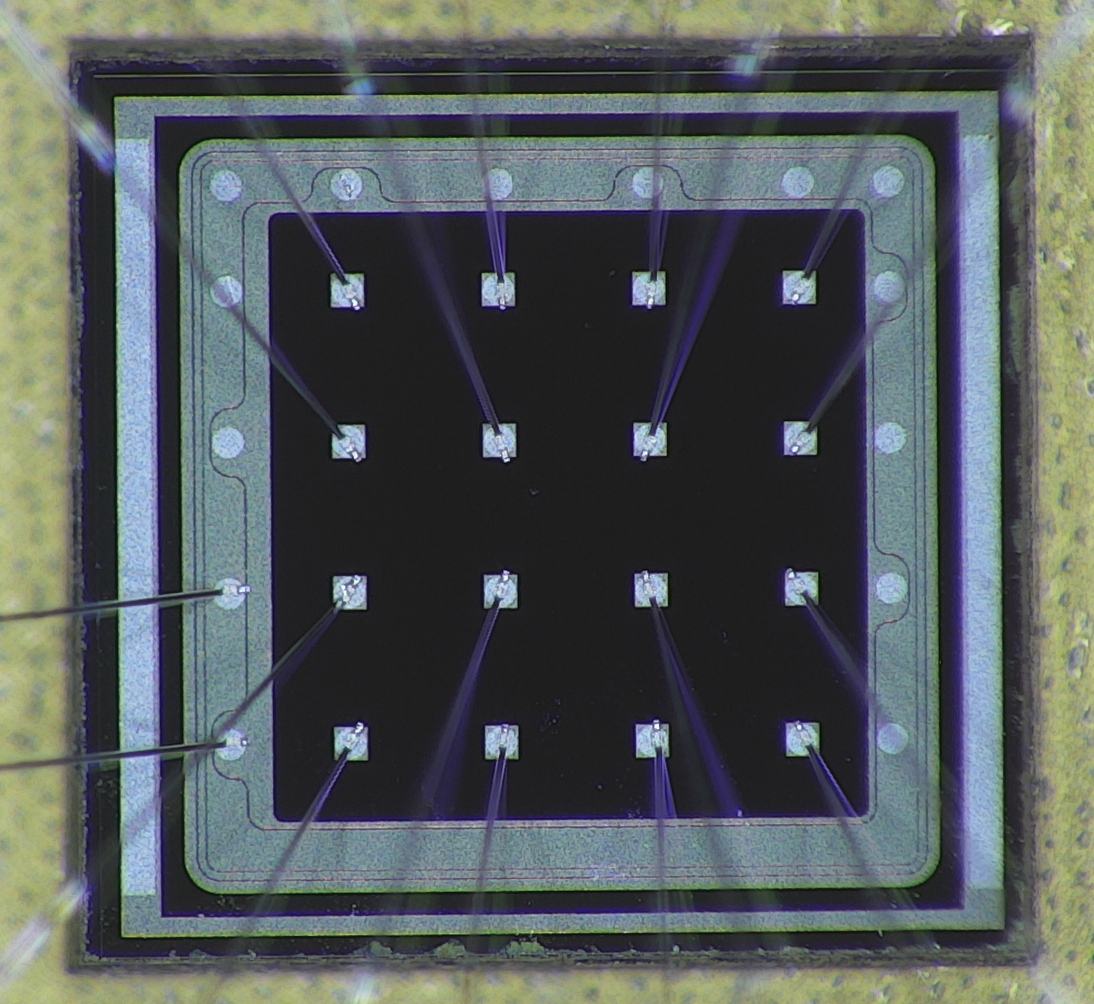}
        \caption{}
        \label{fig:bnl_smallSq}
    \end{subfigure}
    \hfill
    \begin{subfigure}[b]{0.32\textwidth}
    \centering
        \includegraphics[width=\textwidth,height=1.7 in]{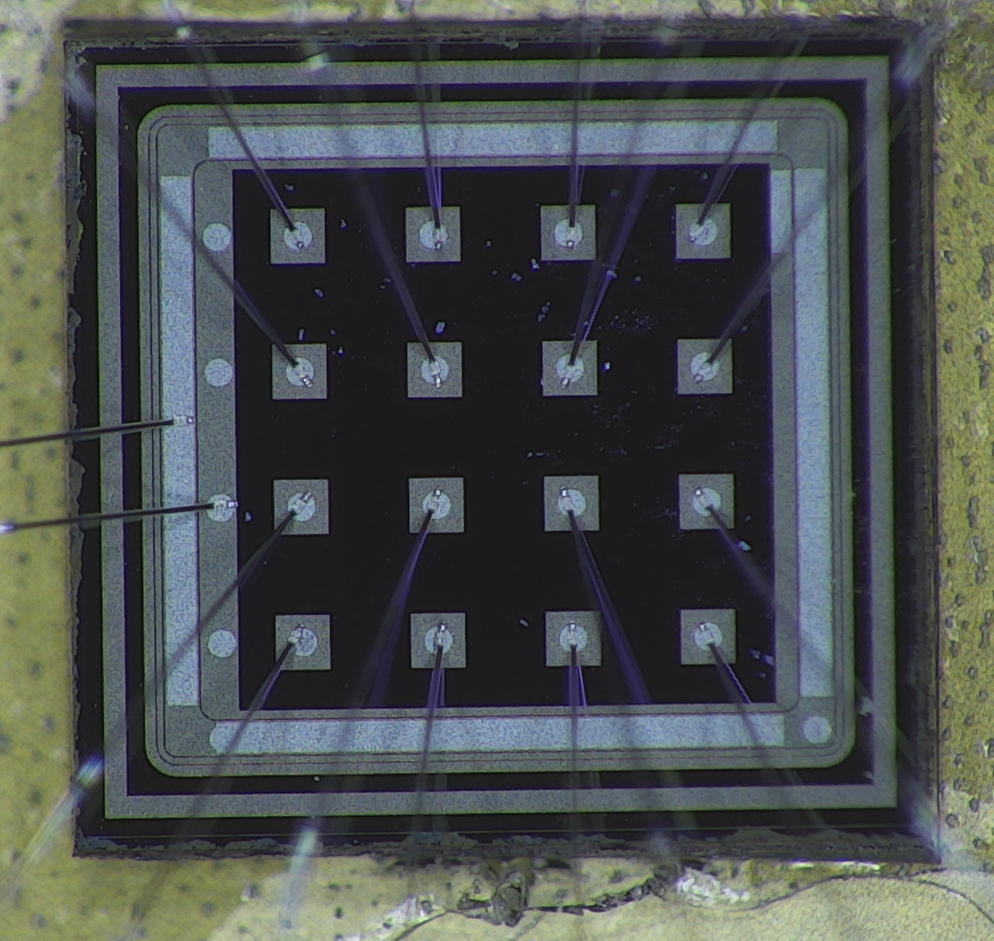}
        \caption{}
        \label{fig:bnl_largeSq}
    \end{subfigure}
     \hfill
    \begin{subfigure}[b]{0.49\textwidth}
        \includegraphics[width=0.6\textwidth,height=1.7 in,right]{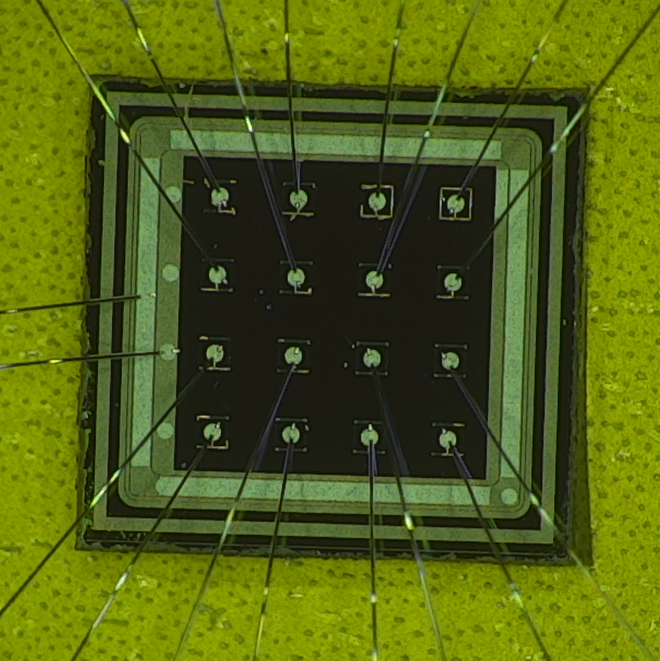}
        \caption{}
        \label{fig:bnl_sqCircle}
    \end{subfigure}
    \hfill
    \begin{subfigure}{0.49\textwidth}
        \includegraphics[width=0.5\textwidth,height=1.7 in]{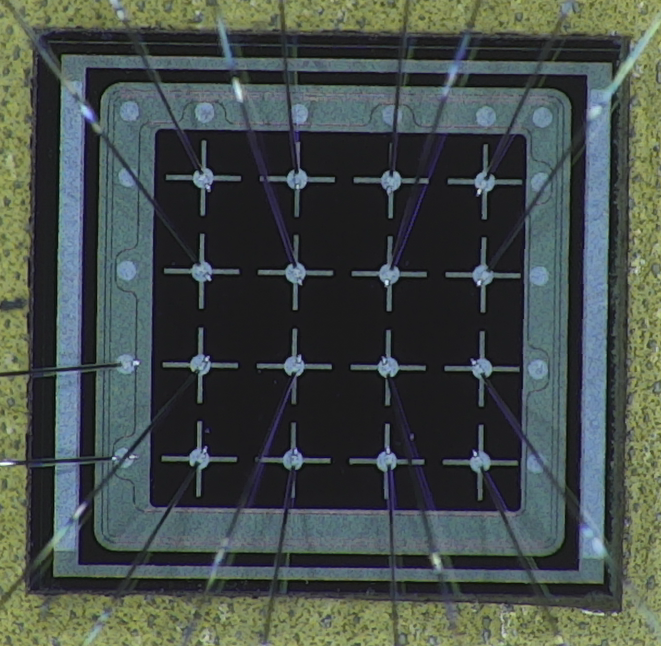}
        \caption{}
        \label{fig:bnl_cross}
    \end{subfigure}
    \caption{Photographs of the BNL AC-LGAD devices tested at FNAL: a) SB1, b) PB1, c) PB2, d) PB3, and e) PB4.}
    \label{fig:bnl_sensor_pics}
\end{figure}

Tables \ref{tab:sensor-info-strips} and~\ref{tab:sensor-info-pixels} summarize the strip and pixel sensors that were studied in this campaign, respectively. 
The sensors can be grouped into four different categories based on their overall geometry: large-pitch strips, small-pitch strips, 2$\times$2 pixel, and 4$\times$4 pixel sensors. 
Small--pitch strip sensors have a pitch of \SI{80}{\um}, and all other remaining sensors have a pitch of \SI{500}{\um}. 
The sensor parameters were carefully chosen to vary in coupling capacitance, $n^{+}$-layer sheet resistance, electrode geometry (both strip length and strip/pixel width) and active thickness. 
 The sheet resistance is defined as the electrical resistivity of a material divided by the sheet thickness. 
The units are described as \SI{}{\ohm/\sq}, to avoid any confusion with the units of resistance. 
The coupling capacitance and the sheet resistance of the $n^{+}$ layer are controlled by varying the thickness of the oxide layer and the doping concentration of the $n^{+}$ layer, respectively. 
We use the following naming convention for our sensors from here onwards:
\begin{itemize}
    \item The first letter shows whether the device is a strip (S) or pixel (P) sensor.
    \item The second letter indicates the sensor manufacturer: H for HPK and B for BNL.
    \item The third letter is an indexing number.
    \item For certain sensors, an extra character is added before the indexing number, to highlight a specific feature, e.g. the letter `N' indicates a narrow strip.
\end{itemize}
Using this naming convention, \textit{PB1} indicates a pixel from BNL and \textit{SHN1} indicates a narrow strip sensor from HPK. 
Figure~\ref{fig:hpk_sensor_pics} and~\ref{fig:bnl_sensor_pics} shows images of some of the tested devices from HPK and BNL, respectively.


\begin{table}[H]
\centering

\begin{tabular}{c| c| c| c| c| c| c| c| c }
 \multicolumn{1}{c|}{Name} & \multicolumn{1}{c|}{Wafer} &  \multicolumn{1}{c|}{Pitch} & \multicolumn{1}{c|}{\begin{tabular}[c]{@{}c@{}} Strip\\length \end{tabular}} & \multicolumn{1}{c|}{\begin{tabular}[c]{@{}c@{}} Metal\\width \end{tabular}} & \multicolumn{1}{c|}{\begin{tabular}[c]{@{}c@{}} Active\\thickness \end{tabular}} & \multicolumn{1}{c|}{Sheet resistance} & \multicolumn{1}{c|}{\begin{tabular}[c]{@{}c@{}} Coupling\\Capacitance \end{tabular}} &  \begin{tabular}[c]{@{}c@{}} Optimal bias\\voltage \end{tabular} \\
&  & \multicolumn{1}{c|}{\si{[\um]}} & \multicolumn{1}{c|}{\si{[\mm]}} & \multicolumn{1}{c|}{\si{[\um]}} &  \multicolumn{1}{c|}{\si{[\um]}} & \multicolumn{1}{c|}{\si{[\Omega / \sq]}} & \multicolumn{1}{c|}{\si{[\pico F / \mm^2]}} &  \si{[\V]} \\
\hline
\hline
\multicolumn{9}{c}{HPK Wide strip}\\ 
\hline
\hline

SH1 & W9 & \multirow{7}{*}{\centering 500} & \multirow{7}{*}{\centering 10} & \multirow{5}{*}{\centering 50} & 20 & 1600 & 600 & 114\\\cline{1-2}\cline{6-9}
SH2 & W4 &  &  &  & \multirow{4}{*}{\centering 50} & \multirow{2}{*}{\centering 400} & 240 & 204\\\cline{1-2}\cline{8-9}
SH3 & W8 &  &  &  &  &  & 600 & 200\\\cline{1-2}\cline{7-9}
SH4 & W2 &  &  &  &  & \multirow{2}{*}{\centering 1600} & 240 & 180\\\cline{1-2}\cline{8-9}
SH5 & W5 &  &  &  &  &  & 600 & 190\\\cline{1-2}\cline{5-9}
SH6 & W9 &  &  & \multirow{2}{*}{\centering 100} & 20 & 1600 & 600 & 112\\\cline{1-2}\cline{6-9}
SH7 & W8 &  &  &  & 50 & 400 & 600 & 208\\

\hline
\hline
\multicolumn{9}{c}{HPK Narrow strip}\\ 
\hline
\hline
SHN1 & WN1 & \multirow{2}{*}{\centering 80} & \multirow{2}{*}{\centering 10} & \multirow{2}{*}{\centering 60} & 20 & 1600 & 240 & 112\\\cline{1-2}\cline{6-9}
SHN2 & WN2 &  &  &  & 50 & 1600 & 240 & 190\\
\hline
\hline
\multicolumn{9}{c}{BNL Wide strip}\\ 
\hline
\hline

SB1 & WB1 & \multirow{2}{*}{\centering 500} & \multirow{2}{*}{\centering 10} & 50 & 50& 1400 & 270 & 170\\ \cline{1-2}\cline{5-9}
SB2 &  WB1 & & & 100 & 50  & 1400 & 270 & 160\\ \cline{1-2}\cline{5-9}
\hline
\end{tabular}

\caption{Summary table with the name, geometrical parameters, and other characteristics of interest of the strip sensors used in this study.}
\label{tab:sensor-info-strips}
\end{table}

\begin{table}[H]
\centering

\begin{tabular}{c| c| c| c| c| c| c| c}
 \multicolumn{1}{c|}{Name} & \multicolumn{1}{c|}{Wafer} &  \multicolumn{1}{c|}{Pitch} & \multicolumn{1}{c|}{\begin{tabular}[c]{@{}c@{}} Metal\\width \end{tabular}} & \multicolumn{1}{c|}{\begin{tabular}[c]{@{}c@{}} Active\\thickness \end{tabular}} & \multicolumn{1}{c|}{Sheet resistance} & \multicolumn{1}{c|}{Capacitance} &  \begin{tabular}[c]{@{}c@{}} Optimal bias\\voltage \end{tabular} \\
&  & \multicolumn{1}{c|}{\si{[\um]}} & \multicolumn{1}{c|}{\si{[\um]}} &  \multicolumn{1}{c|}{\si{[\um]}} & \multicolumn{1}{c|}{\si{[\Omega / \sq]}} & \multicolumn{1}{c|}{\si{[\pico F / \mm^2]}} &  \si{[\V]} \\
\hline
\hline
\multicolumn{8}{c}{HPK 2 x 2 Square pixel}\\ 
\hline
\hline
PH1 & WP1 & \multirow{3}{*}{\centering 510} & \multirow{3}{*}{\centering 500} & 20 & 1600 & 600 & 105\\\cline{1-2}\cline{5-8}
PH2 & WP2 &  &  & 30 & 1600 & 600 & 140\\\cline{1-2}\cline{5-8}
PH3 & WP3 &  &  & 50 & 1600 & 600 & 190\\
\hline
\hline
\multicolumn{8}{c}{HPK 4 x 4 Square pixel}\\ 
\hline
\hline
PH4 & W11 & \multirow{4}{*}{\centering 500} & \multirow{3}{*}{\centering 150} & 20 & 400 & 600 & 116\\\cline{1-2}\cline{5-8}
PH6 & W8 &  &  & \multirow{2}{*}{\centering 50} & 400 & 600 & 200\\\cline{1-2}\cline{6-8}
PH7 & W5 &  &  &  & 1600 & 600 & 185\\\cline{1-2}\cline{4-8}
\hline
\hline





 

\multicolumn{8}{c}{BNL 4 x 4 Square pixel}\\ 

 \hline
 \hline%
PB1 & WP4 &  \multirow{2}{*}{\centering 500} & 100 & 30 &1400 & 695 & 115\\\cline{1-2}\cline{4-8}


PB2 & WP4 & & 200 & 30 & 1400 & 695 & 115\\ 
 \hline
 \hline

\multicolumn{8}{c}{BNL 4 x 4 Squared Circle pixel}\\ 

 \hline
 \hline 
PB3 & WP4  & 500 & 110(*) & 30 &1400 & 695 &110\\ 
 \hline
 \hline
 
 \multicolumn{8}{c}{BNL 4 x 4 Cross pixel}\\ 

 \hline
 \hline
PB4 & WP4 & 500 & 400$\times$25(**) & 30 & 1400 & 695 & 115\\ 
 \hline
\end{tabular}

\caption{Summary table with the name, geometrical parameters, and other characteristics of interest of the pixel sensors used in this study. (*) PB3 sensor has a circular pad of 110~\si{\um} diameter at the center of a square frame with a side of 200~\si{\um} and width of 10\si{\um} per side. (**) PB4 sensor pads form a cross shape with two metal strips that are 400~\si{\um} long and 25~\si{\um} wide (See Figure~\ref{fig:bnl_cross}).}
\label{tab:sensor-info-pixels}
\end{table}

\subsection{\textbf{Reconstruction techniques}}\label{sec:reco}
In this section, we summarize the reconstruction methods employed to extract the proton time of arrival and hit location on the sensor. These techniques remain very similar to what was described in previous reports \cite{Madrid:2022rqw}.

\subsubsection{Time reconstruction}\label{sec:timereco}
The reconstructed hit time is the difference between the proton mean arrival time on the sensor and an external time of reference given by the MCP-PMT, for events where the signals surpass a \SI{15}{\mV}  threshold. The mean hit time is defined as the timestamp measured in the channel using a constant fraction discriminator (CFD) algorithm at 50$\%$ level to the leading edge of the signal. On top of this, we implemented two correction methods to the timestamp measurement that showed great improvements in the overall performance results.

Given the large length of these sensors, a position-dependent delay is introduced in the mean arrival times of the signal based on the impinging location of the proton. 
To correct for this, the first strategy uses the external tracker to determine the proton hit position as a function of $\it{x}$ and $\it{y}$, and creates a reference map of correction values depending on the location of the hit. 
The second correction strategy utilizes the signals from the leading and sub-leading channels on the sensor and the \textit{multi-channel timestamp} as defined in \cite{Madrid:2022rqw}, is given by:
\begin{equation}\label{eqn:weighted-time}
    t = \frac{a_1^2t_1 + a_2^2t_2}{a_1^2 + a_2^2},
\end{equation}
where subscript 1 (2) refers to the leading (subleading) channel: $t_{1(2)}$ is the tracker-corrected time of arrival and $a_{1(2)}$ is the amplitude of the leading (subleading) channel respectively.

The jitter from one channel is given by $\sigma_{\rm{jitter}} = \rm{N}/\frac{dV}{dt}$, where $N$ is the baseline noise and $\frac{dV}{dt}$ is the signal slew rate, for events throughout the sensor. 
For the noise measurement, we calculate the RMS of the samples from each waveform, using the portion before the start of the signal. We then compile these values for each event, taking the mean of this distribution as the noise level.
However, since the aim is to calculate the contribution of the jitter to the time resolution arising from multiple channels, a ``weighted jitter'' method is implemented as follows:
\begin{equation}\label{eqn:weighted-jitter}
    \sigma_{t,\rm{jitter}} = \sqrt{\frac{a_1^4\sigma^2_{t_1,\rm{jitter}} +    a_2^4\sigma^2_{t_2,\rm{jitter}}}{(a_1^2 + a_2^2)^2}
    }
\end{equation}
where $\sigma_{t,\rm{jitter}}$ is the weighted jitter from the AC-LGAD signal, and $\sigma_{t_1,\rm{jitter}}$ and $\sigma_{t_2,\rm{jitter}}$ are the jitter from the leading and subleading channels, respectively. 

All time resolution numbers quoted in this paper subtract the MCP reference contribution of \SI{10}{\ps} in quadrature.

\subsubsection{Position reconstruction}\label{sec:posreco}
Spatial reconstruction of the proton hit position benefits from the signal sharing among adjacent channels in the AC-LGAD sensors. 
Our position reconstruction strategy employs all selections as required for the time reconstruction for accepting hits (including the \SI{15}{\mV}  threshold), in addition to selections that determine the feasibility of an interpolated reconstruction using multiple neighboring channels. 
These interpolation techniques were previously introduced as \textit{one-} and \textit{two-strip} reconstruction in \cite{Madrid:2022rqw}. 
As this paper also includes pixel sensor results, these interpolation methods are henceforth renamed in a more general way as \textit{one-} and \textit{two-channel }reconstruction.\par

The driving factor for the performance of the spatial resolution of these sensors lies in the efficiency of their two-channel reconstruction for signal proton hits. 
A profile utilizing the signal shared between two adjacent neighboring channels can be defined as the amplitude fraction \textit{f} ($f = a_1/(a_1+a_2)$, where $a_1$ and $a_2$ are leading and sub-leading channel amplitudes) as a function of the distance from the center position of the leading channel.
One can then fit this characteristic curve for each sensor to determine the proton hit position relative to the center of the leading amplitude channel, as presented in Figure~\ref{fig:strip-res_cap-AmpEffFit} center. 
This method can be used in events that satisfy two conditions: the two leading channels must be neighbors with their amplitudes above the noise threshold; and the amplitude fraction must be smaller than some value, above which most of the signal is constrained in a single metalized channel. 
Events that fail either condition are instead classified as one-channel events and are assigned a position at the center of the leading channel given that the leading channel passes the amplitude threshold. 
When a proton strikes near the center of the gap region, the majority of events fall in the two-channel category. 
However, signals from protons that strike very close to the metalized strip fall mostly in the one-channel category, resulting in significantly degraded resolution approaching the limit of the metal width/$\sqrt{12}$.

\begin{figure}[H]
\centering
\hspace{20mm}\includegraphics[width=0.80\textwidth]{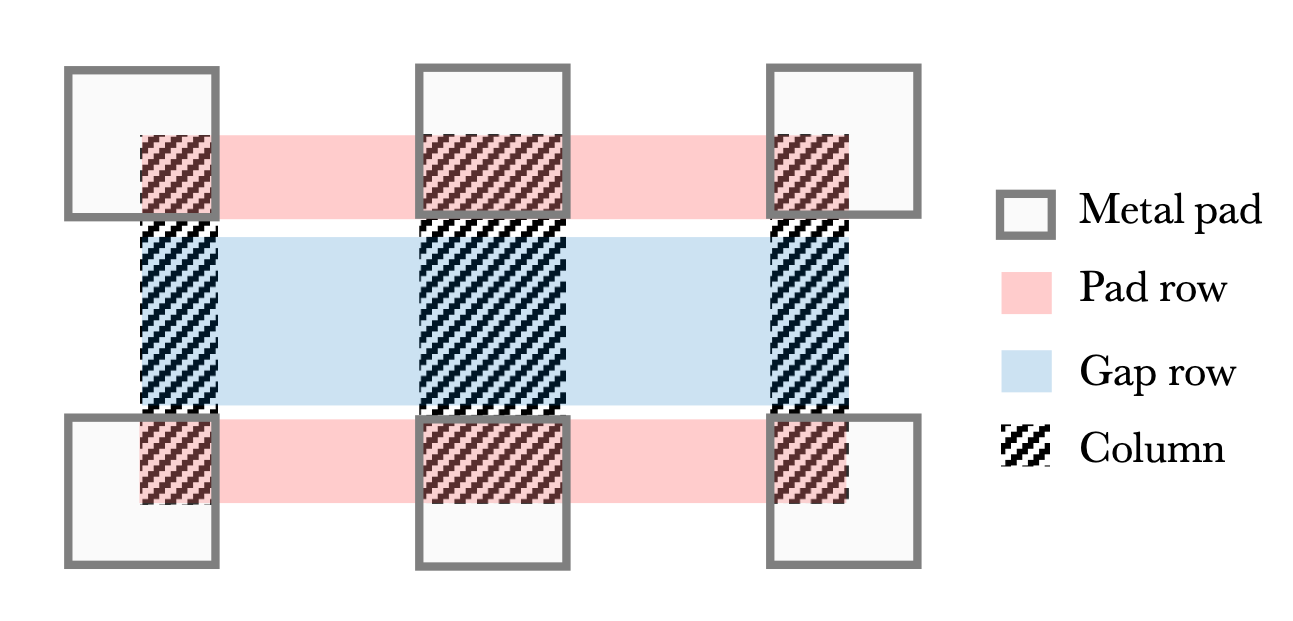}
 \caption{Diagram of a pixel sensor with 2$\times$3 pads, where different regions are shown with different colors.}
\label{fig:HPK_PadGapDiagram}
\end{figure}

While the above method is directly applicable on strip sensors, pixel sensors need to be treated somewhat differently, since their specific geometry permits signal sharing in more than one direction. 
In order to keep using a position reconstruction technique similar to the strip sensors, the signal-sharing profile along the x-direction for pixel sensors was characterized in terms of neighboring rows and columns of metal pads. 
The signals from neighboring pads are collected individually, and then combined in the offline data analysis to determine the signal-sharing profile to reconstruct the $x$ coordinate of the proton hit. 
This same technique can be used to also reconstruct the $y$ position of the proton hit, by considering all events from metal pads in the same row. Figure~\ref{fig:HPK_PadGapDiagram} shows a diagrammatic representation of \textit{rows} and \textit{columns} for an HPK 4x4 pixel sensor where only 2x3 pads are read out. 
The two-channel equivalent reconstruction, or the \textit{two-column reconstruction}, for the pixel sensors, is then defined as follows: the \textit{leading column} contains the metal pad with the largest signal amplitude; the \textit{sub-leading column} is required to be adjacent to the leading column and should contain a metal pad with the next-largest signal.


All position resolution numbers quoted in this paper remove the tracker reference contribution of \SI{5}{\um} in quadrature.

\section{Results}\label{sec:results}

The sensors presented in Tables~\ref{tab:sensor-info-strips} and \ref{tab:sensor-info-pixels} were grouped into subsets with common characteristics to clarify the impact of geometrical or composition-related parameters on each sensor's performance. 
We divide our findings into three broad categories: long strip sensors, pixel sensors from HPK, and pixel sensors from BNL. 
The following sections discuss the results from each of these categories.

\subsection{\textbf{Long strip sensors}}\label{sec:results-long-strips}
In the following paragraphs, we study the effect of varying metal electrode width (50 vs. 100 \si{\um}), $n^{+}$ sheet resistance (400 vs. 1600 \si{\Omega / \sq}), coupling capacitance of the AC electrodes (240 vs. 600 pF/\si{\mm}$^{2}$), and the active thickness of the sensor (20 vs. 50 \si{\um}) on the performance of 1~\si{\cm} long strip sensors.
In addition, the results for 80-\si{\um}-pitch devices (small pitch) with the same active thickness variation is also reported. 

\subsubsection{Metal width}
\begin{figure}[H]
    \centering
    \includegraphics[width=0.49\textwidth]{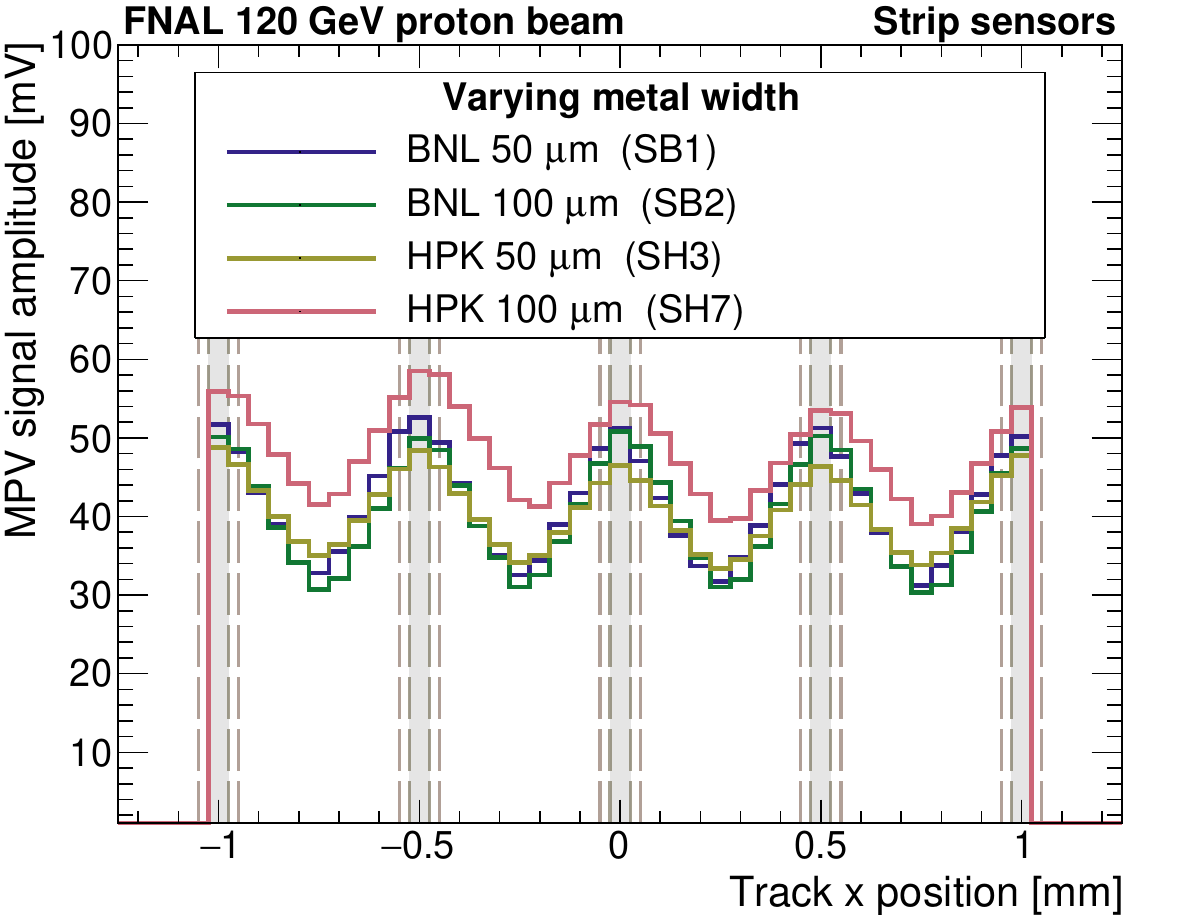}
    \includegraphics[width=0.49\textwidth]{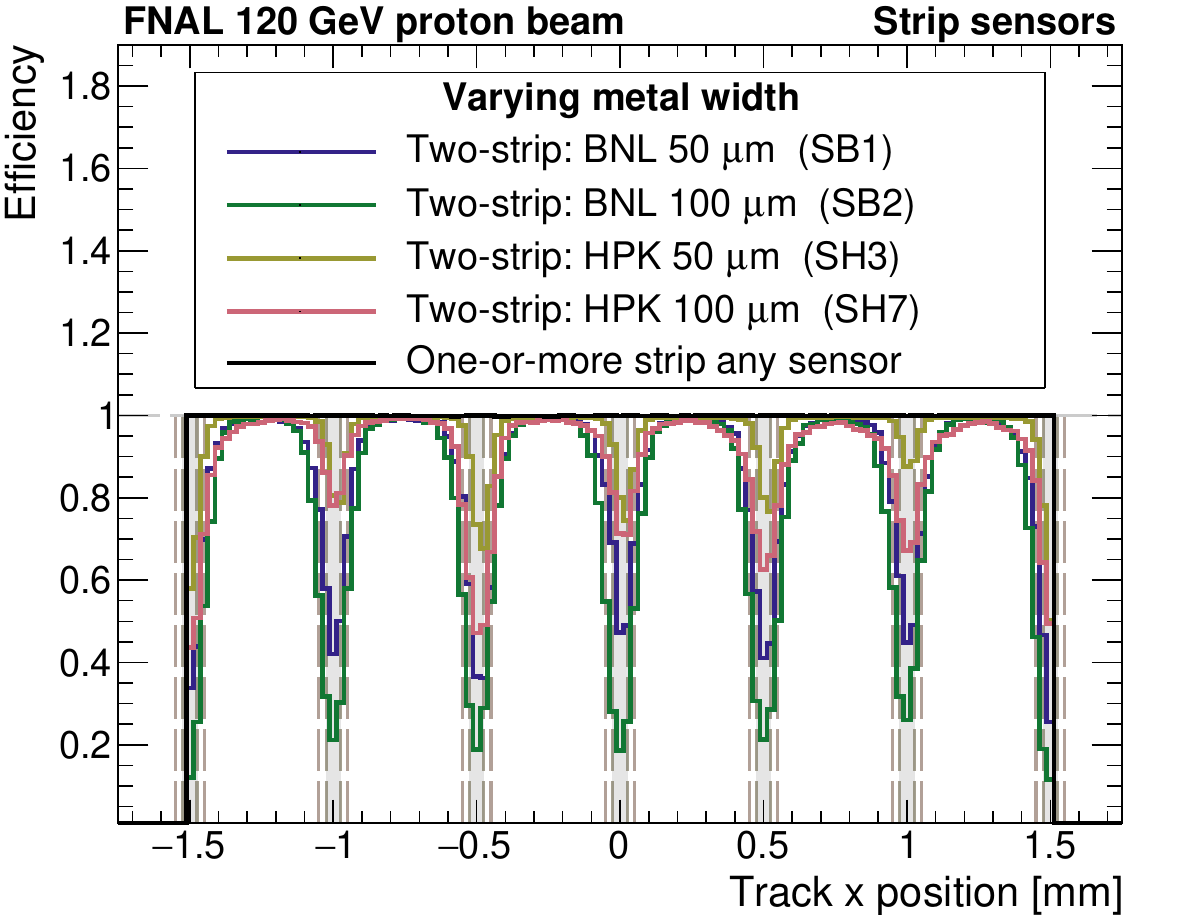}
    \caption{MPV amplitude for the leading strip as a function of the $x$-position (left) and efficiency as a function of the $x$-position (right) for strip sensors of different metal widths. The HPK sensors presented have a \SI{50   }{\um} active thickness, \SI{400}{\Omega / \sq} sheet resistance, and \SI{600}{\pico F / \mm^2} coupling capacitance. The BNL sensors have a \SI{50}{\um} active thickness, \SI{1400}{\Omega / \sq} sheet resistance, and \SI{270}{\pico F / \mm^2} coupling capacitance. The grey regions indicate the metal strips and the dashed lines represent the edges of each strip corresponding to different strip widths.}
    \label{fig:strip-metalwidth-AmpEffFit}
\end{figure}

The impact of varying metal widths (50 vs 100 \si{\um}) on BNL and HPK sensors while keeping all other sensor properties constant is shown in Figure~\ref{fig:strip-metalwidth-AmpEffFit}. 
The most probably value (MPV) of the signal amplitude for the leading strip and the hit reconstruction efficiency as functions of the reference tracker position ($x$) are shown.
The overall amplitude is observed to be consistent between the four devices with the HPK device with the largest metal electrode (SH7) having a slightly higher values.
From Figure~\ref{fig:strip-metalwidth-AmpEffFit} (right), we note that a 50-\si{\um}-metal-width sensor shows an improvement in two-strip reconstruction efficiency as compared to the 100-\si{\um}-wide strip sensor (in agreement with the prediction from~\cite{Madrid:2022rqw}), while achieving similar signal amplitudes. 
The one or more strip reconstruction efficiency is unity for all sensors uniformly across the surface indicating that these devices produced a measurable signal for every proton hit. 

\subsubsection{Sheet resistance and coupling capacitance}\label{sec:strip_res_cc}
\begin{figure}[H]
    \centering
    \includegraphics[width=0.32\textwidth]{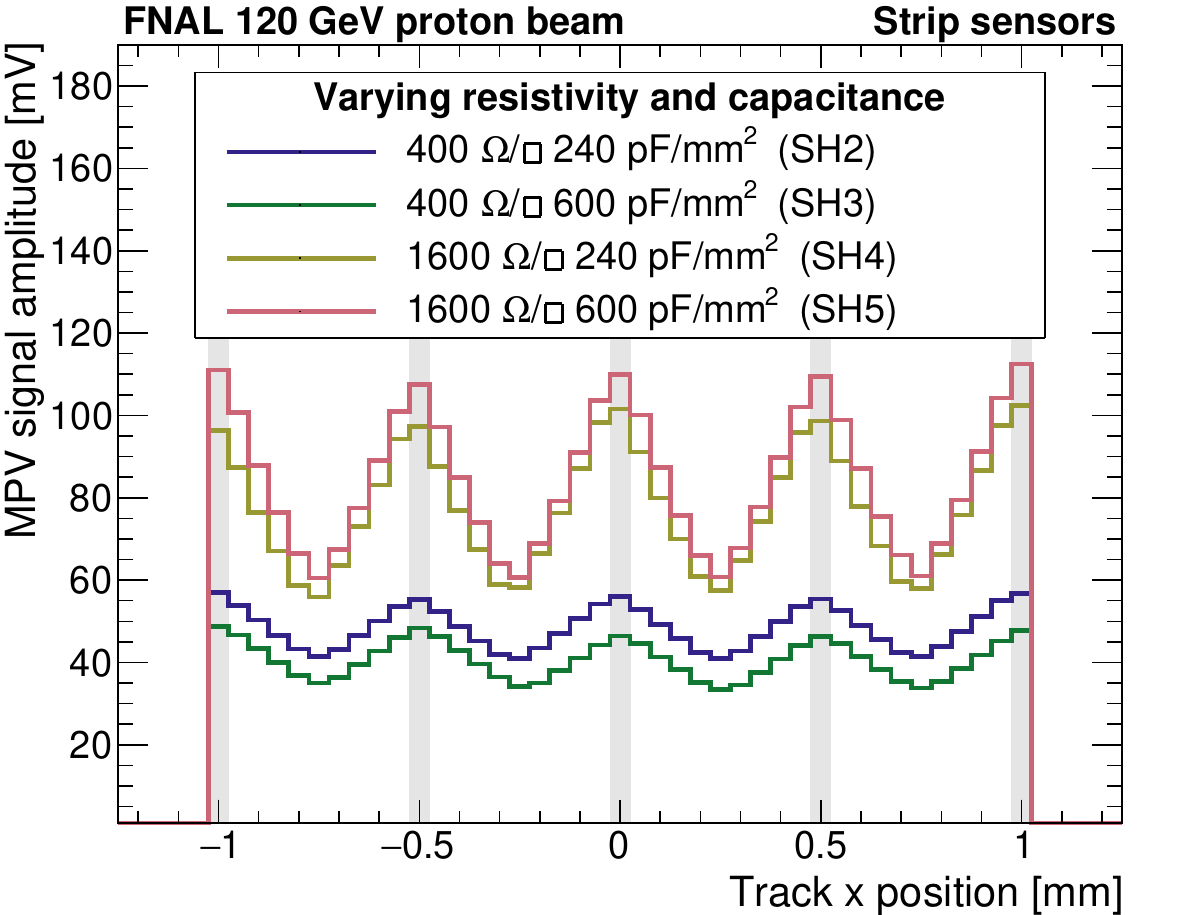}
    \includegraphics[width=0.32\textwidth]{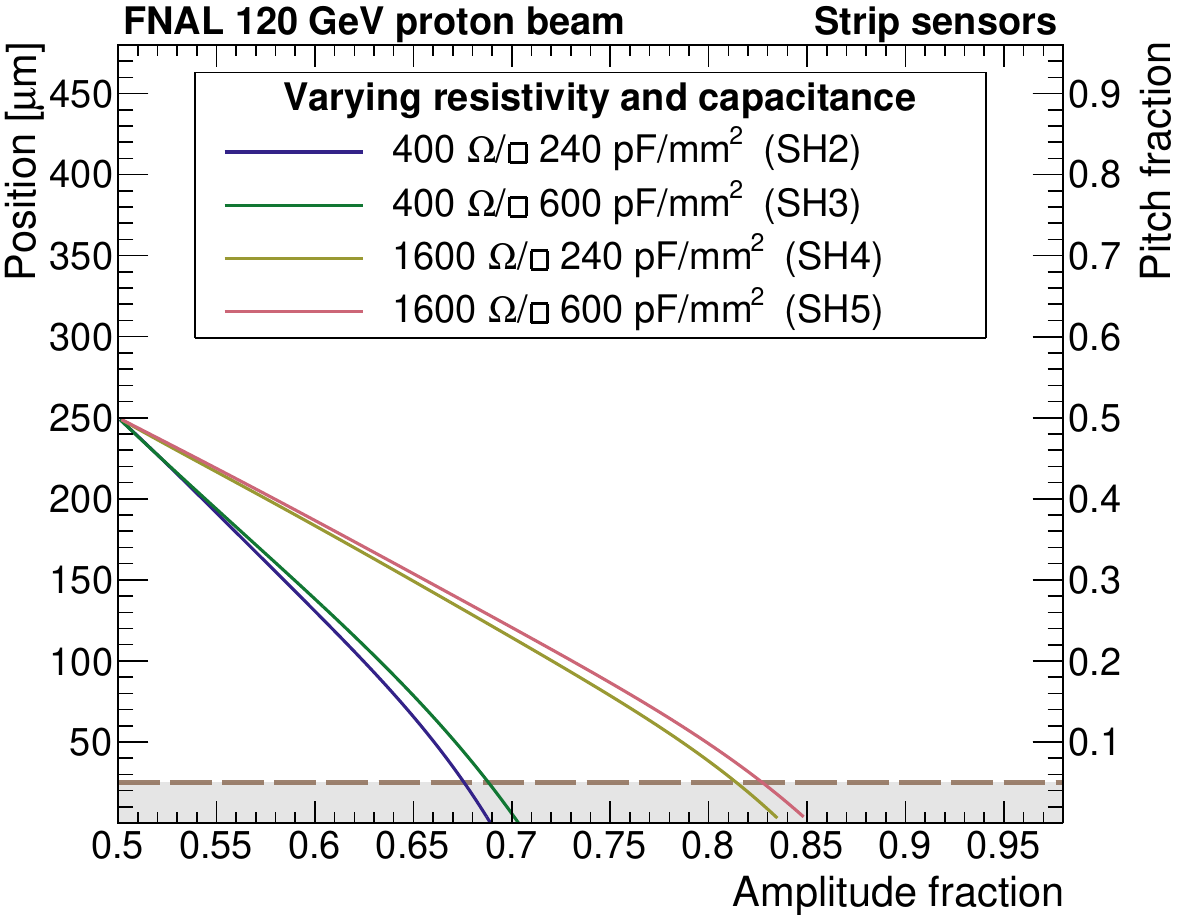}
    \includegraphics[width=0.32\textwidth]{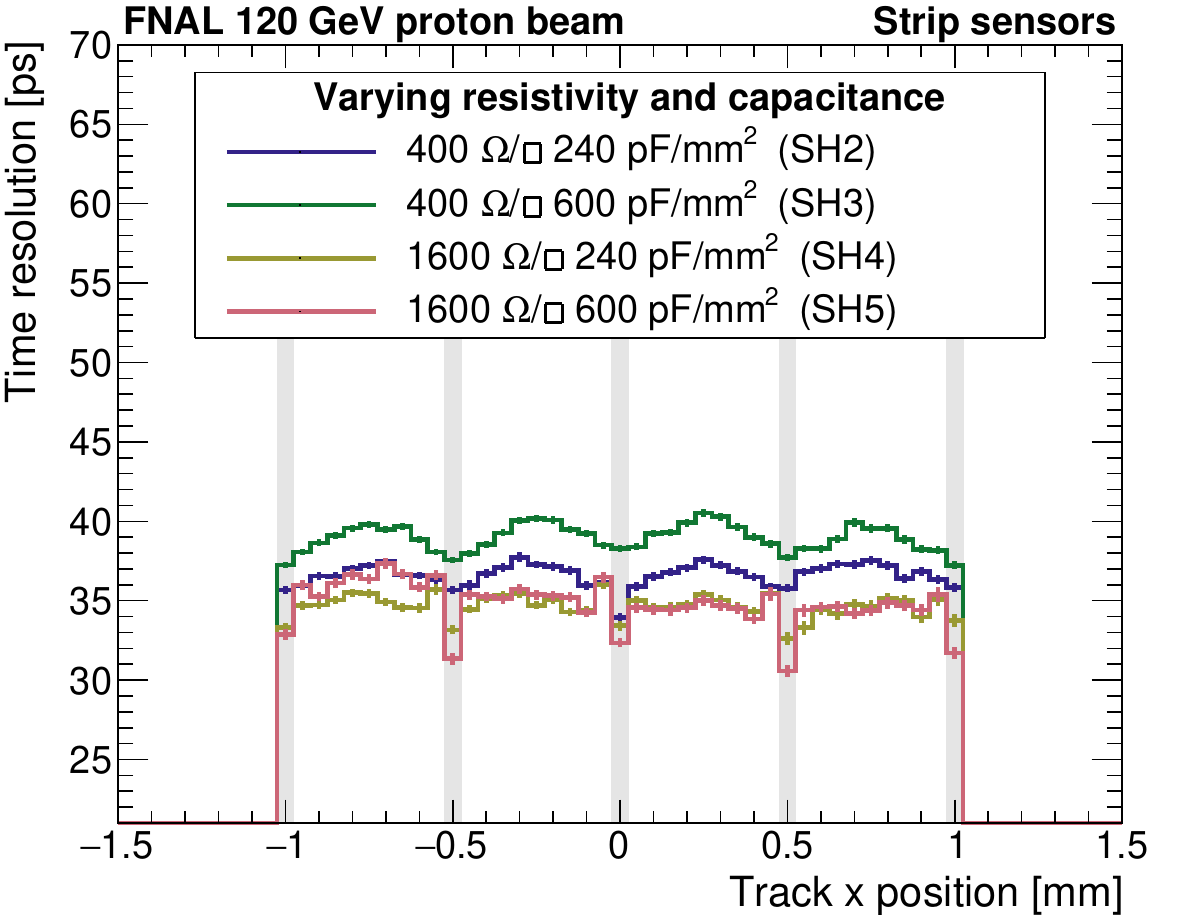}
    \caption{MPV amplitude of the leading strip as a function of $x$ position (left), position reconstruction fit results (center), and time resolution as a function of $x$ position (right) for HPK strip sensors of different coupling capacitance and sheet resistance values. The sensors presented here have a \SI{50}{\um} active thickness and a \SI{50}{\um} strip width.}
    \label{fig:strip-res_cap-AmpEffFit}
\end{figure}

A set of four HPK sensors with combinations of varying $n^{+}$ sheet resistance (400 vs 1600 \si{\Omega / \sq}) and coupling capacitance (240 vs 600 pF/\si{\mm}$^{2}$) were chosen to study how these two quantities impact the sensor performance. 
With increasing $n^{+}$ sheet resistance, we see signals more concentrated in the strips closest to the particle hit position, particularly the leading strip, as can be seen in Figure~\ref{fig:strip-res_cap-AmpEffFit} (left). 
However, no clear trend is observed in the signal amplitude as a function of the coupling capacitance. 
An increase in the signal amplitude is expected to provide better overall time resolution values, and this is observed for the HPK sensors in Figure~\ref{fig:strip-res_cap-AmpEffFit} (right). 
However, a factor 2 increase in the amplitude of the HPK 1600 \si{\Omega / \sq} sensors as compared to the HPK 400 \si{\Omega / \sq} sensors only results in a $\sim$3-5 ps overall improvement in the time resolution values. 
This can be attributed to the fact that the jitter in these sensors is subdominant to the Landau time resolution contribution, around 30 ps for 50-\si{\um}-thick LGADs. 
Once signal size is adequate to obtain low jitter, further increase in signal amplitude does not improve the total time resolution. 
The fits to the amplitude fraction in Figure~\ref{fig:strip-res_cap-AmpEffFit} (center) show that increased coupling capacitance and decreased $n^{+}$ sheet resistance values correspond to more diffuse charge sharing among strips (less rapidly varying amplitude fraction), as expected. 
The charge sharing is affected more strongly by the sheet resistance than by the coupling capacitance. The high contrast between leading and sub-leading strips in sensors with higher sheet resistance allows for more precise spatial interpolation for higher $n^{+}$ sheet resistance values. 
These results are also in agreement with~\cite{KITA2023168009,bishop2024longdistancesignalpropagationaclgad}. 
The overall spatial and temporal resolution numbers for the four sensors are quoted in Table~\ref{tab:summary_strips}.

\subsubsection{Active thickness}
It is a common expectation that LGADs with an active thickness of \SI{50}{\um} can reach a time resolution close to \SI{30}{\ps}, dominated by variation in drift time due to Landau fluctuations in the ionization profile for charged particles~\cite{Sadrozinski_2018}.
DC-LGADs with thinner active thickness have been show to exhibit smaller Landau contributions to the time resolution and faster slew rates at the expense of smaller initial ionization~\cite{zhao_comparison_2019}.

We study the impact of thickness in AC-LGAD sensors, considering the performance of devices with 50 and \SI{20}{\um} active thickness.
Since AC-LGADs divide signals among multiple channels, in certain cases their performance can be more sensitive than DC-LGADs to the reduced total ionization in thinner devices.

\begin{figure}[H]
    \centering
    \includegraphics[width=0.32\textwidth]{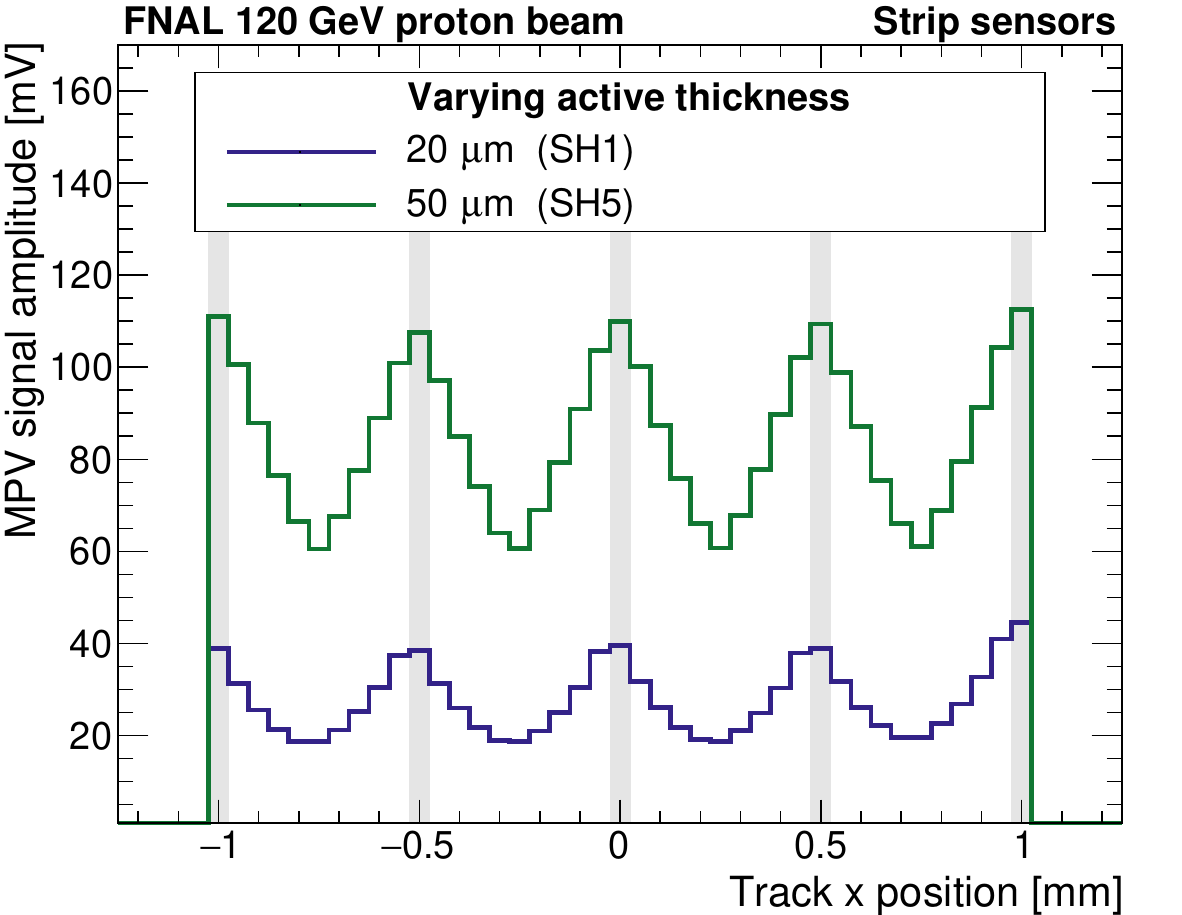}
    \includegraphics[width=0.32\textwidth]{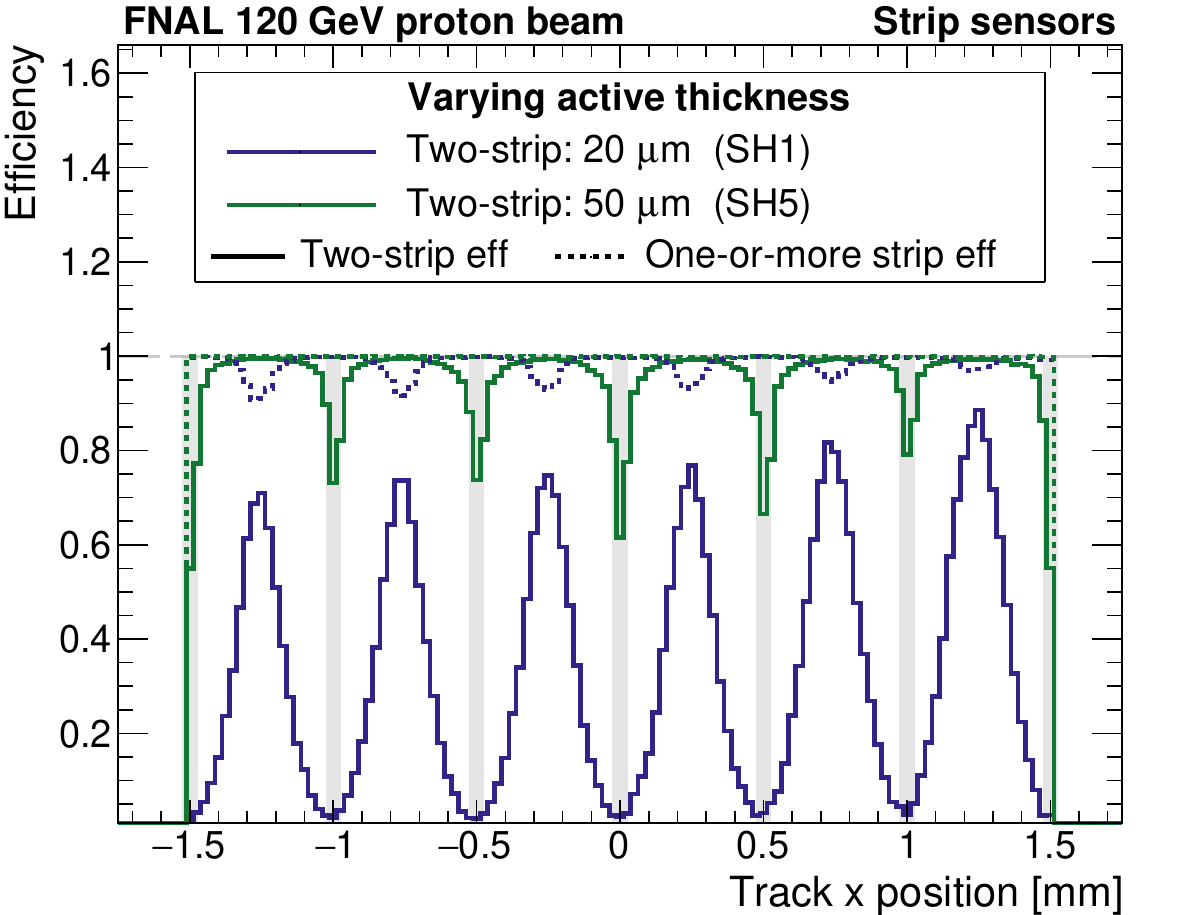}
    \includegraphics[width=0.32\textwidth]{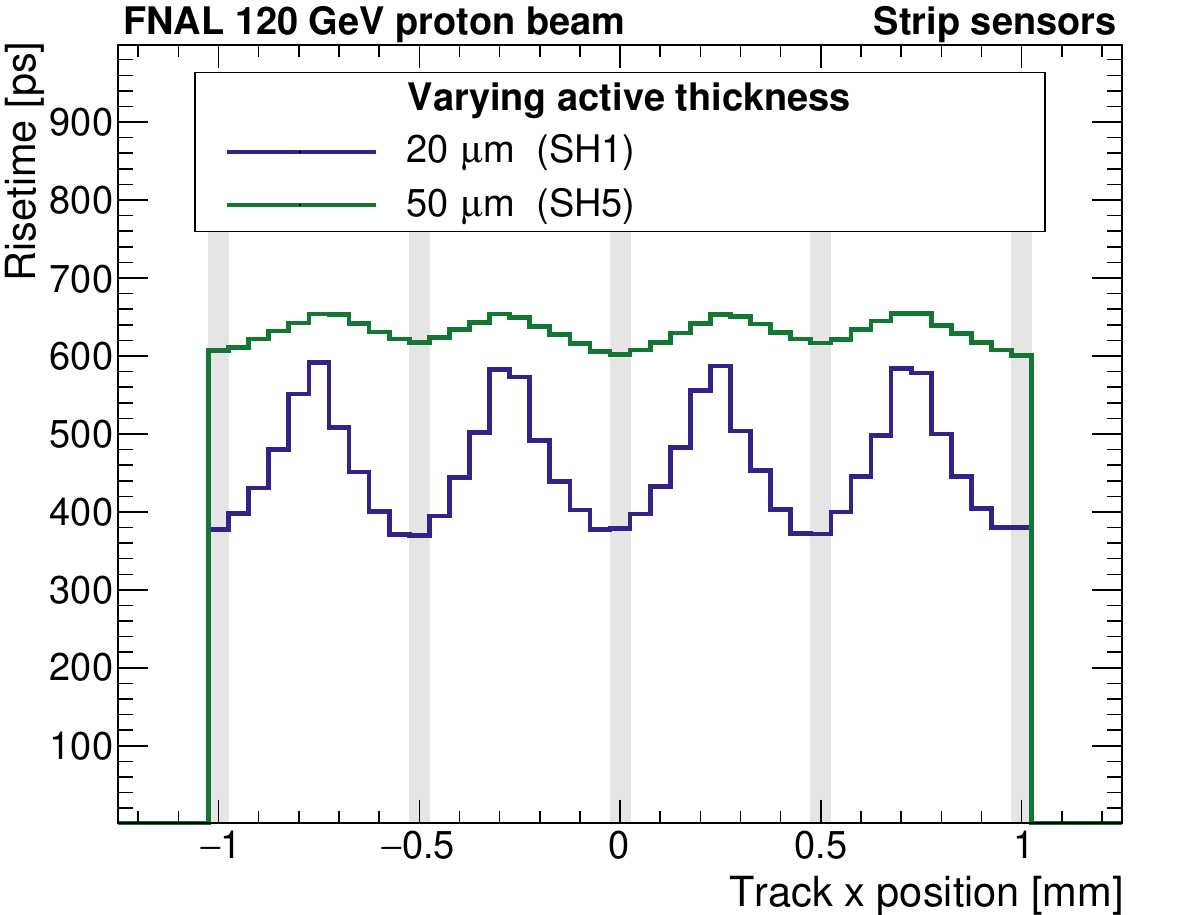}
    \caption{MPV amplitude of leading strip (left),  efficiency (center) and risetime (right) as functions of the $x$ position for strip sensors of different active thicknesses. The sensors presented here are from HPK with a \SI{500}{\um} pitch, \SI{50}{\um} strip width, \SI{1600}{\Omega / \sq} sheet resistance, and \SI{600}{\pico F / \mm^2} coupling capacitance.}
    \label{fig:strip-act_thick-AmpEffRise}
\end{figure}

We note in the left plot of Figure~\ref{fig:strip-act_thick-AmpEffRise} a substantial decrease in the amplitude when moving from \SI{50}{\um} to \SI{20}{\um} active thickness.
As a result and shown in the middle plot of the same figure, the efficiency to find a signal in the subleading channel for the \SI{20}{\um} device decreases substantially and frequently the position interpolation based on charge sharing is not possible.
The signal rise-time is also observed to be dependent on the active thickness: faster signal rise-times are better achieved in thinner sensors, as compared to slower rise-times in thicker sensors, as can be seen from Figure~\ref{fig:strip-act_thick-AmpEffRise}(right). 
Additionally, the small risetime in the thinner sensor enhances the visibility of subtle variations in risetime induced by propagation across the resistive layer, leading to larger risetime from signals originating in the center of the gaps.

\begin{figure}[H]
    \centering
    \includegraphics[width=0.49\textwidth]{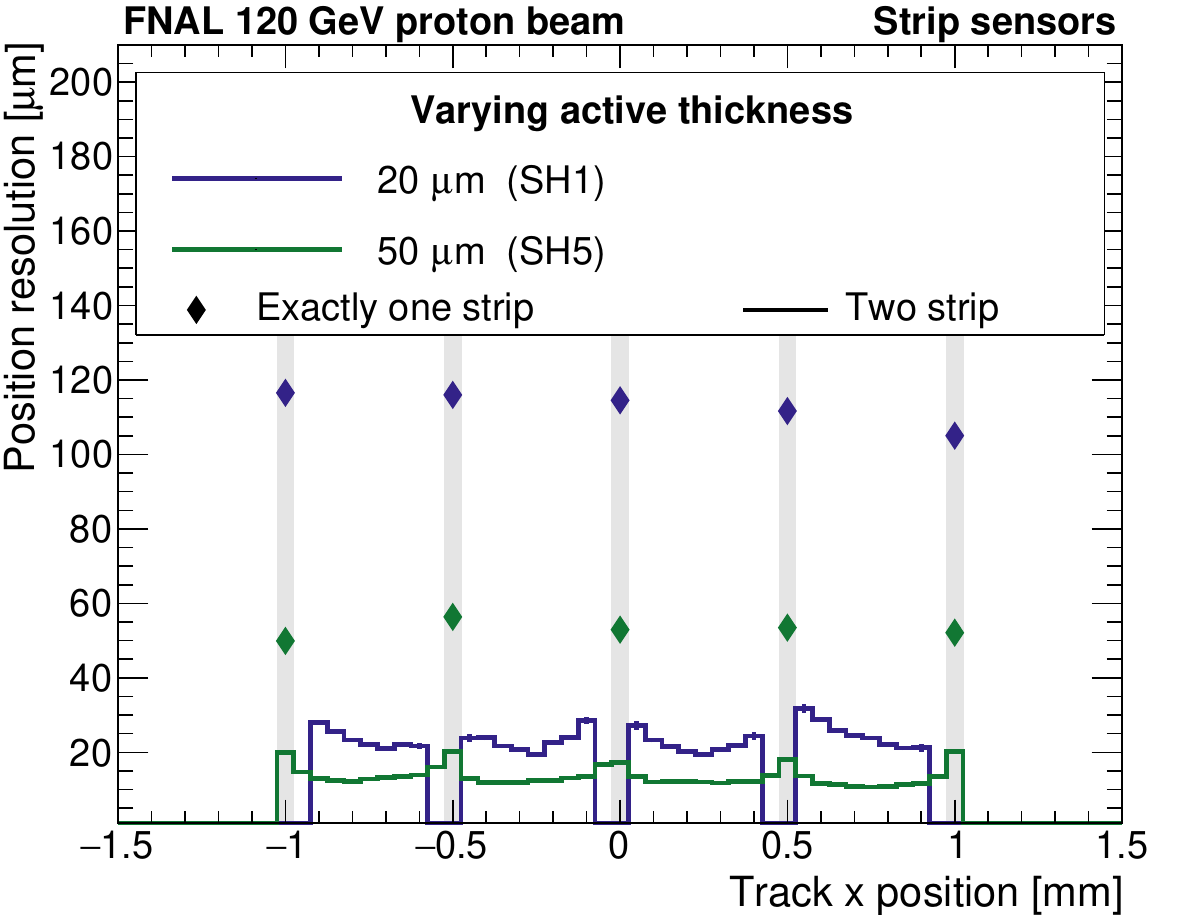}
    \includegraphics[width=0.49\textwidth]{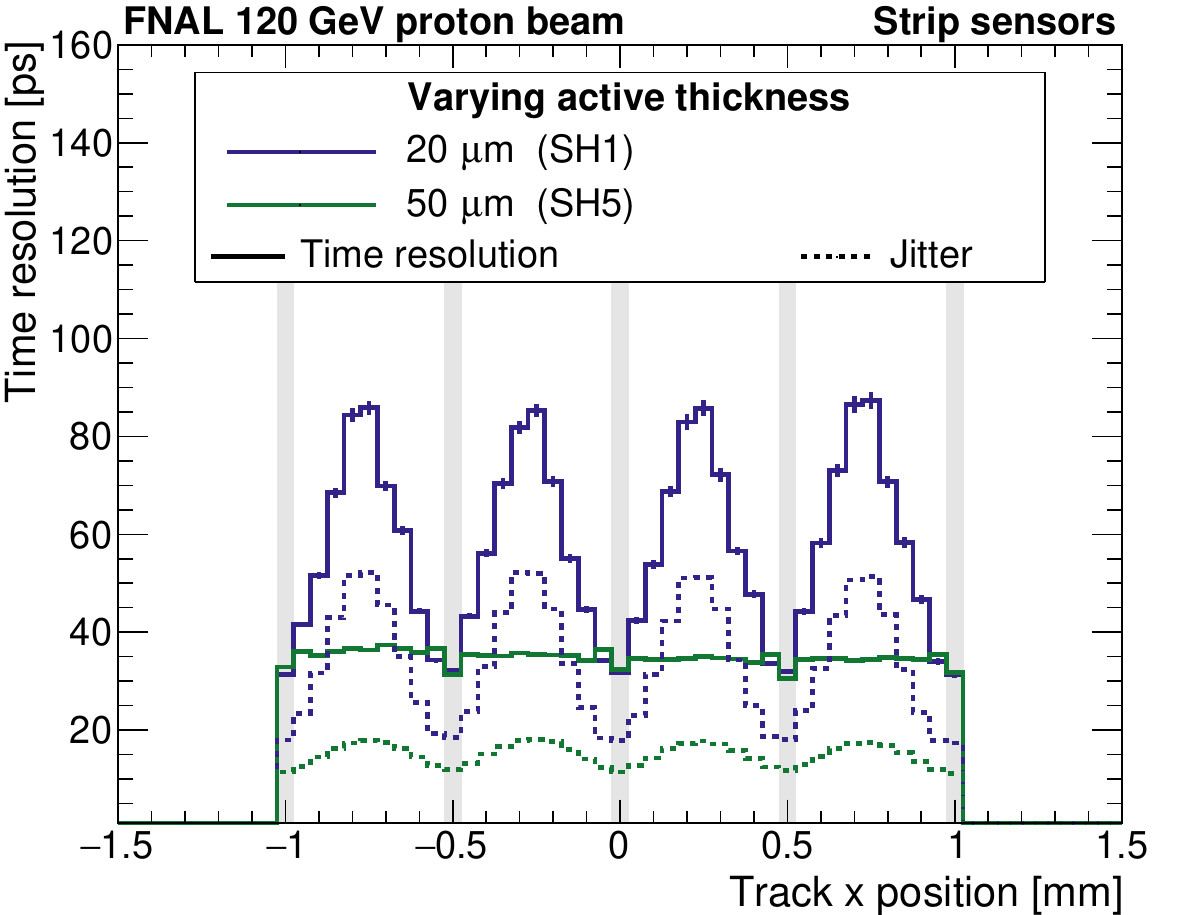}
    \caption{Position (left) and time (right) resolution as functions of $x$ position for strip sensors of different active thicknesses. The sensors presented have \SI{500}{\um} pitch, \SI{50}{\um} strip width, \SI{1600}{\Omega / \sq} sheet resistance, and \SI{600}{\pico F / \mm^2} coupling capacitance. Spatial resolution values have a tracker contribution of \SI{5}{\um} removed in quadrature. Time resolution values have a reference contribution of \SI{10}{\ps} removed in quadrature.}
    \label{fig:strip-act_thick-Resolutions}
\end{figure}

The weighted jitter, defined in~\ref{sec:timereco}, can be used as an estimate of the non-Landau component of the time resolution and in the context of the active thickness study. As visible in Figure~\ref{fig:strip-act_thick-Resolutions}, the jitter component is found to be larger for the thinner sensor due to its dramatically reduced signal amplitudes. 
Thus, despite having a smaller Landau contribution and slightly faster risetime, the 20-\si{\um}-thick strip sensor yields significantly worse time resolution than its 50-\si{\um} counterpart.

We also note here that the 50-\si{\um}-thick sensor has a better overall two-strip reconstruction efficiency and higher signal amplitudes in both the metal strips and gap regions (Figure~\ref{fig:strip-act_thick-AmpEffRise}), leading to a better observed spatial resolution throughout (see Figure~\ref{fig:strip-act_thick-Resolutions}). 
The overall spatial and temporal resolution numbers for the two sensors are quoted in Table~\ref{tab:summary_strips}.

\subsubsection{Small pitch}
\begin{figure}[H]
    \centering
    \includegraphics[width=0.49\textwidth]{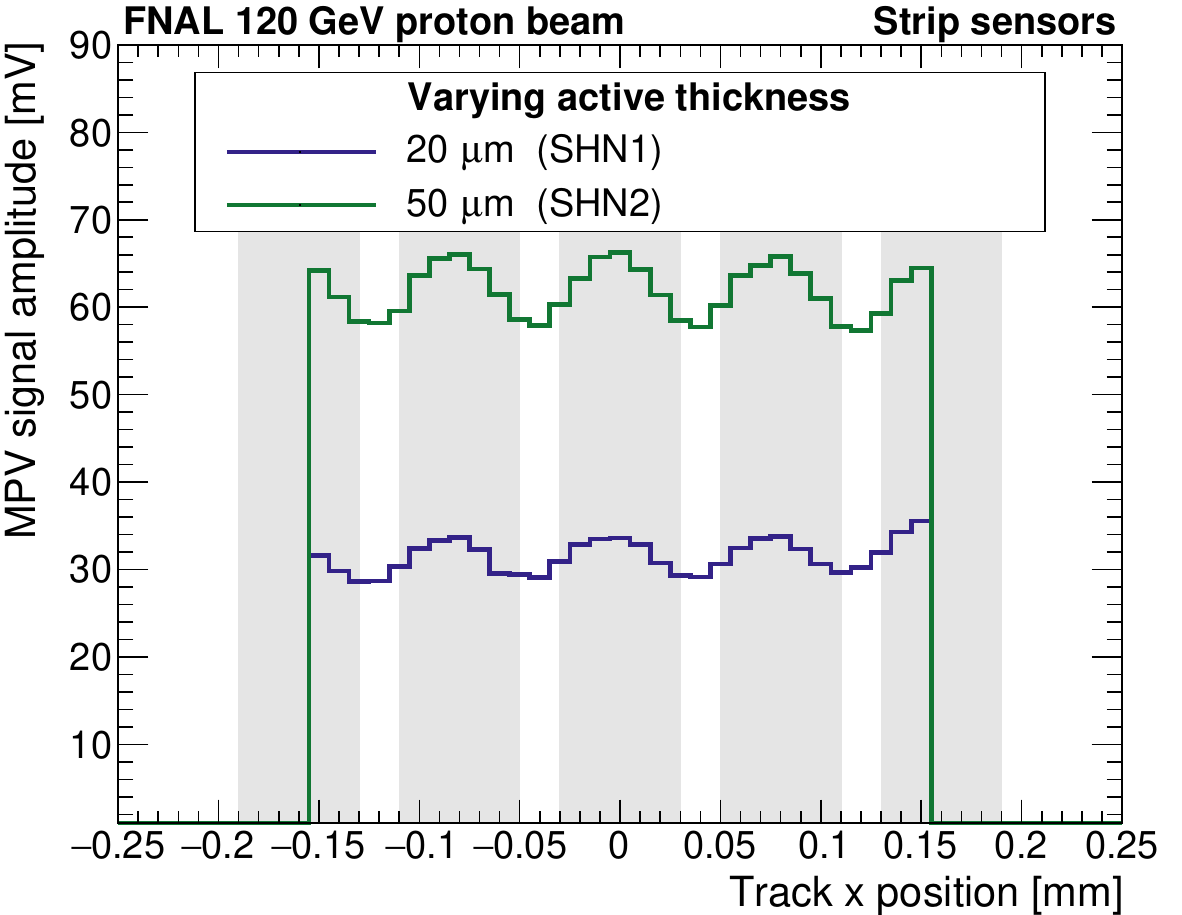}
    \includegraphics[width=0.49\textwidth]{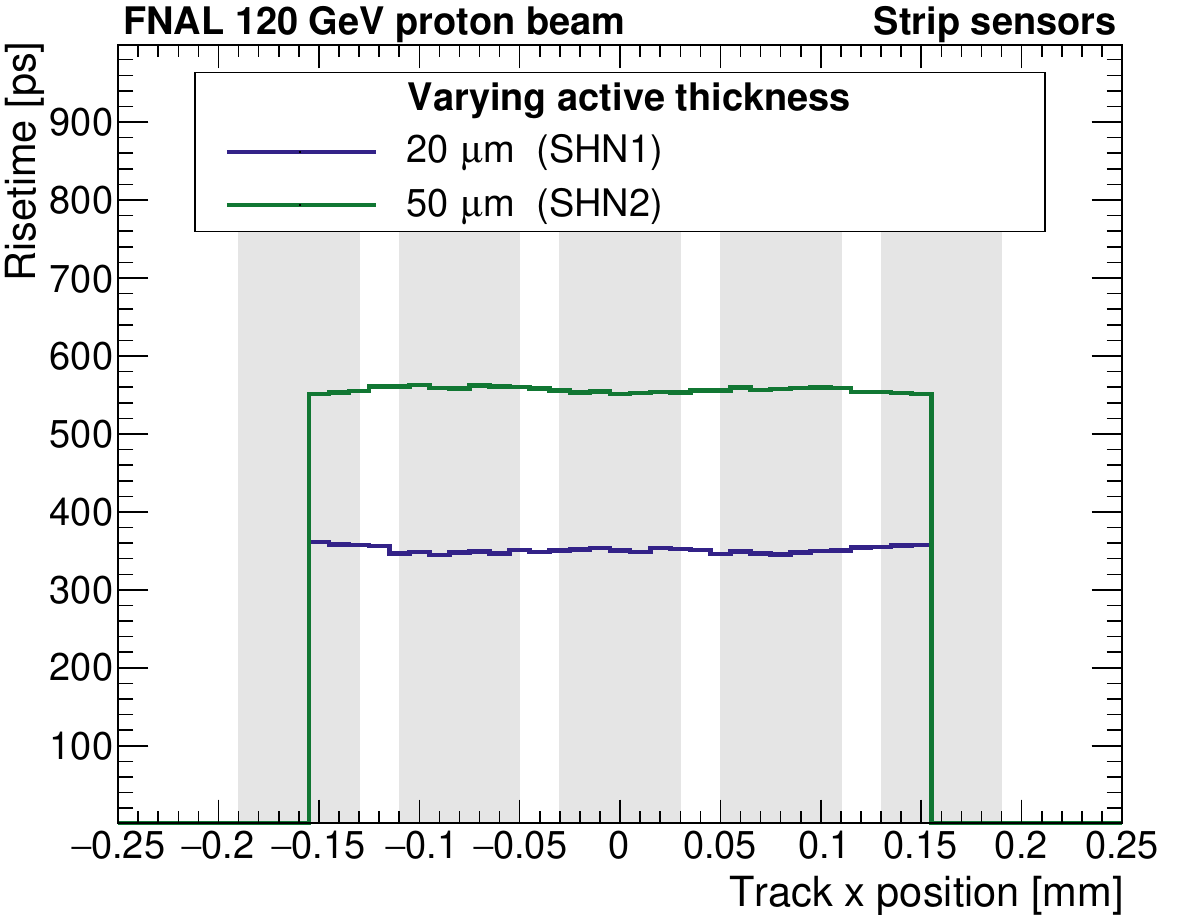}
    \caption{MPV amplitude for the leading strip (left) and risetime (right) as functions of the $x$ position for strip sensors of different active thickness. The sensors presented have \SI{80}{\um} pitch, \SI{60}{\um} strip width, \SI{1600}{\Omega / \sq} sheet resistance, and \SI{240}{\pico F / \mm^2} coupling capacitance.}
    \label{fig:strip-narrow-AmpEffRise}
\end{figure}

\begin{figure}[H]
    \centering
    \includegraphics[width=0.49\textwidth]{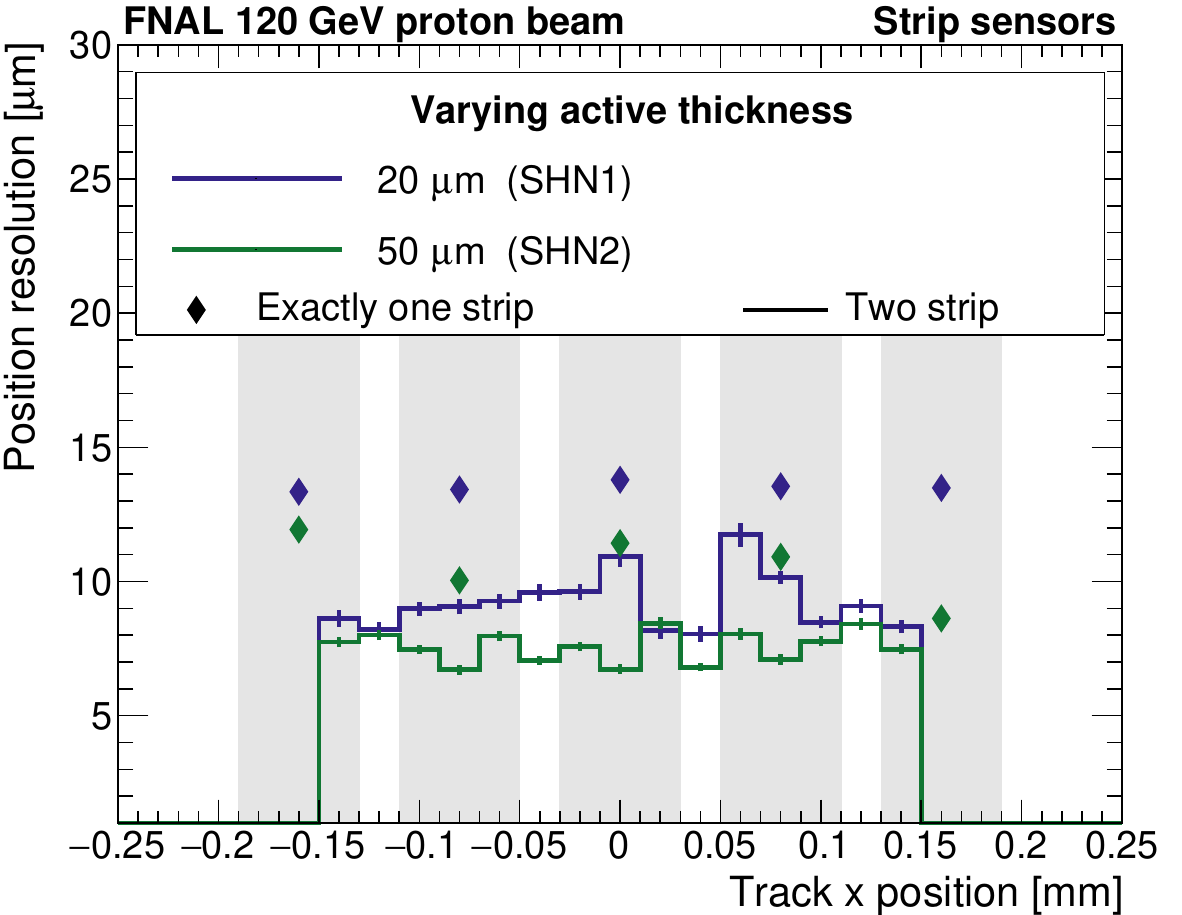}
    \includegraphics[width=0.49\textwidth]{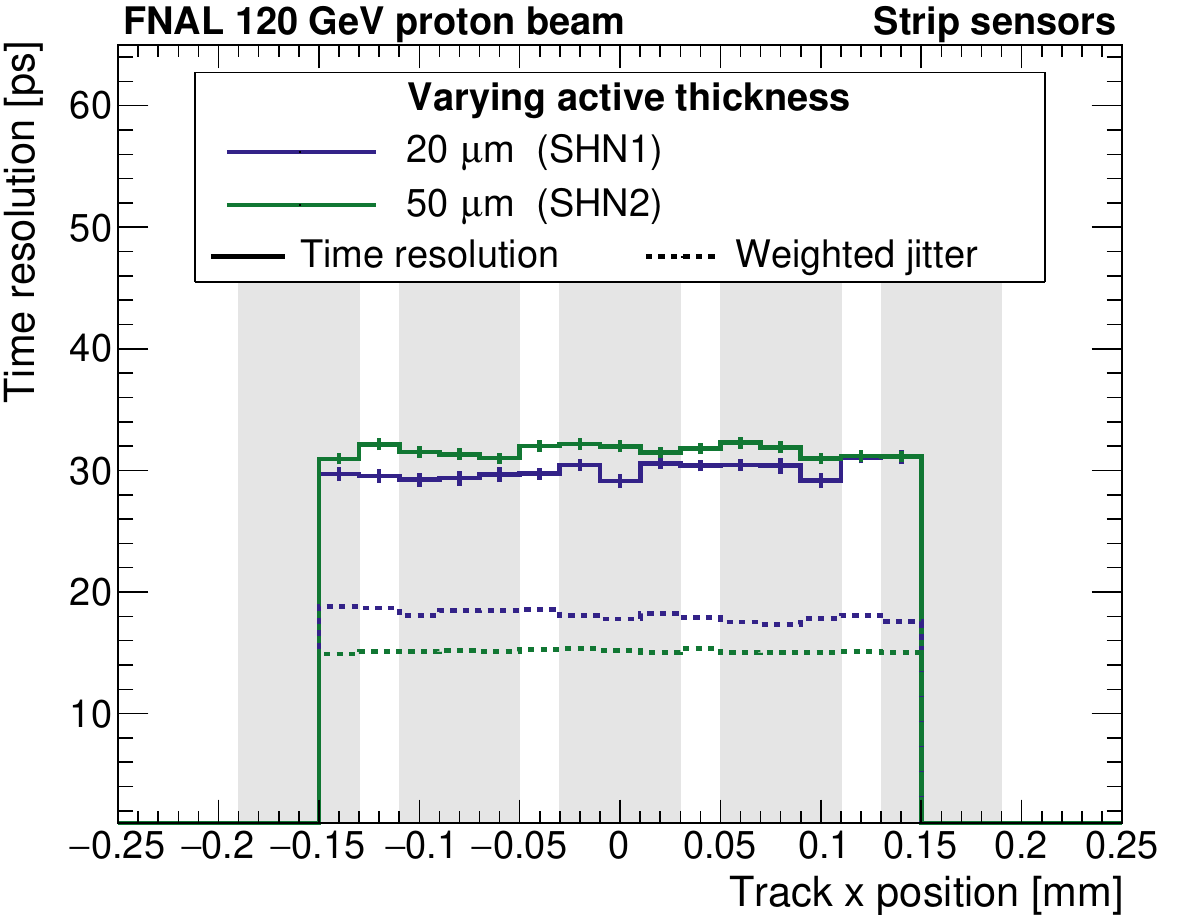}
    \caption{Position (left) and time (right) resolution as functions of the $x$ position for strip sensors of different active thickness. The sensors presented have \SI{80}{\um} pitch, \SI{60}{\um} strip width, \SI{1600}{\Omega / \sq} sheet resistance, and \SI{240}{\pico F / \mm^2} coupling capacitance. The time resolution plot on the right also shows the weighted jitter, which was defined in \ref{sec:timereco}. Spatial resolution values have a tracker contribution of \SI{5}{\um} removed in quadrature. Time resolution values have a reference contribution of \SI{10}{\ps} removed in quadrature.}
    \label{fig:strip-narrow-Resolutions}
\end{figure}

A set of two HPK sensors with a narrow pitch of 80 \si{\um}, with 60 \si{\um}  metal width and varying active thicknesses (20 and 50 \si{\um}) were chosen to study how a narrow pitch can affect the sensor performance. 
These sensors benefit from the more uniform signal amplitude and risetime values throughout the sensor, and the variations in these quantities between the metal strips and gaps are much smaller, as can be seen in Figure~\ref{fig:strip-narrow-AmpEffRise}. 
This results in a more uniform spatial and temporal resolution throughout the sensor, and the narrow pitch sensors perform slightly better (Figure~\ref{fig:strip-narrow-Resolutions}) as compared to their coarser (500-\si{\um})-pitch counterparts (Figure~\ref{fig:strip-act_thick-Resolutions}). 
We note that the 50-\si{\um}-thick sensor has higher signal amplitude and risetime values and lower jitter contribution when compared to the 20-\si{\um}-thick sensor. Considering the larger Landau contribution in the thicker sensor, in this case the total time resolution comes out to be roughly the same for both sensors. 
The overall spatial and temporal resolution numbers for the two sensors are summarized in Table~\ref{tab:summary_strips}.

\subsection{\textbf{HPK pixel sensors}}\label{sec:results-pixels-HPK}

Table~\ref{tab:sensor-info-pixels} summarizes all the HPK pixel sensors tested in this campaign. 
Among them, two types of HPK pixel geometries, the  2x2 pad (Fig.~\ref{fig:hpk_2x2}) and the 4x4 pad (Fig.~\ref{fig:hpk_4x4}) configurations, were studied. 
The performance of these sensors was systematically assessed exploring variations in active thickness, sheet resistance, metal pad width, and bias voltages. 
Section~\ref{sec:thickness_4x4_hpk} discusses the performance of the 4x4 pad geometry sensors. 
The performance of the 2x2 pad sensors is discussed in Section~\ref{sec:2x2_hpk}. 

\begin{figure}[H]
    \centering
    \includegraphics[width=0.49\textwidth]{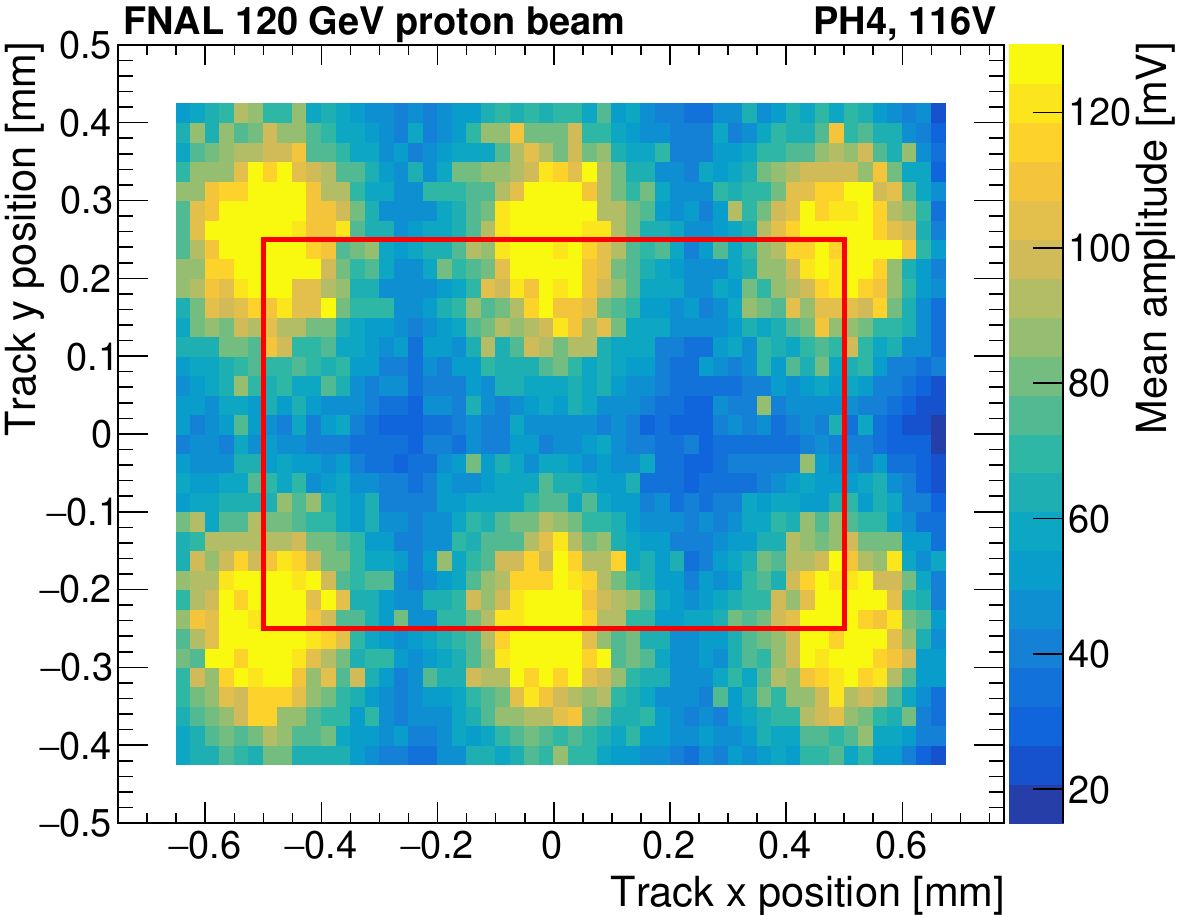}
    \includegraphics[width=0.49\textwidth]{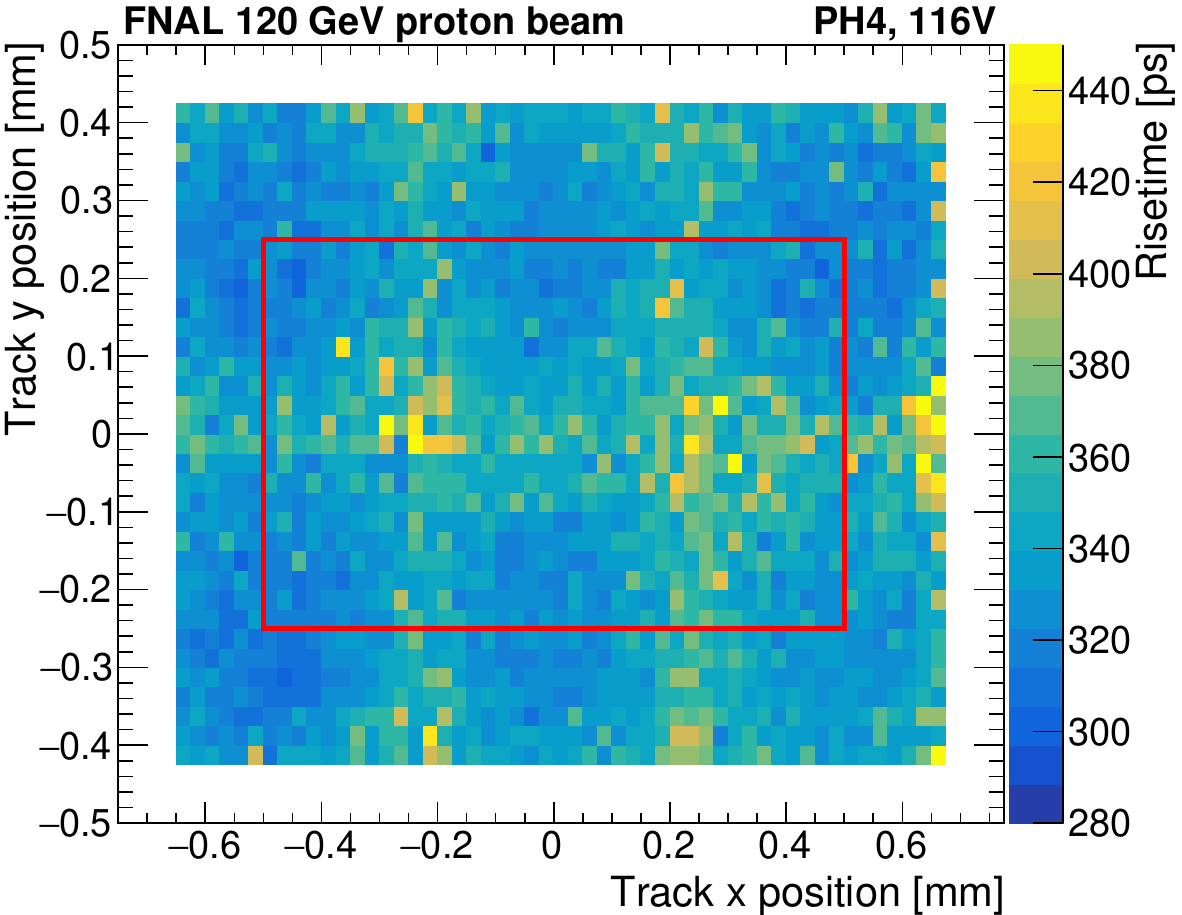}

    \caption{(Left) Mean amplitude map as a function of tracker $x$ and $y$ hit position for HPK-produced 20-\si{\um}-thick sensor with $500\times 500$ \si{\um} pitch and $150$ \si{\um} of metal pad width. This sensor has a sheet resistance of \SI{400} {\si{\Omega / \sq}} and a capacitance of \SI{600}{\pico F / \mm^2}. (Right) Risetime map as a function of tracker $x$ and $y$ hit position for the same pixel sensor. The area bounded by red lines is considered the region of interest for studying the performance of this sensor.
    }
    \label{fig:pixelHPK-Amp}
\end{figure}

Due to constraints in the multi-channel readout capabilities, we were only able to read out six neighboring metal pads for any given 4x4 pixel sensor. All other channels are grounded. As an example, we refer to some results of the PH4 sensor's response to explain the characteristics of the pixel analysis. Figure~\ref{fig:pixelHPK-Amp} (left) shows the map of maximum amplitude among the six connected channels for sensor \textit{PH4} and reveals a distinct signal pattern.
Signals originating under the metal pads are generally higher, in contrast to the gap regions where signal sizes are lower.
This spatial variation provides insights into other sensor response characteristics, such as the signal risetime, as shown in the right plot of Figure~\ref{fig:pixelHPK-Amp}.
Metal pads exhibit a slightly faster response, suggesting a lower jitter contribution to the time resolution, while gap regions suffer from a larger jitter due to smaller signal sizes.
For every 4x4 pixel sensor, we define a region of interest, as shown by the red rectangular box connecting the six metal pad centers. This selection is designed to exclude effects arising from non-read or missing channels at the periphery. The behavior in this region better represents the performance expected for the interior channels within a large pixel array, as would be used in a full-scale detector.

\begin{figure}[H]
    \centering
    
    \includegraphics[width=0.24\textwidth]{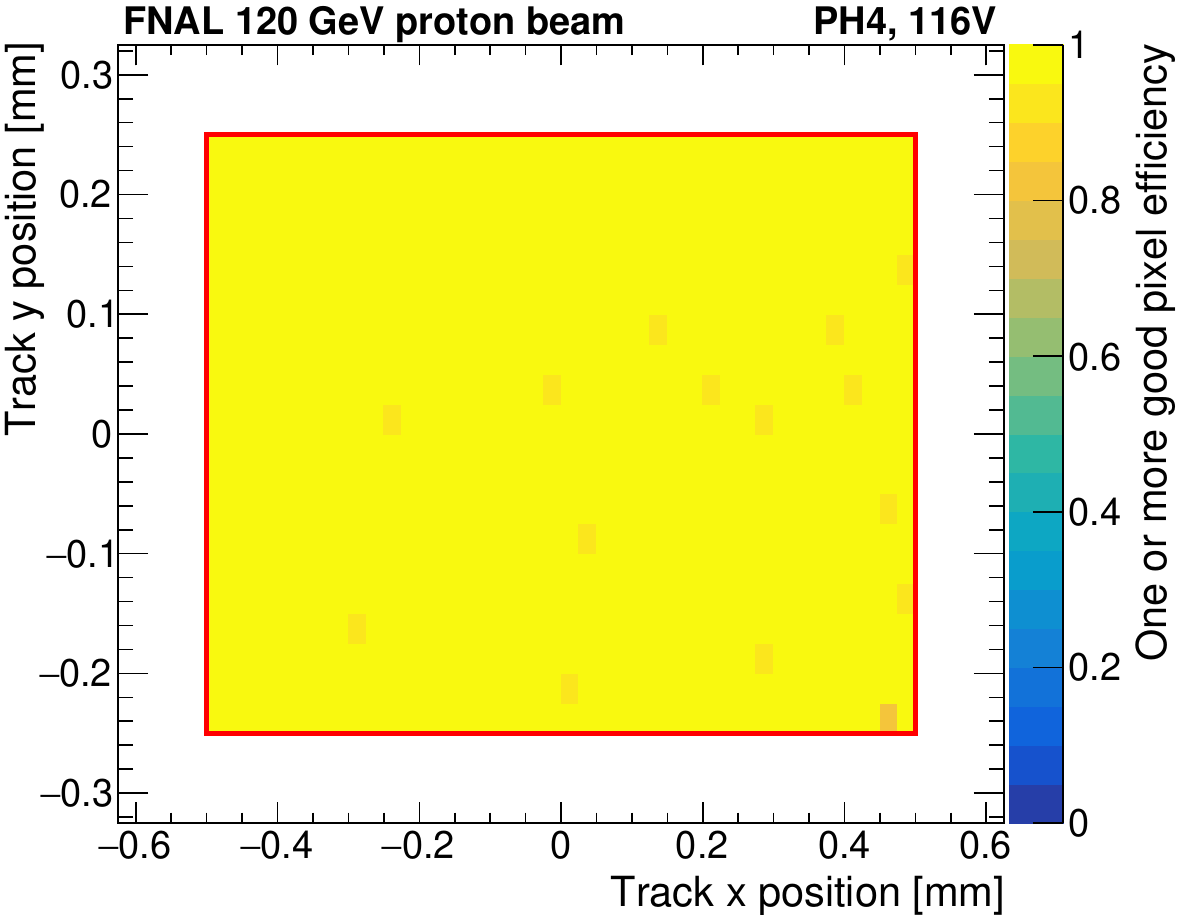}
    \includegraphics[width=0.24\textwidth]{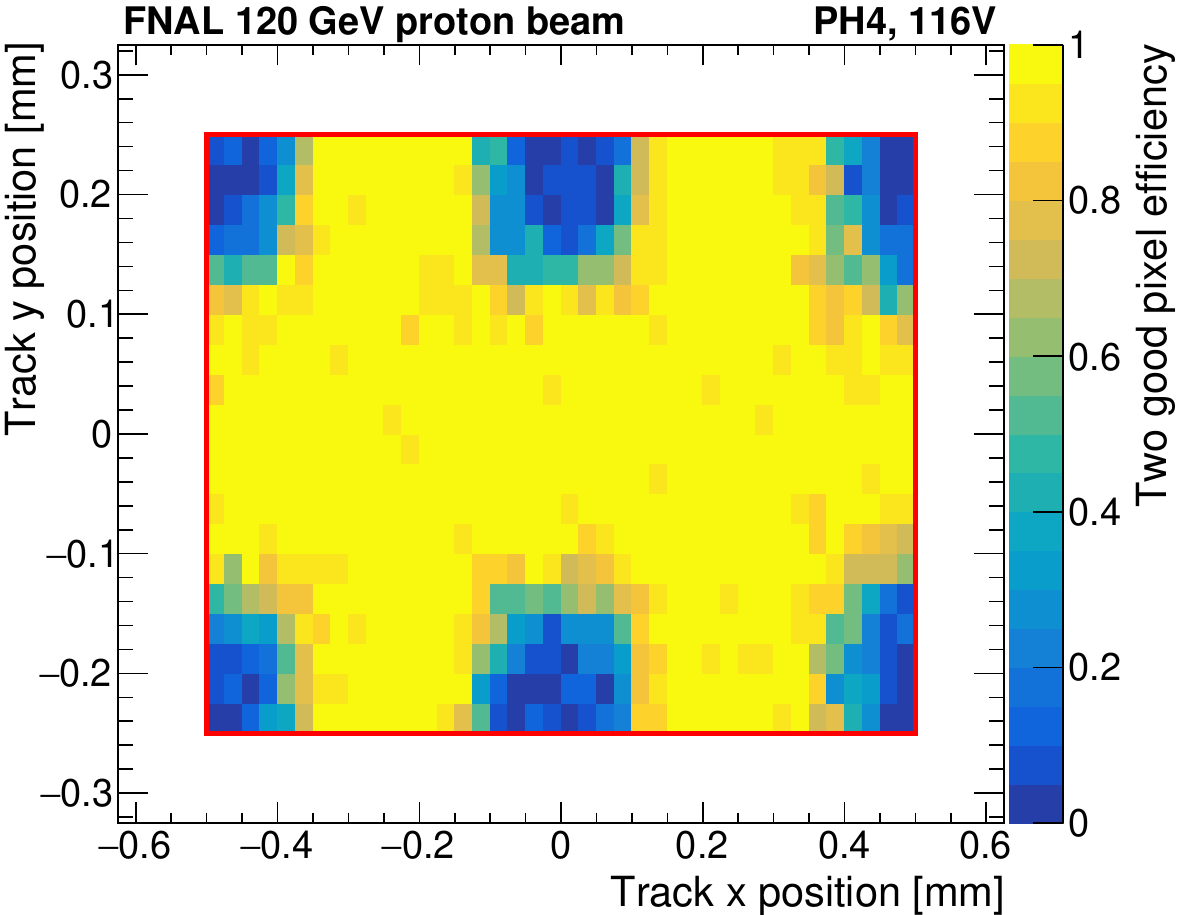}
    \includegraphics[width=0.24\textwidth]{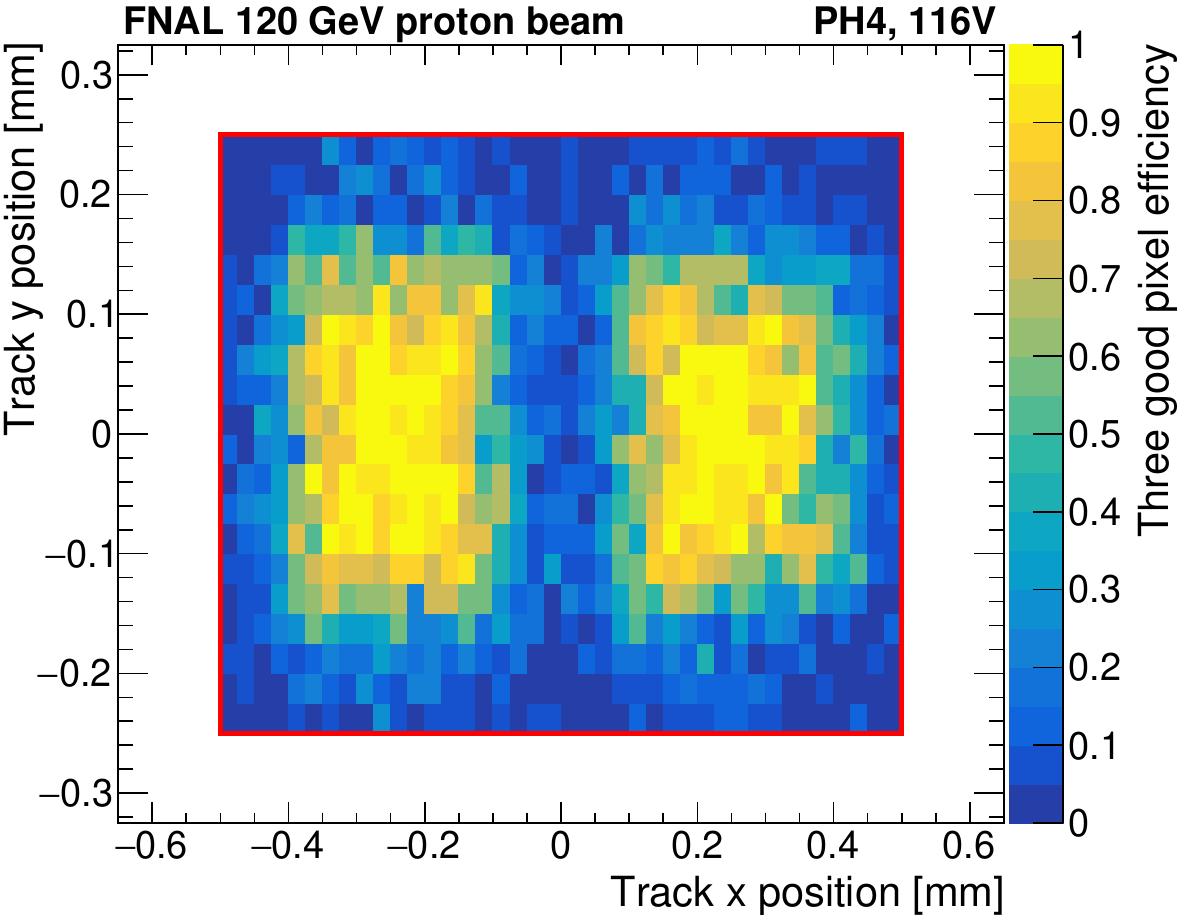}
    \includegraphics[width=0.24\textwidth]{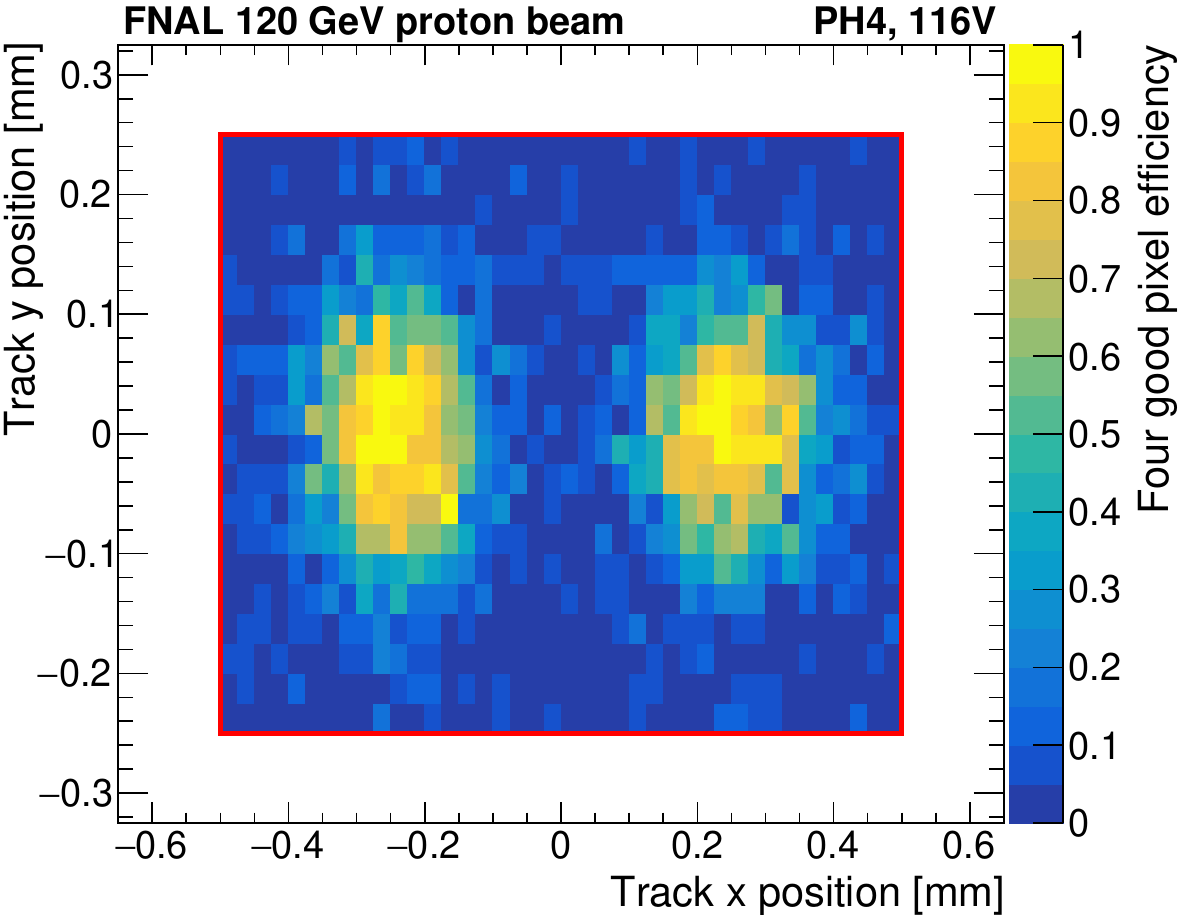}
    \caption{Efficiency maps as a function of tracker $x$ and $y$ hit position for the HPK 20-\si{\um}-thick sensor with $500\times 500$ \si{\um} pitch and $150$ \si{\um} of metal pad width. This sensor has a sheet resistance of \SI{400} {\si{\Omega / \sq}} and a capacitance of \SI{600}{\pico F / \mm^2}. (Left most) One-or-more good pixel efficiency map. (Second from left) Two good pixel efficiency map. (Third from left) Three good pixel efficiency map. (Right) Four good pixel efficiency map.
    }
    \label{fig:pixelHPK-Eff}
\end{figure}

The two-good-pixel efficiency shown in Figure~\ref{fig:pixelHPK-Eff} (second-from-left) indicates where the two-or-more channel position reconstruction method can be applied, as described in Section~\ref{sec:reco}.

The 4$\times$4 pixel sensors are a 100\% efficient for reconstructing any proton hits across the surface, as shown in Fig.~\ref{fig:pixelHPK-Eff} (left). There is a significant drop in two-hit efficiency near the metal pads, leaving the one-channel reconstruction as the unique possibility.
Note that the overall two-channel reconstruction efficiency decreases with an increasing metal pad width. The probability of reconstructing a hit seen by three (Fig.~\ref{fig:pixelHPK-Eff} third-from-left) or four (Fig.~\ref{fig:pixelHPK-Eff} right) surrounding pixels becomes increasingly small and is restricted to the gap areas surrounded by three or more pixels. 

\subsubsection{Active thickness and sheet resistance}\label{sec:thickness_4x4_hpk}
\begin{figure}[H]
    \centering
    \includegraphics[width=0.32\textwidth]{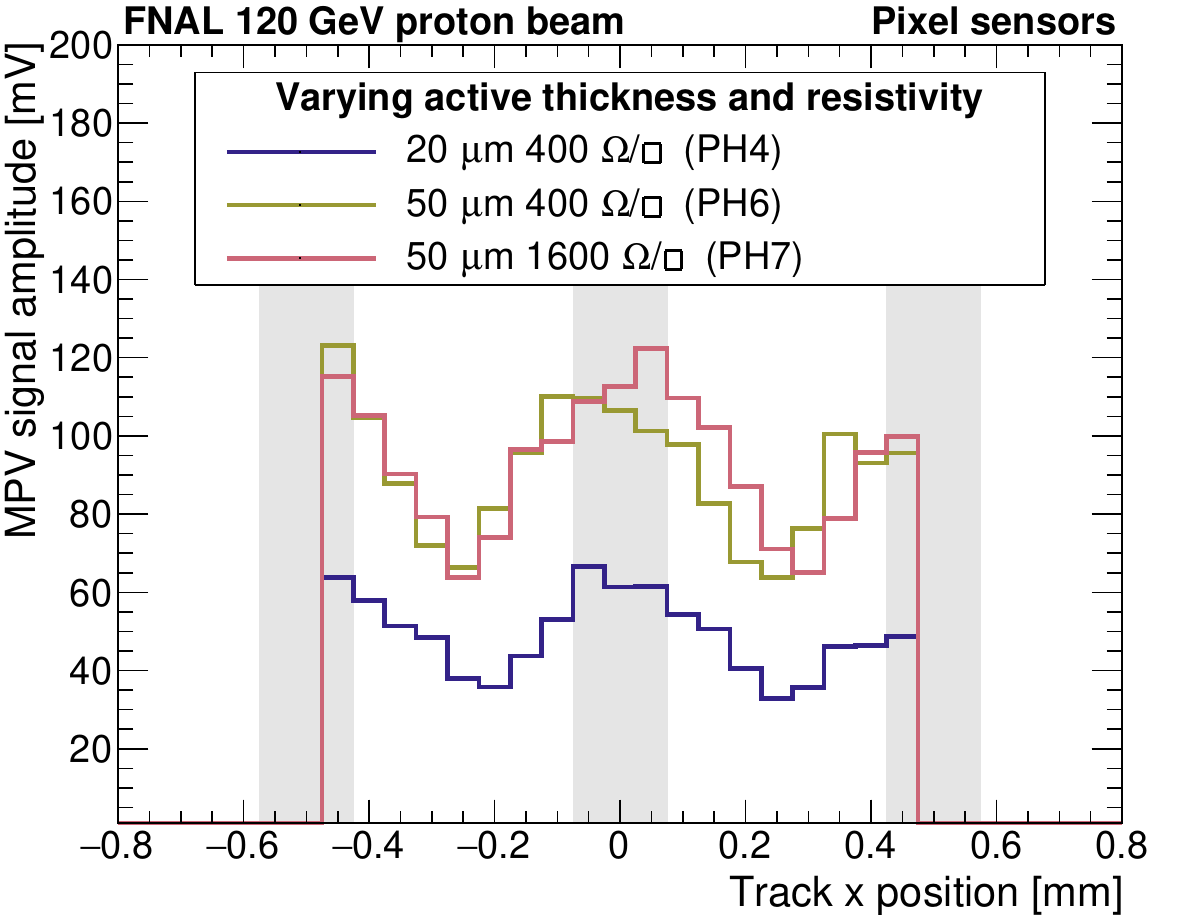}
    \includegraphics[width=0.32\textwidth]{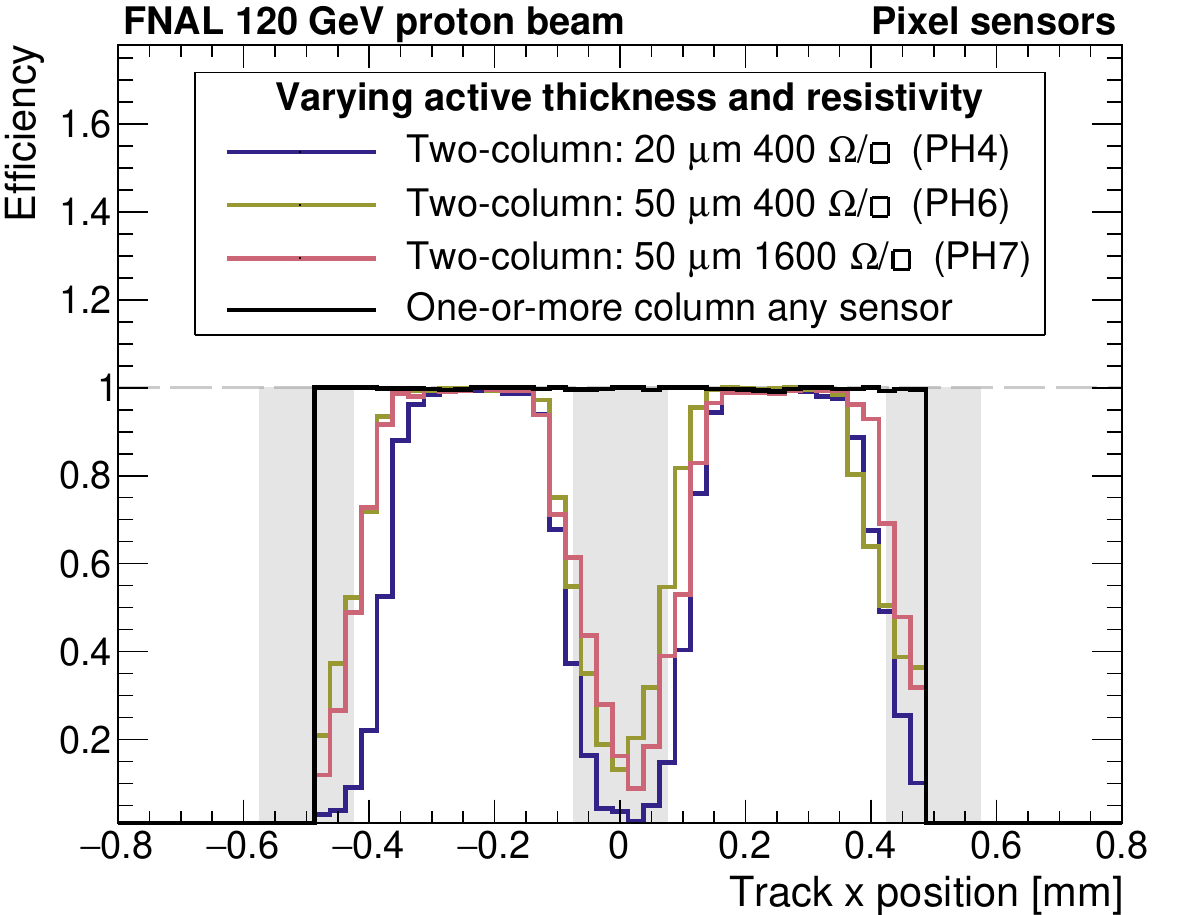}
    \includegraphics[width=0.32\textwidth]{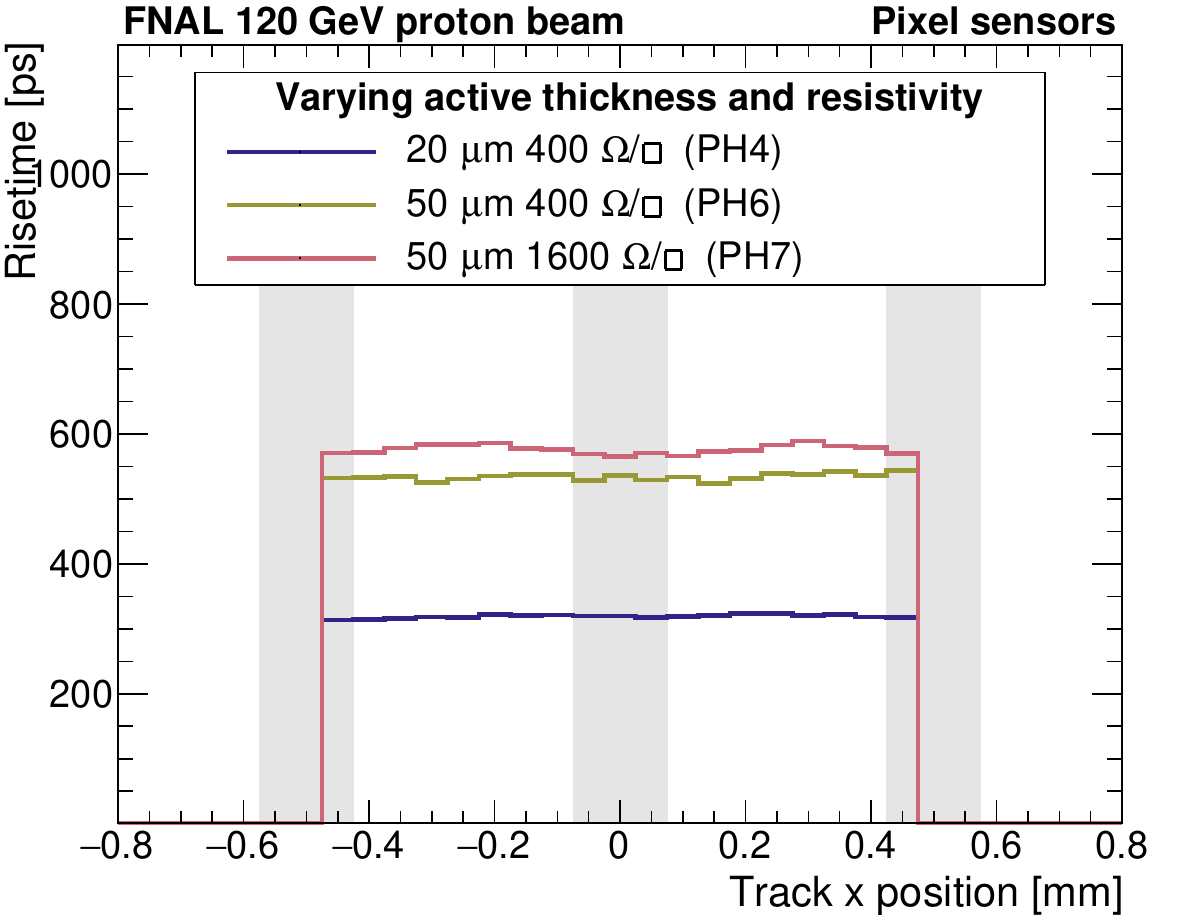}
    \caption{MPV amplitude (left), one/two-column efficiency (center), and signal rise time (right) per column as a function of tracker $x$ position for pixel sensors of different active thicknesses and sheet resistance. The sensors presented have \SI{500}{\um} pitch, \SI{150}{\um} metal width, and \SI{600}{\pico F / \mm^2} coupling capacitance.}
    \label{fig:pixelHPK-act_thick_and_res-AmpEffRise}
\end{figure}

The performance of 4x4 pixel sensors was studied by varying the active thickness of the sensor and the sheet resistance of the $n^+$ layer while keeping the capacitance of the dielectric layer constant. 
All other geometric parameters, such as pitch and metal width, were kept constant. 
The MPV signal size per column (defined in Sec~\ref{sec:posreco}) for a 50 \si{\um} sensor is nearly double that of a sensor with 20 \si{\um} thickness as shown in Figure~\ref{fig:pixelHPK-act_thick_and_res-AmpEffRise} (left). 
Relative to the case of strip sensors  (Sec~\ref{sec:strip_res_cc}), the increased $n^{+}$ sheet resistance has a generally smaller impact on the primary signal amplitude in pixel sensors. This can be understood considering the radial propagation of signals that originate in gaps between channels. In the case of long strips, some portion of the signal propagates at oblique angles, traveling several \si{\milli\meter} through the resistive layer before reaching an electrode. For pixels, the signals that arrive in the leading several channels can only propagate up to a few hundred \si{\um} and therefore experience significantly less resistive attenuation.

The two-column efficiency is close to unity in the gaps and close to zero at the center of the metal pads as shown in Figure~\ref{fig:pixelHPK-act_thick_and_res-AmpEffRise} (center). 
There is no significant difference in efficiency among sensors with different sheet resistances, with only a slight variation observed between sensors with two different thicknesses, attributed to variations in signal sizes depending on the active thickness. 
A notable difference in risetime, nearly two times faster for the 20-\si{\um}-thick sensors, is observed in Figure~\ref{fig:pixelHPK-act_thick_and_res-AmpEffRise} (right).
A relatively uniform risetime is observed across the surface for both the 50-\si{\um}-thick and 20-\si{\um}-thick sensors with the 400 \si{\Omega / \sq} sheet resistance. 
A uniform overall difference in risetime ($\sim$ 50 ps) is observed when comparing the two 50-\si{\um}-thick sensors, with the signal being slower for the sensor with the higher sheet resistance. \par



\begin{figure}[H]
    \centering
    \includegraphics[width=0.49\textwidth]{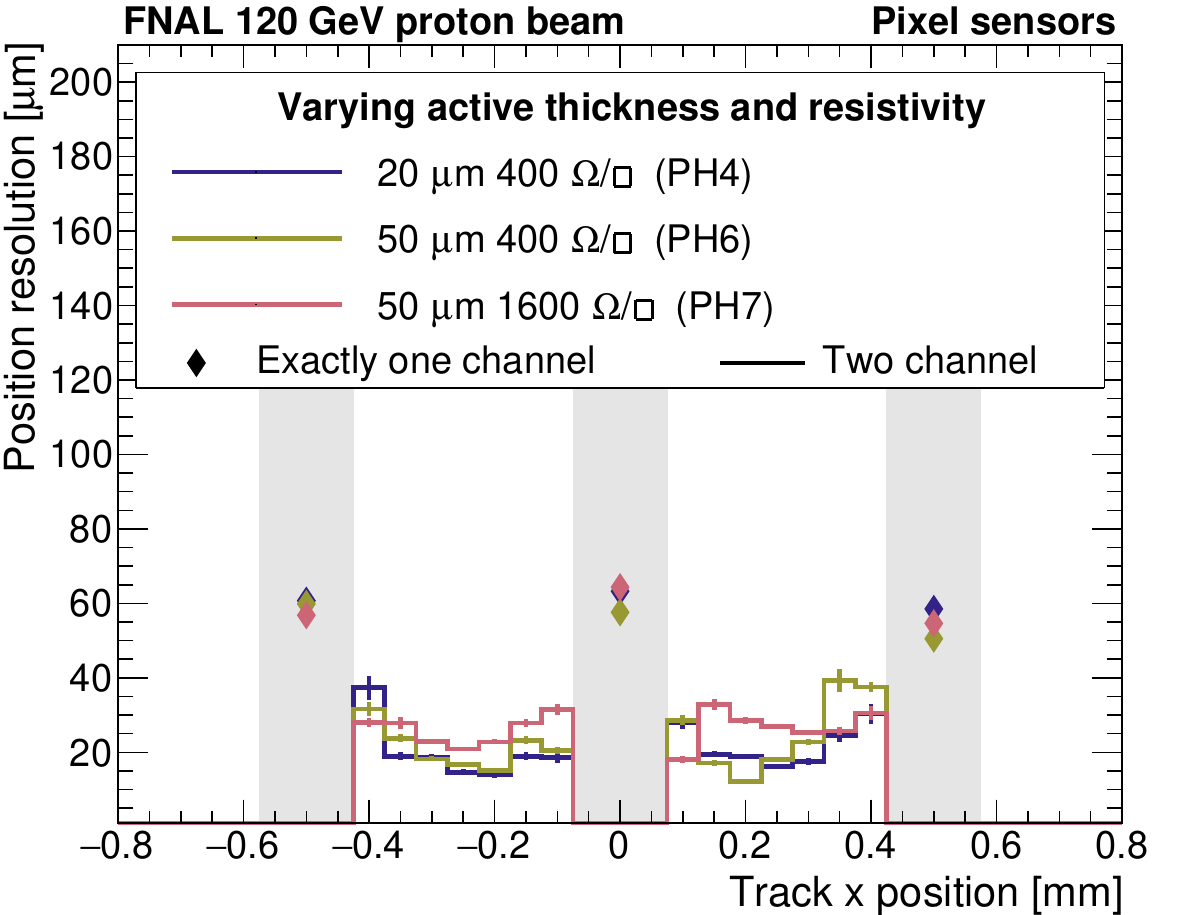}
    \includegraphics[width=0.49\textwidth]{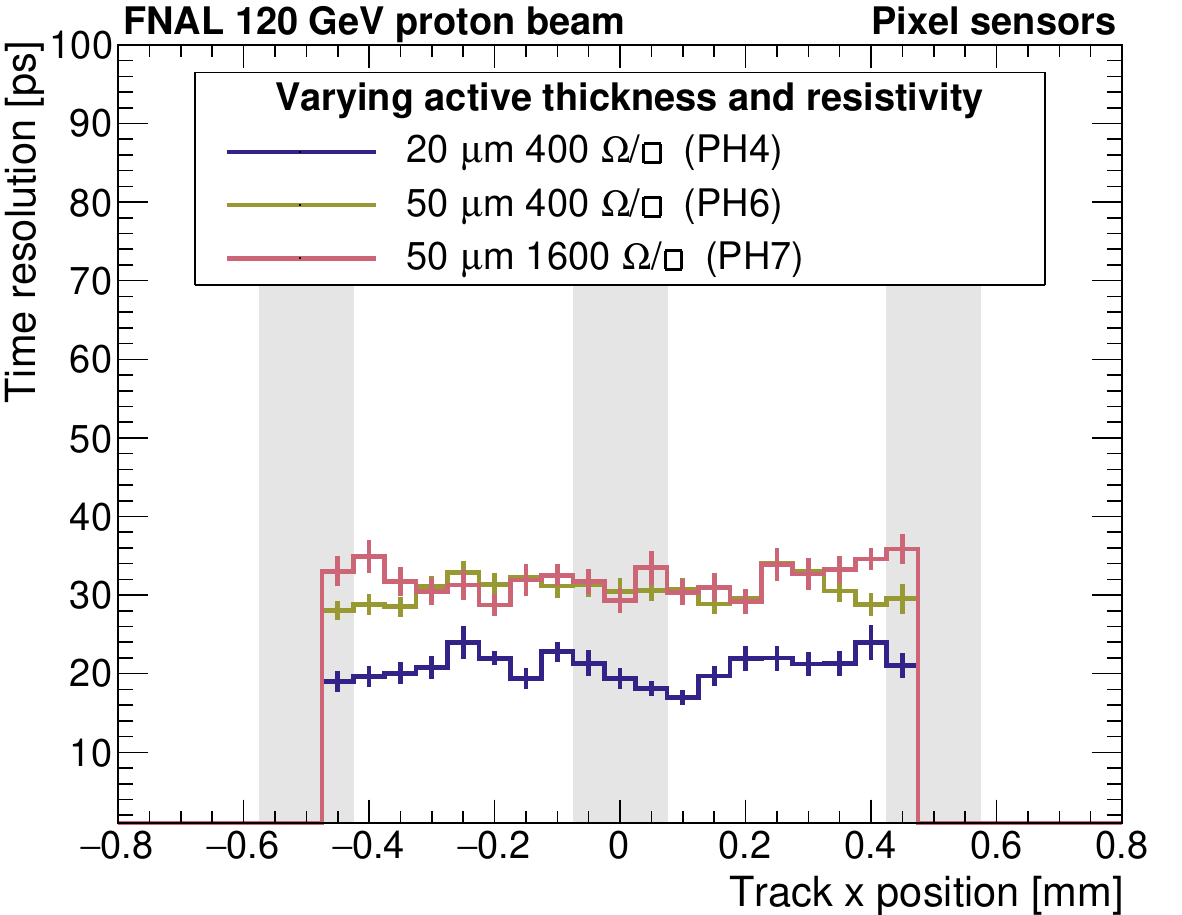}
    \caption{Position (left) and time (right) resolution as a function of tracker $x$ position for pixel sensors of different active thicknesses and sheet resistance. The sensors presented have \SI{500}{\um} pitch, \SI{150}{\um} metal width, and \SI{600}{\pico F / \mm^2} coupling capacitance. The spatial resolution values have a reference contribution of \SI{5}{\um} from the tracker removed in quadrature. The time resolution values have a reference contribution of \SI{10}{\ps} from the MCP removed in quadrature.}
    \label{fig:pixelHPK-act_thick_and_res-Resolutions}
\end{figure}

Figure~\ref{fig:pixelHPK-act_thick_and_res-Resolutions} (left) displays the reconstructed spatial resolution from two-column reconstruction in the gap and one-column reconstruction in the metal regions. 
The two-column resolution for all sensors is approximately 20 \si{\um} with small variations, with the best values obtained at the gap centers. 
However, the one-column resolution characteristic of the metal pad centers ranges from 50 - 80 \si{\um}. 

The reconstructed time resolution of the 4x4 pixel sensors is shown in Figure~\ref{fig:pixelHPK-act_thick_and_res-Resolutions} (right).  
A nearly uniform time resolution of approximately 30 ps is observed across the surface for 50-\si{\um}-thick sensors, with no substantial variations observed for varying $n^{+}$ sheet resistance. 
For the 20-\si{\um}-thick sensor, the time resolution is fairly uniform around 20~ps.
The result for the \SI{20}{\um} device is a strong indication that for pixelated devices decreasing the active thickness is an effective means of improving the time resolution.

\subsubsection{Bias scan}\label{sec:2x2_hpk}
A crucial aspect for understanding the sensor timing performance is the dependence on the operating bias voltage. 
We performed a bias voltage scan for three HPK 2$\times$2 pixel sensors with \SI{510}{\um} pitch, \SI{500}{\um} metal width, \SI{1600}{\ohm / \sq} sheet resistance, \SI{600}{\pico F / \mm^2} coupling capacitance and different active thicknesses (20, 30 and 50 \SI{}{\um}). 
As these sensors are nearly fully metalized, there is  little signal sharing and the multi-channel spatial reconstruction is frequently not possible. 
As result, we will restrict the discussion in this section to only the timing performance of these sensors.

\begin{figure}[H]
    \centering
    \includegraphics[width=0.49\textwidth]{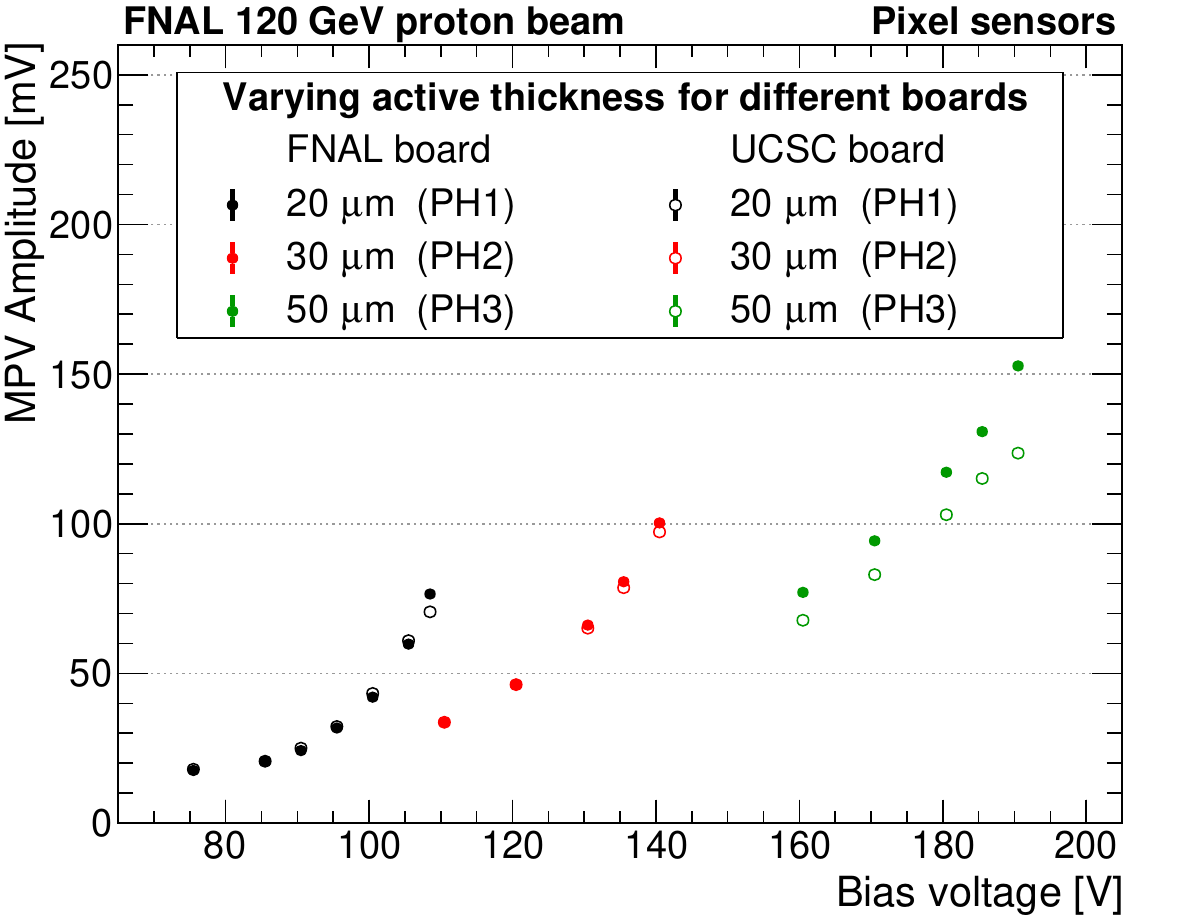}
    \includegraphics[width=0.49\textwidth]{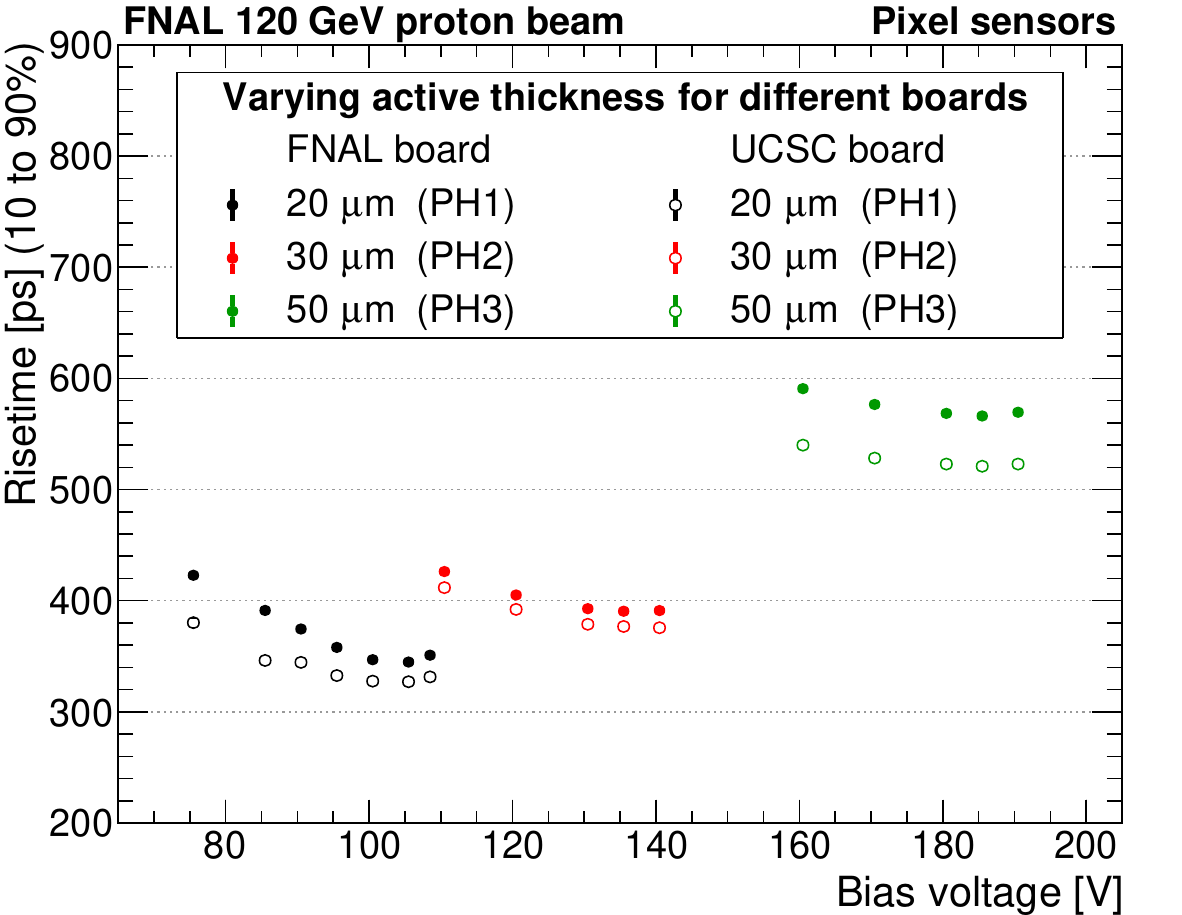}

    \caption{The observed MPV amplitude (left) and risetime values (right) for HPK 2$\times$2-pixel sensors with varying active thicknesses at different bias voltages. The sensors presented have \SI{510} {\um} pitch, \SI{500} {\um} metal width, \SI{1600} {\Omega / \sq} sheet resistance, and \SI{600} {\pico F / \mm^2} coupling capacitance. The UCSC boards allow for only single-channel readout, and thus all results shown here are from a single channel only. }
    \label{fig:pixelHPK-bias_scan1}
\end{figure}

\begin{figure}[H]
    \centering
    \includegraphics[width=0.49\textwidth]{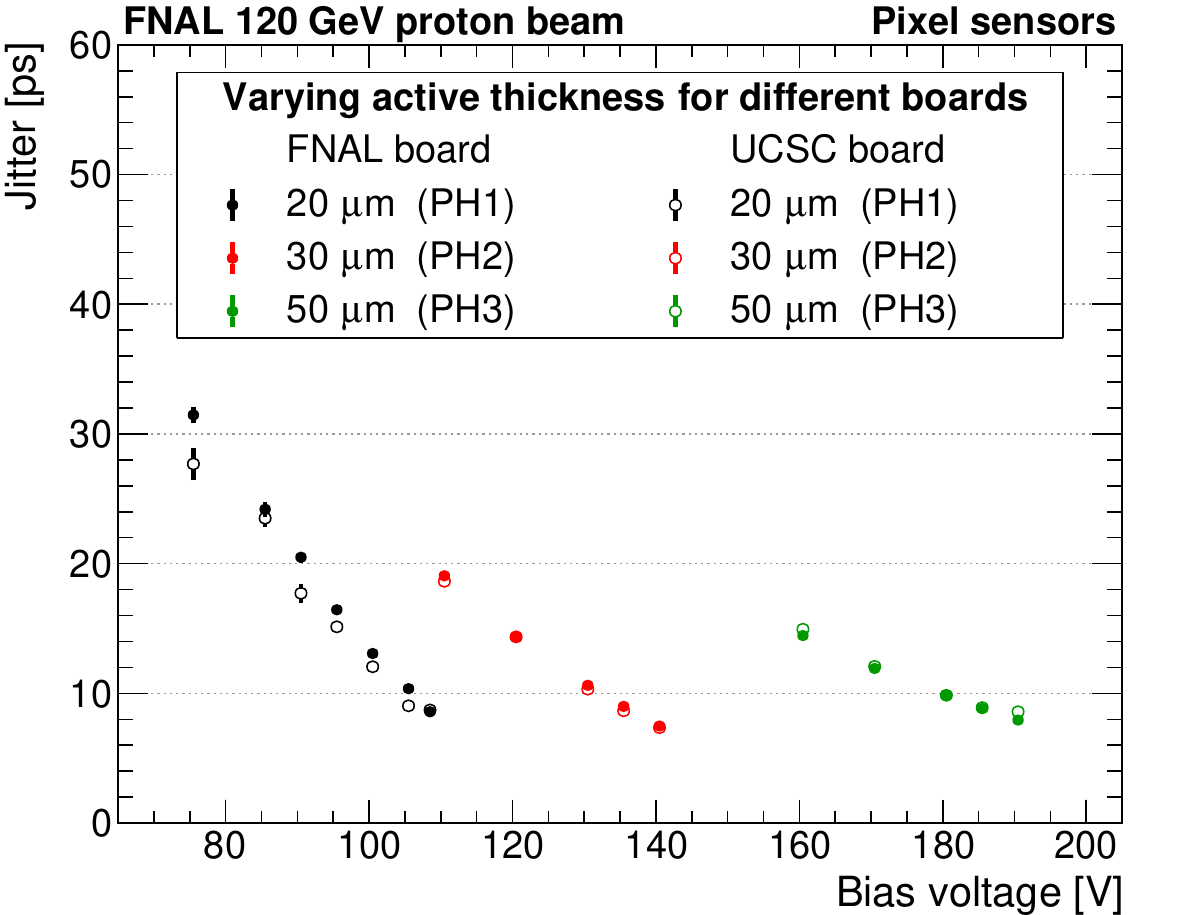}
    \includegraphics[width=0.49\textwidth]{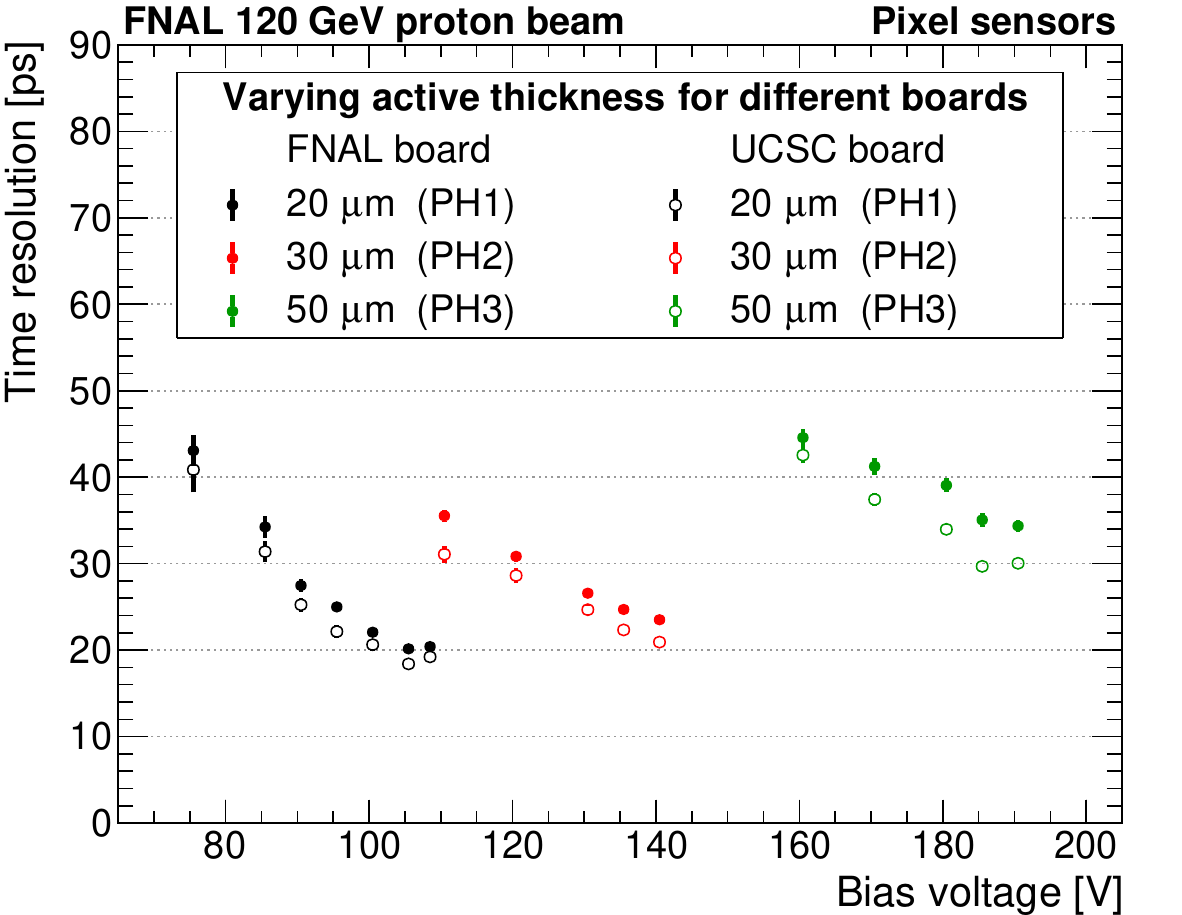}

    \caption{The observed jitter (left) and time resolution values (right) for HPK 2$\times$2-pixel sensors with varying active thicknesses at different bias voltages. The sensors presented have \SI{510}{\um} pitch, \SI{500}{\um} metal width, \SI{1600}{\Omega / \sq} sheet resistance, and \SI{600}{\pico F / \mm^2} coupling capacitance. The UCSC boards allow for only single-channel read-out, and thus all results shown here are from a single channel only. }
    \label{fig:pixelHPK-bias_scan2} 
\end{figure}

To compare performances in different boards, these sensors were moved between both the Fermilab 16-channel readout board and the single-channel UCSC readout board. The results obtained from both boards are displayed in Figures~\ref{fig:pixelHPK-bias_scan1} and \ref{fig:pixelHPK-bias_scan2}. 
The signal amplitude increases monotonically with the sensor thickness and with the applied bias voltages as illustrated in Figure~\ref{fig:pixelHPK-bias_scan1} (left). 
We see similar amplitude values for each sensor on both boards, except for the \SI{50} {\um} sensor, where we observed higher amplitude values using the Fermilab board. 
As these sensors are predominantly metalized, the amplitude variation between the metal and gap regions was not displayed separately. 

The operational bias voltage ranges vary based on the active thickness of the pixel sensors and increase with increasing active thickness. 
In this study it is ensured that the maximum operating voltage (optimal bias) for each sensor is maintained a few Volts below the breakdown threshold, as listed in Table~\ref{tab:sensor-info-pixels}. 
The signal risetime variations are shown in Figure~\ref{fig:pixelHPK-bias_scan1} (right) and decrease with increasing operating voltage. 
At the optimal bias voltage, the 20-\si{\um}-thick sensor exhibits faster signal rise-times than the other two variations. 
Due to their slightly higher bandwidth, the UCSC boards provided faster signal rise-times as compared to the Fermilab board for all three sensors.

A prominent reduction in the signal jitter ($\sigma_{\rm{jitter}} = \rm{Noise}/\frac{dV}{dt}$) is observed with increasing bias voltages, as shown in Figure~\ref{fig:pixelHPK-bias_scan2} (left). 
Irrespective of the active thickness, the jitter contribution converges to around 8 ps at the optimal bias voltage for all sensors indicating large amplitudes for each device. 
We also achieve very similar values of jitter using both boards. 
The overall timing resolution for these sensors also exhibits an improvement with increasing bias voltage, reaching approximately $\sim$ 20, 24, and 35 ps (using  Fermilab board) for thicknesses of 20, 30, and 50 \si{\um}, respectively, as shown in Figure~\ref{fig:pixelHPK-bias_scan2} (right). 
The UCSC boards provide slightly better time resolution due to slight differences in noise and bandwidth.
The plateau in time resolution at the highest bias voltages for each sensor suggests the sensors are very close to breakdown. 
The 20-\si{\um}-thick sensor achieves the best timing performance among all sensors. 

\begin{figure}[H]
    \centering
    \includegraphics[width=0.32\textwidth]{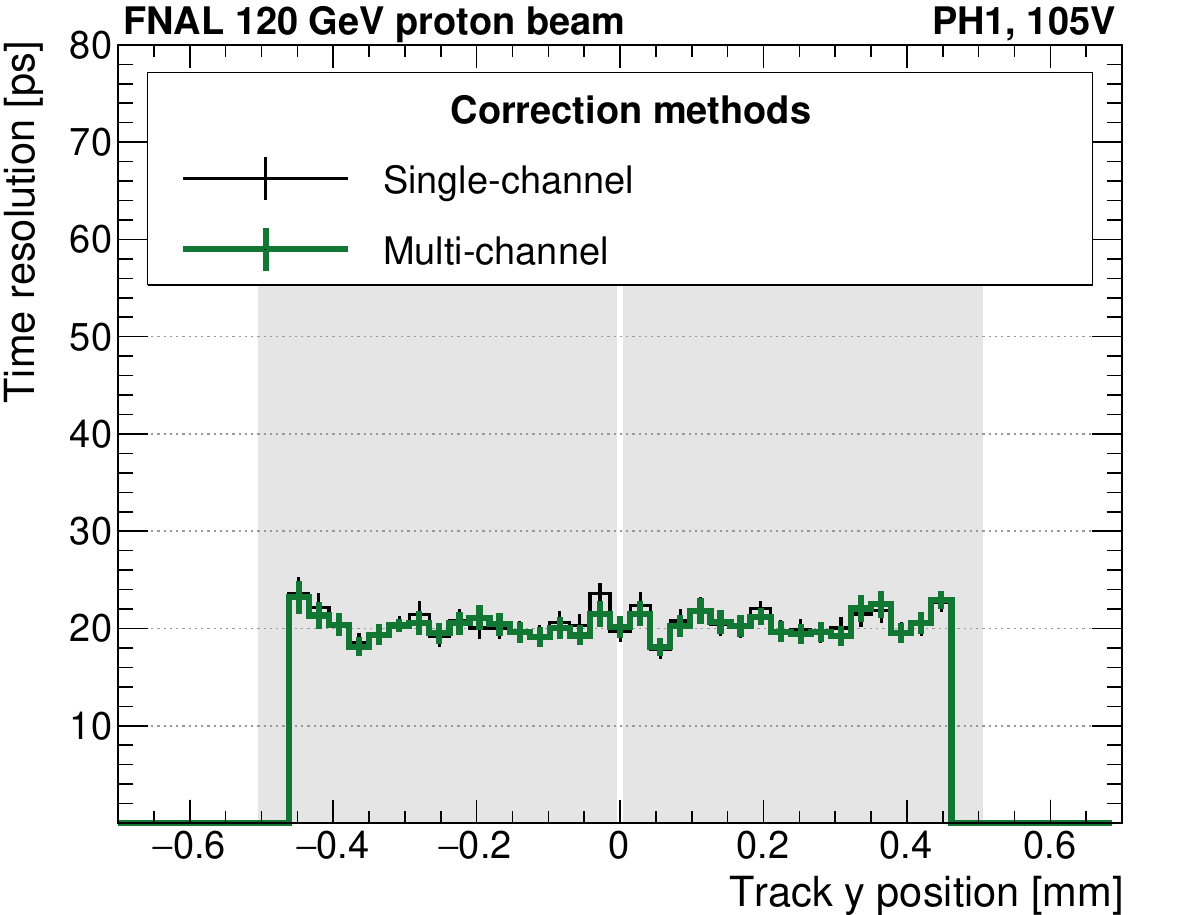}
    \includegraphics[width=0.32\textwidth]{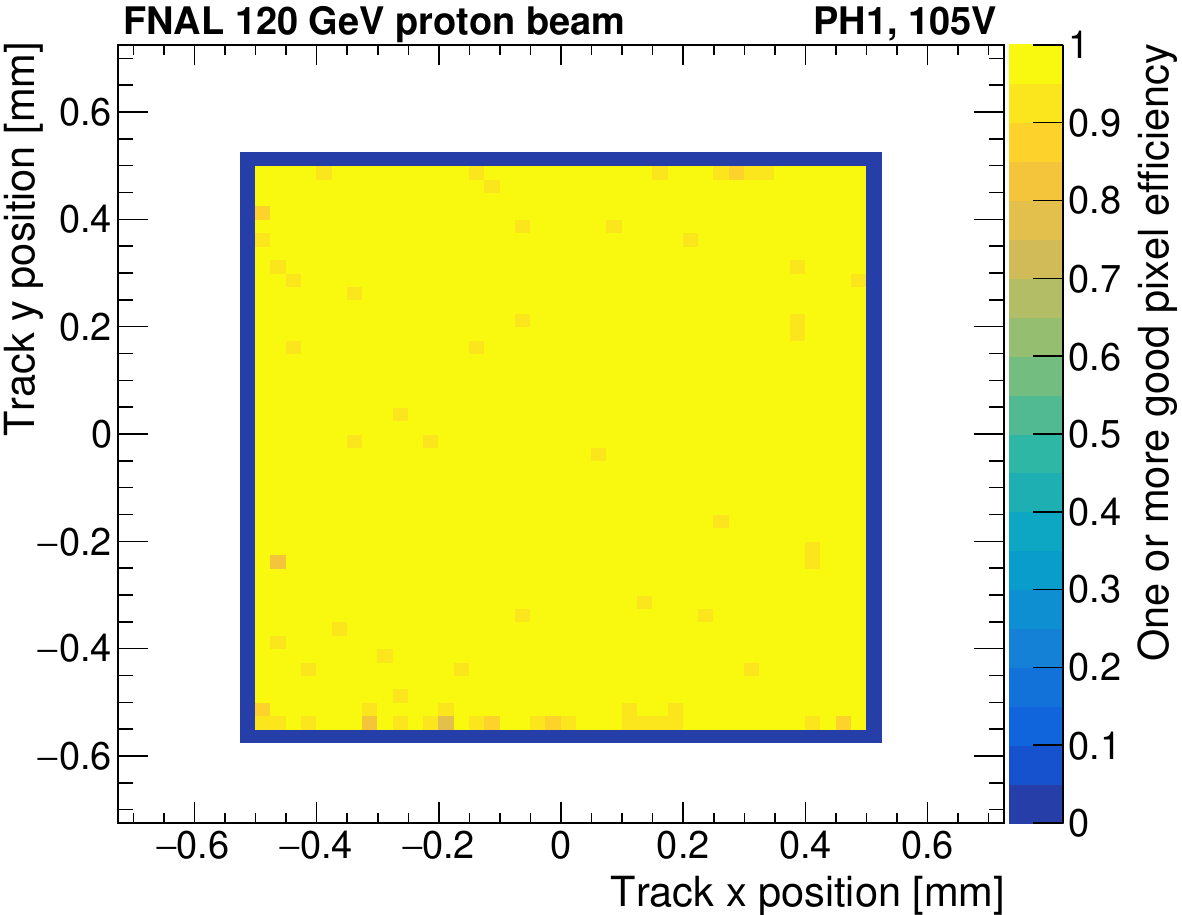}
    \includegraphics[width=0.32\textwidth]{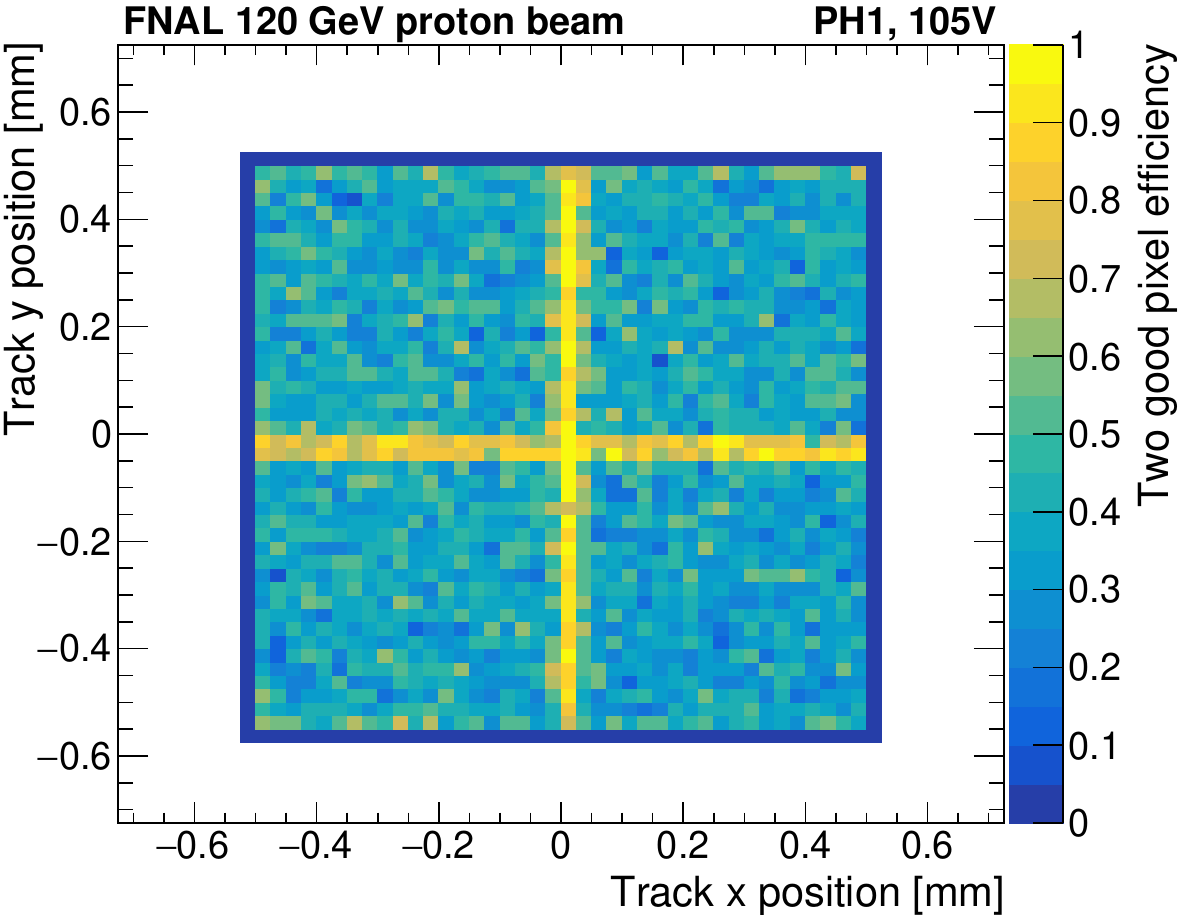}
\caption{Performance of the HPK 2$\times$2 pixel sensor with \SI{20}{\um} active thickness (PH1), \SI{510}{\um} pitch, \SI{500}{\um} metal width, \SI{1600}{\Omega / \sq} sheet resistance, and \SI{600}{\pico F / \mm^2} coupling capacitance. (Left) Time resolution as a function of tracker $y$ position, (center) one-channel and (right) two-channel hit reconstruction efficiencies as a function of tracker $x$ and $y$ position. These results were obtained with the Fermilab 16-channel board.}
    \label{fig:pixelHPK-bias_scan-BestSensor}
\end{figure}

More detailed performance results for the 20-\si{\um}-thick HPK 2$\times$2 sensor using the Fermilab board (as it can read out all 4 channels) are shown in Figure~\ref{fig:pixelHPK-bias_scan-BestSensor}. 
The sensor exhibits a fairly uniform timing resolution of approximately $\sim$ 20 ps across the entire surface. 
While the one-channel hit reconstruction efficiency is 100$\%$ across the full surface, the two-channel hit reconstruction efficiency suffers from the small gap size of the sensor (50 \si{\um}). 
This feature limits the spatial reconstruction capabilities of this sensor.
However, with the device uniformly reaching \SI{20}{ps} time resolution it stands as an ideal candidate for future timing detectors.

\subsection{\textbf{BNL pixel sensors}}\label{sec:results-pixels-BNL} 
\begin{figure}[H]
    \centering
    \includegraphics[width=0.24\textwidth]{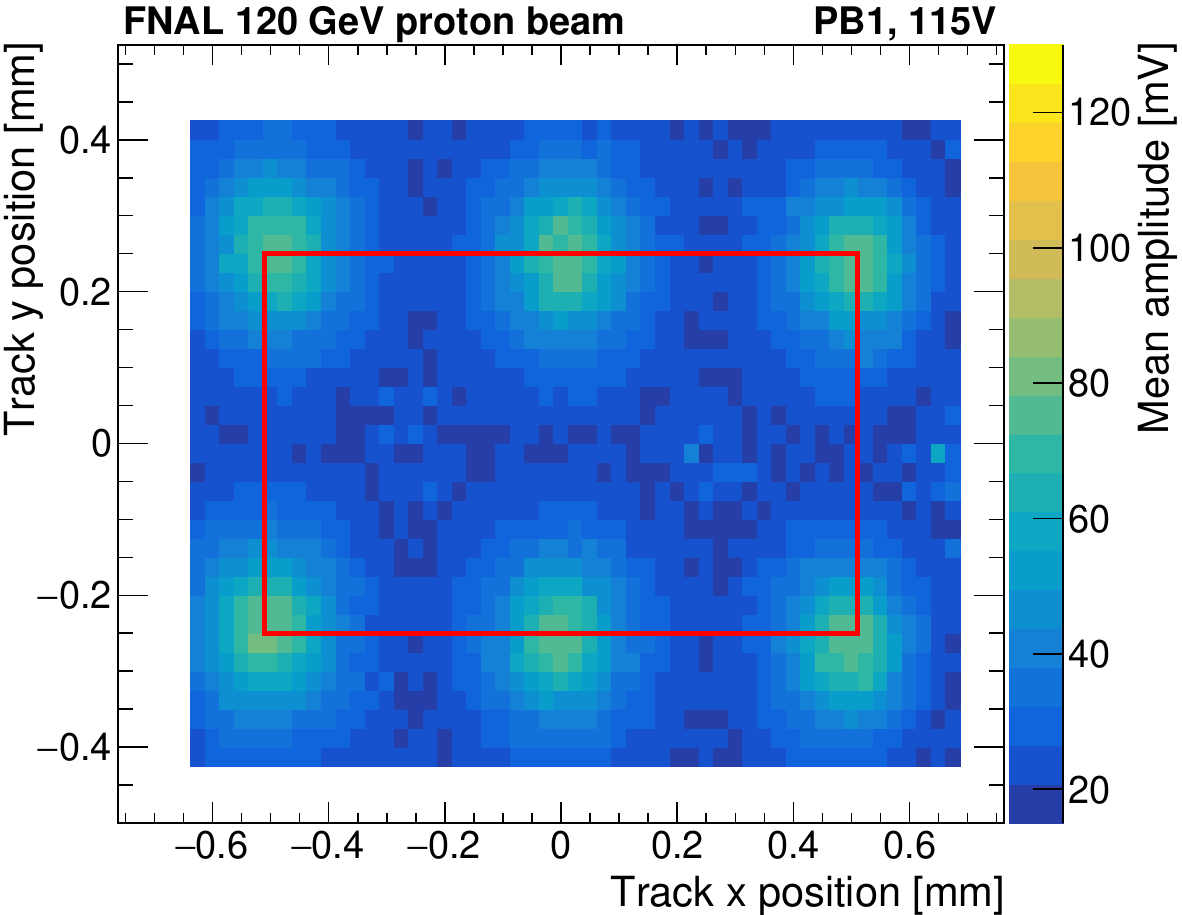}
    \includegraphics[width=0.24\textwidth]{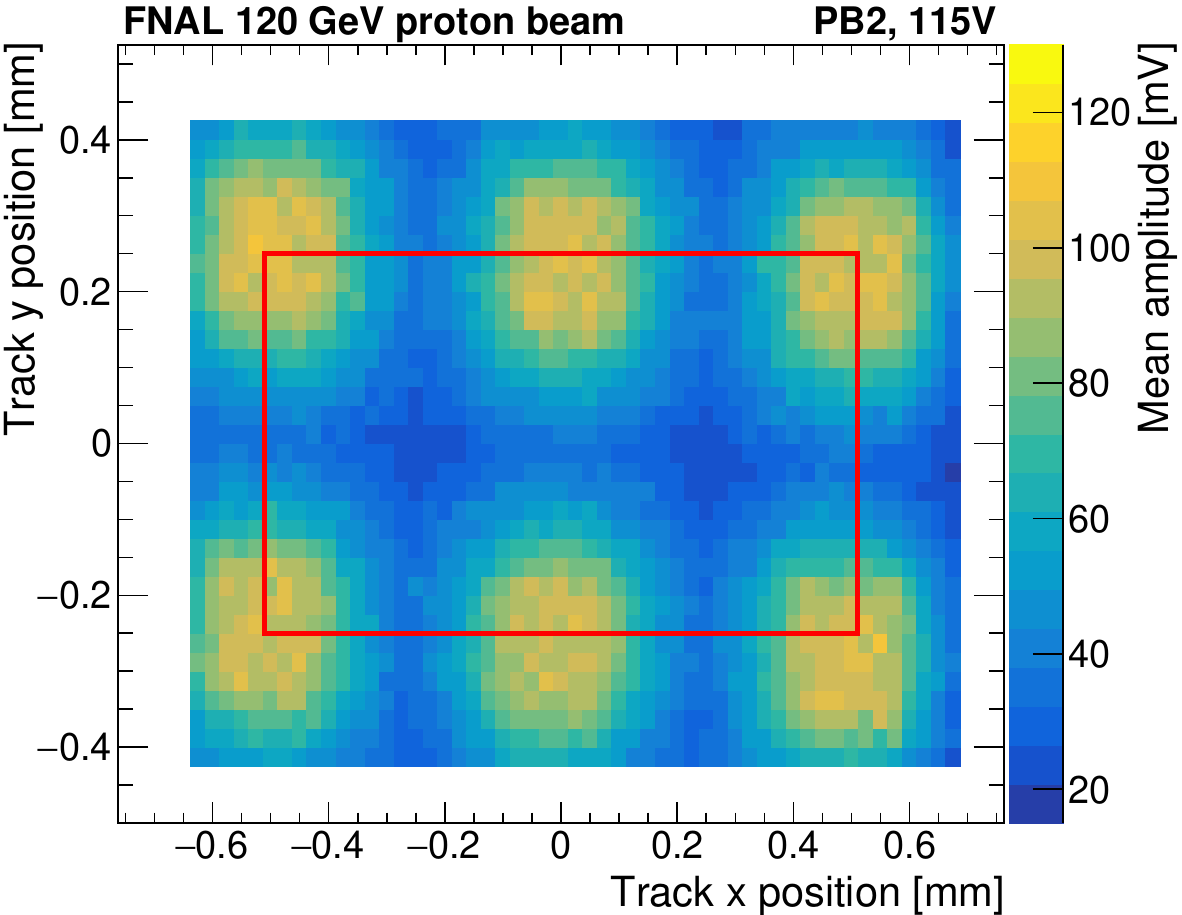}
    \includegraphics[width=0.24\textwidth]{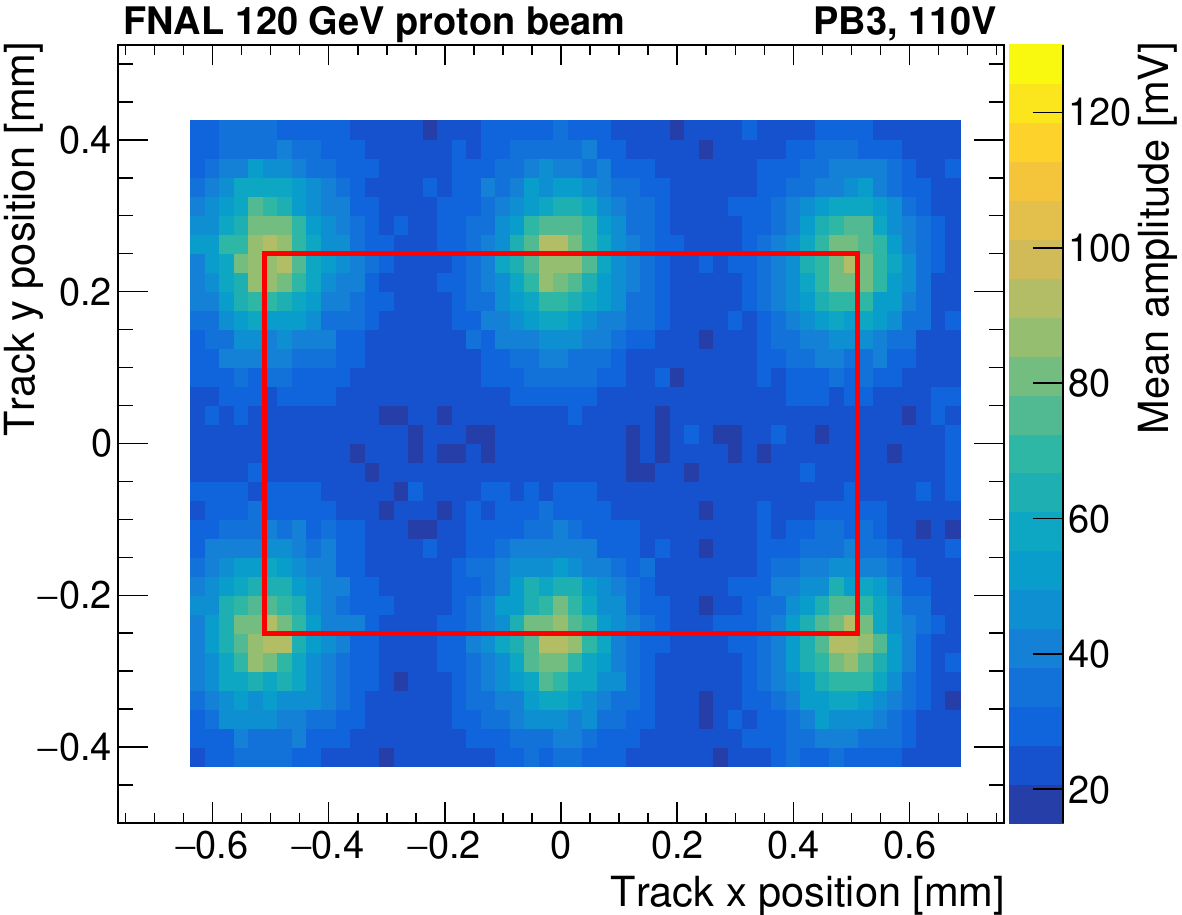}
    \includegraphics[width=0.24\textwidth]{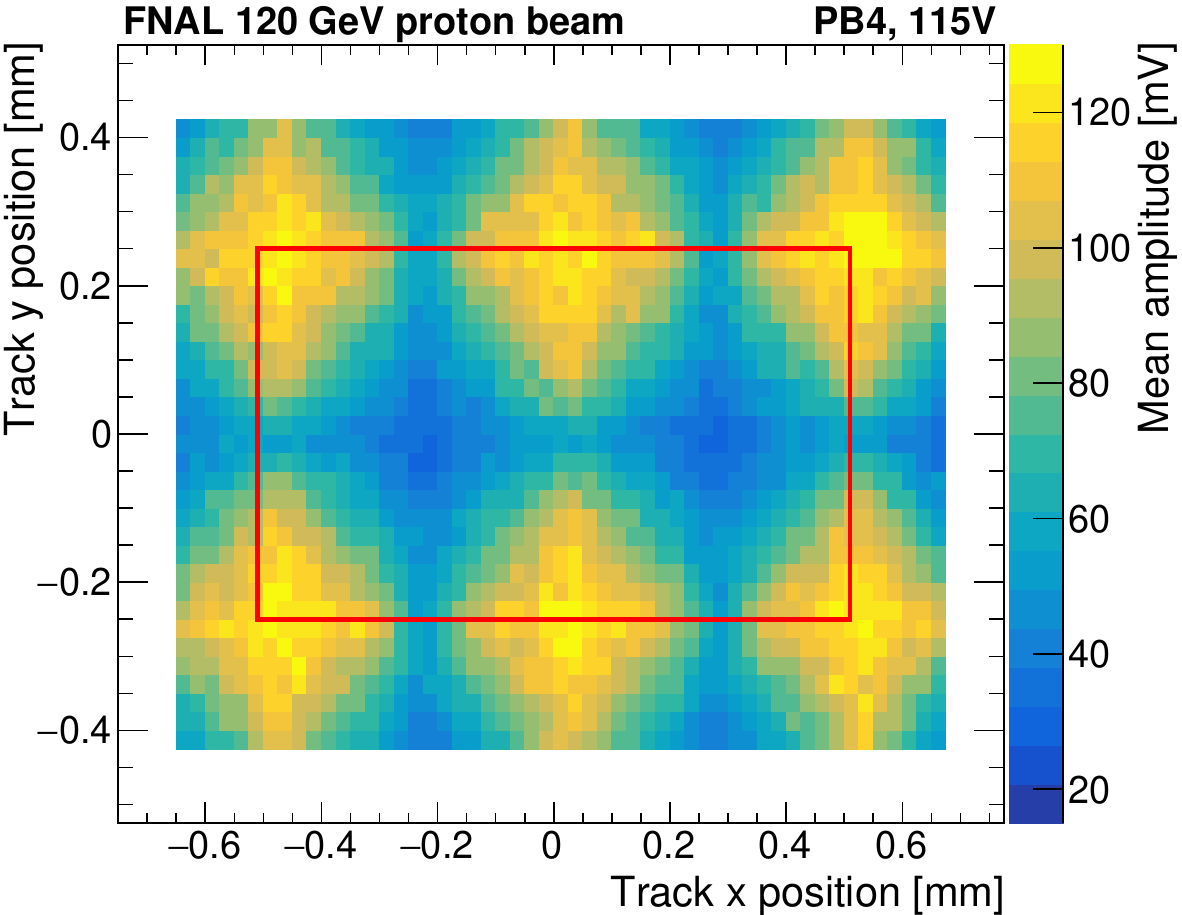}
    \caption{Mean maximum signal amplitudes as a function of the $x$ and $y$ position coordinates for the BNL 4x4 pixel sensors with \SI{30}{\um} active thickness. The metal electrode geometry variations include small squares (leftmost), large squares (second from the left), squared circles (third from the left), and crosses (rightmost). The area within the red lines represents the region of interest.}
    \label{fig:pixelBNL-Amp}
\end{figure}

\begin{figure}[H]
    \centering
    \includegraphics[width=0.24\textwidth]{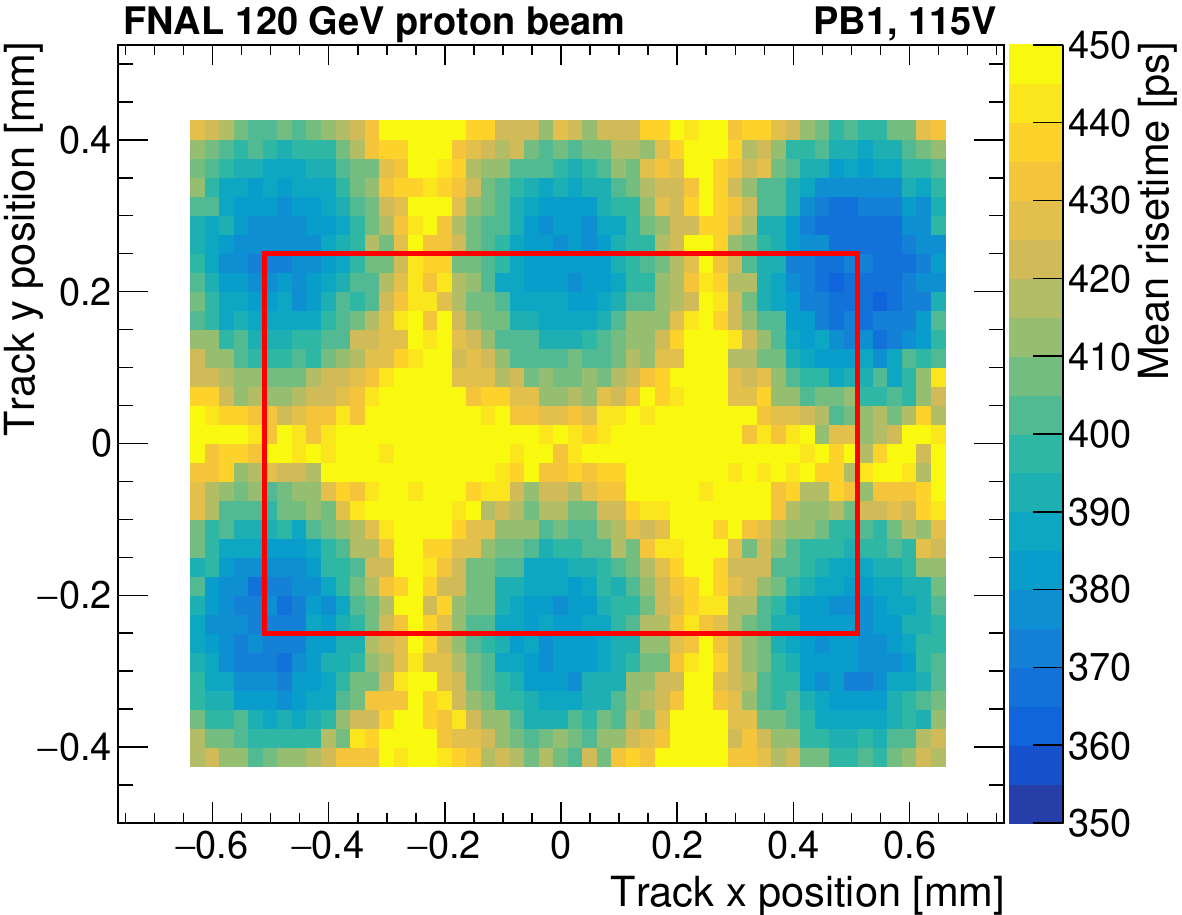}
    \includegraphics[width=0.24\textwidth]{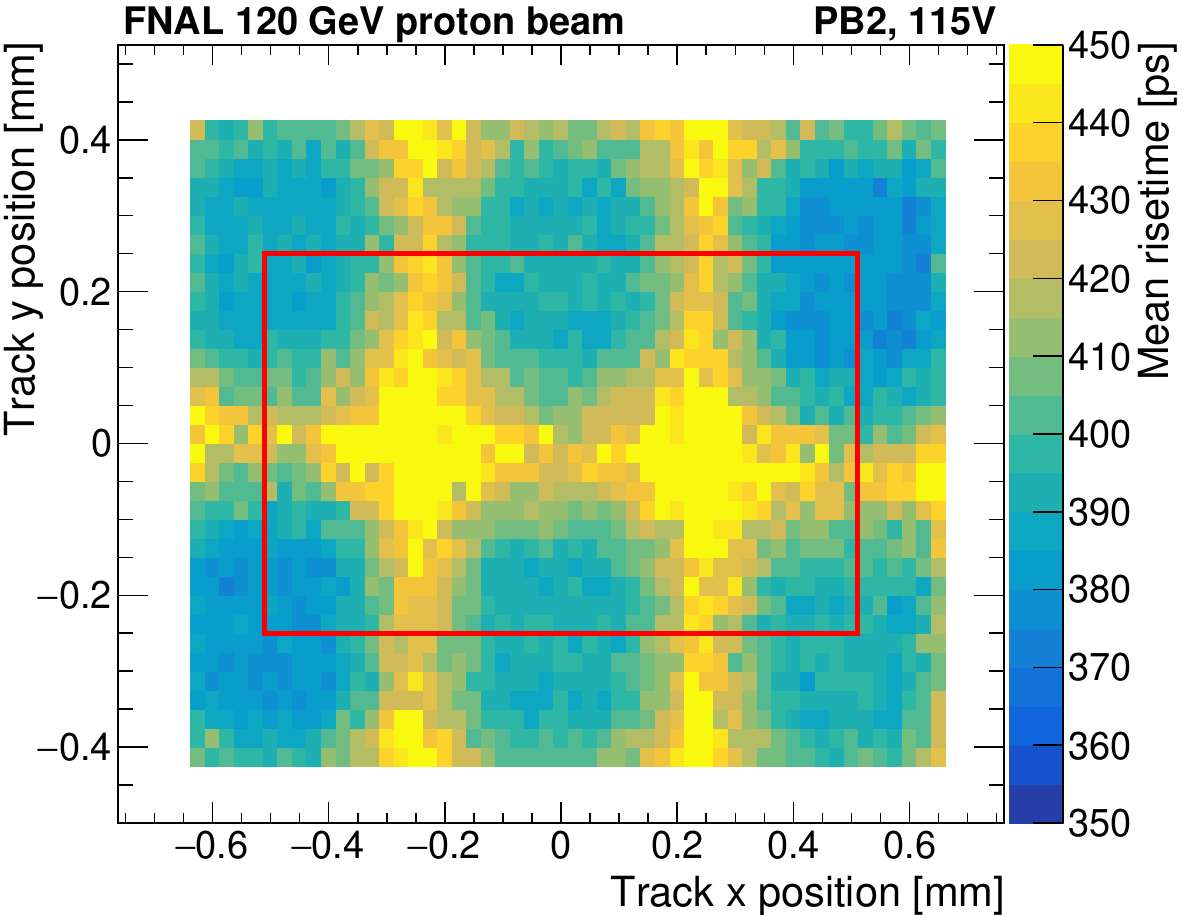}
    \includegraphics[width=0.24\textwidth]{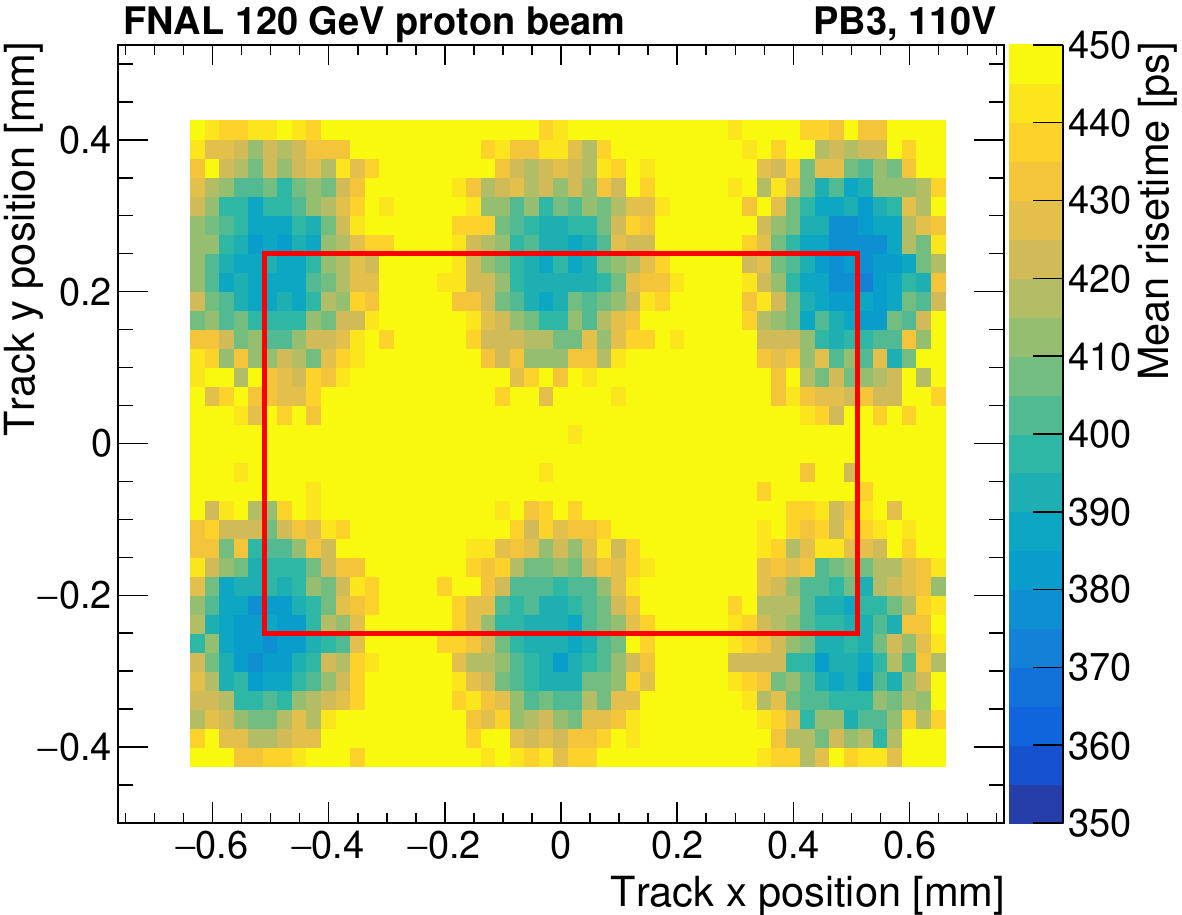}
    \includegraphics[width=0.24\textwidth]{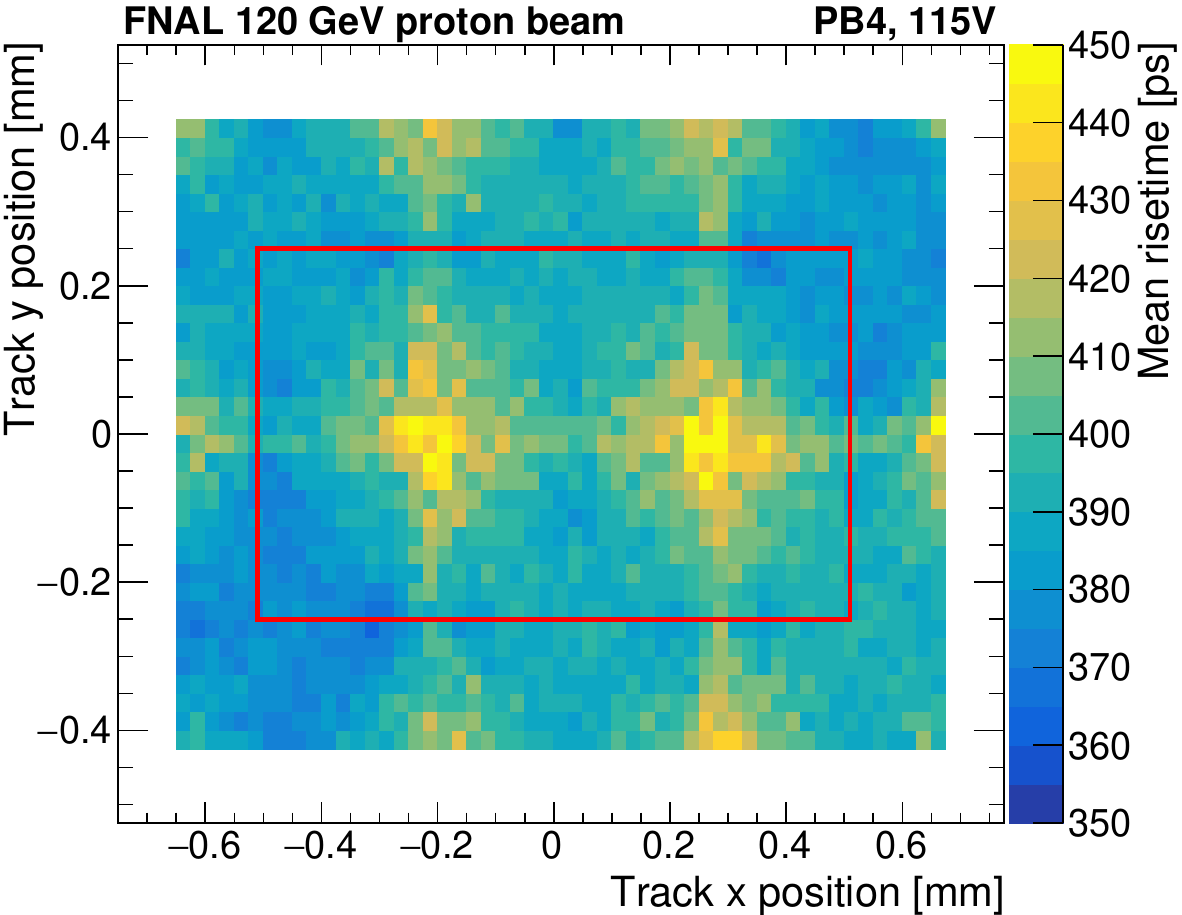}
    \caption{Mean signal rise times as a function of the $x$ and $y$ position coordinates for the BNL 4x4 pixel sensors with \SI{30}{\um} active thickness. The metal electrode geometry variations include small squares (leftmost), large squares (second from the left), squared circles (third from the left), and crosses (rightmost). The area within the red lines represents the region of interest.}
    \label{fig:pixelBNL-Risetime}
\end{figure}

\begin{figure}[H]
    \centering
    \includegraphics[width=0.24\textwidth]{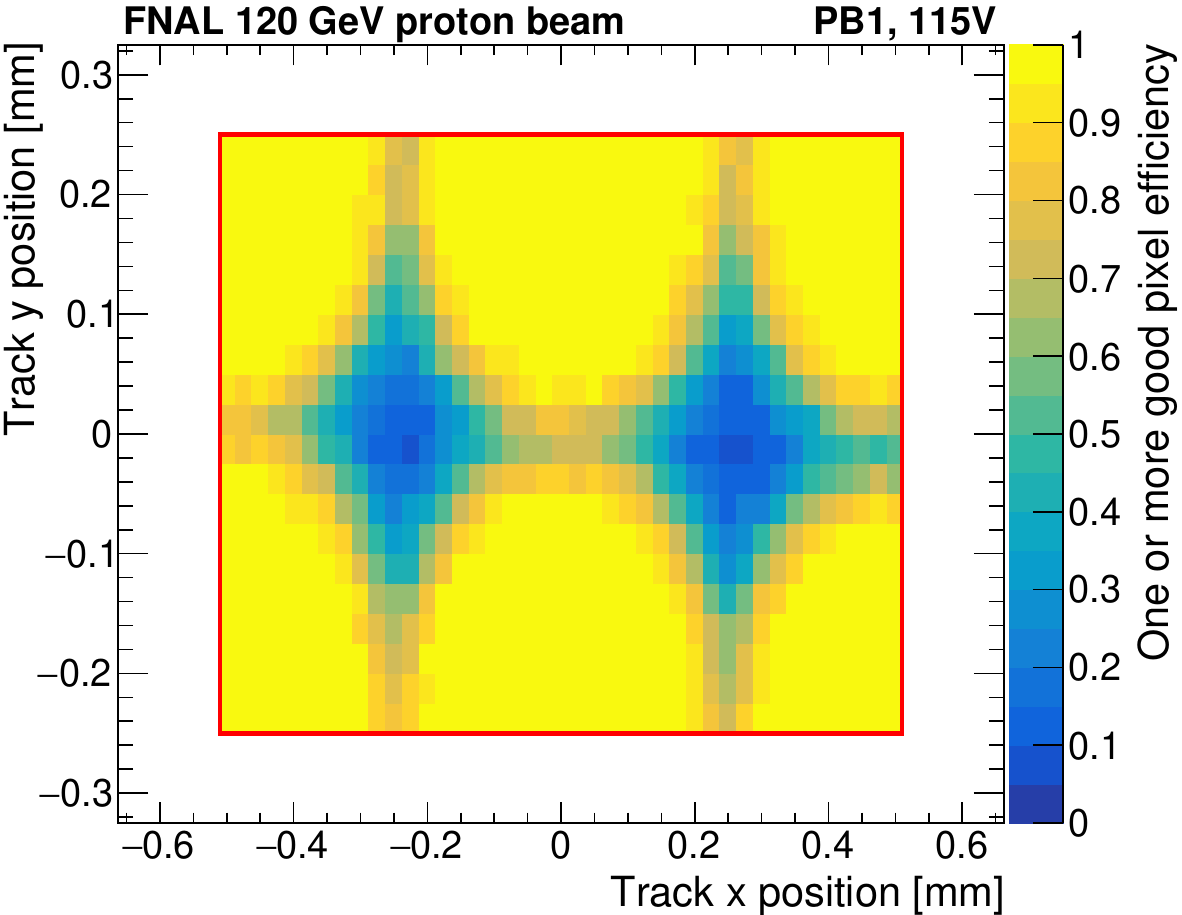}
    \includegraphics[width=0.24\textwidth]{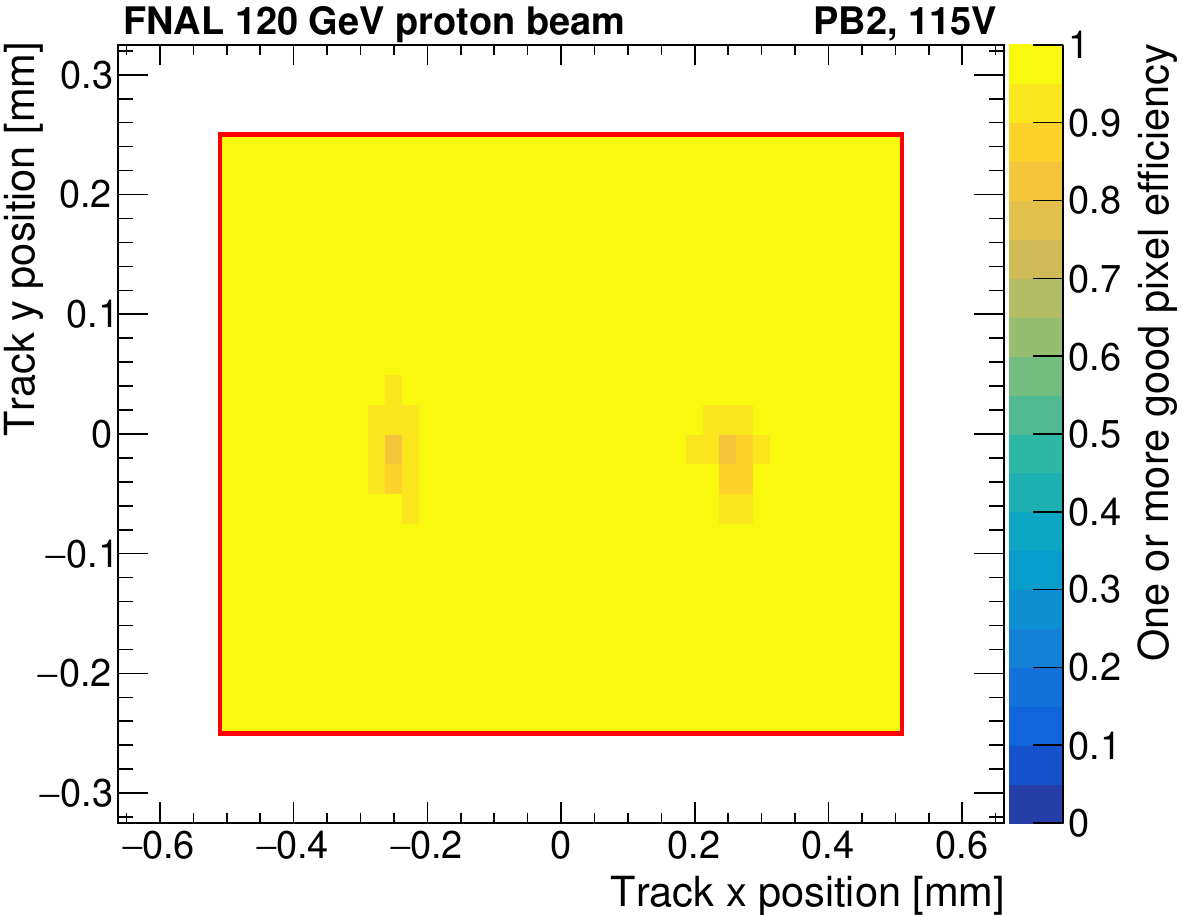}
    \includegraphics[width=0.24\textwidth]{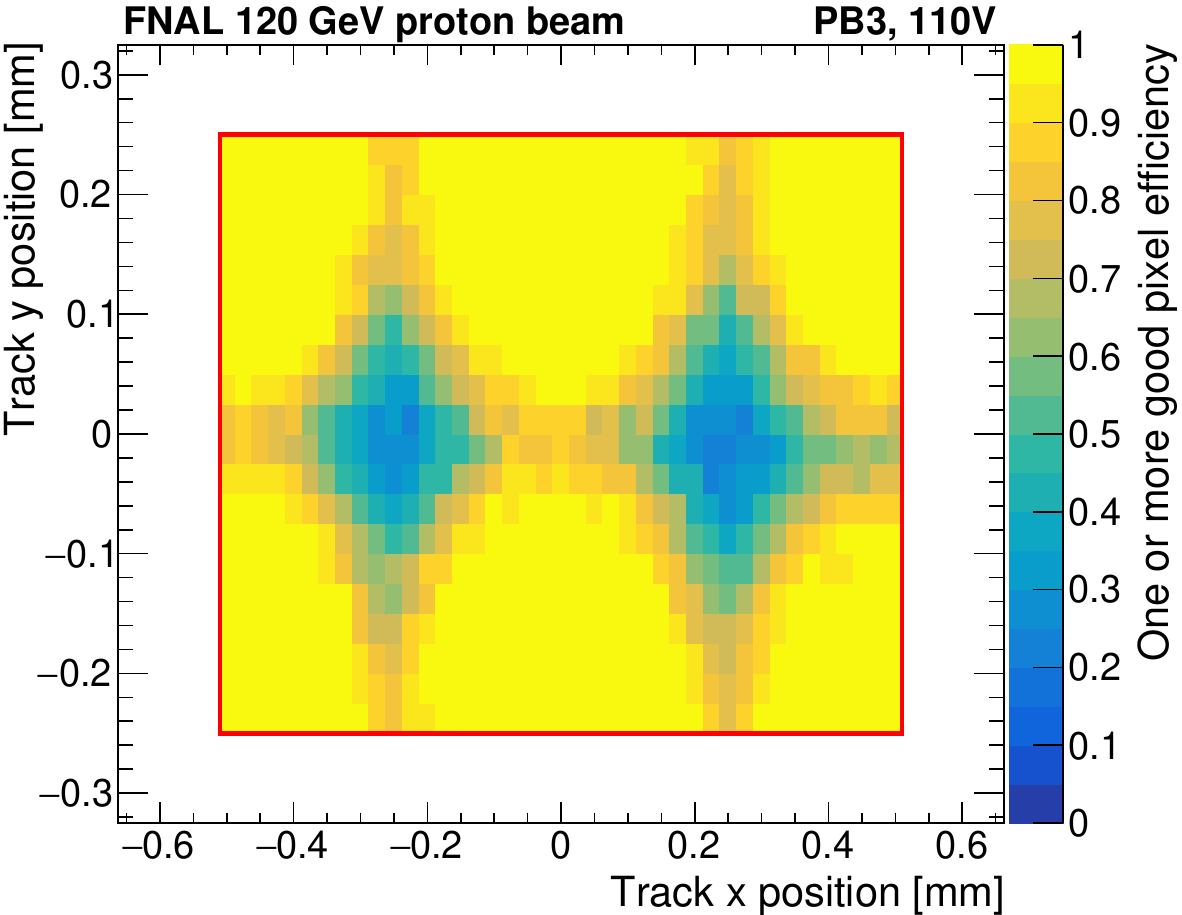}
    \includegraphics[width=0.24\textwidth]{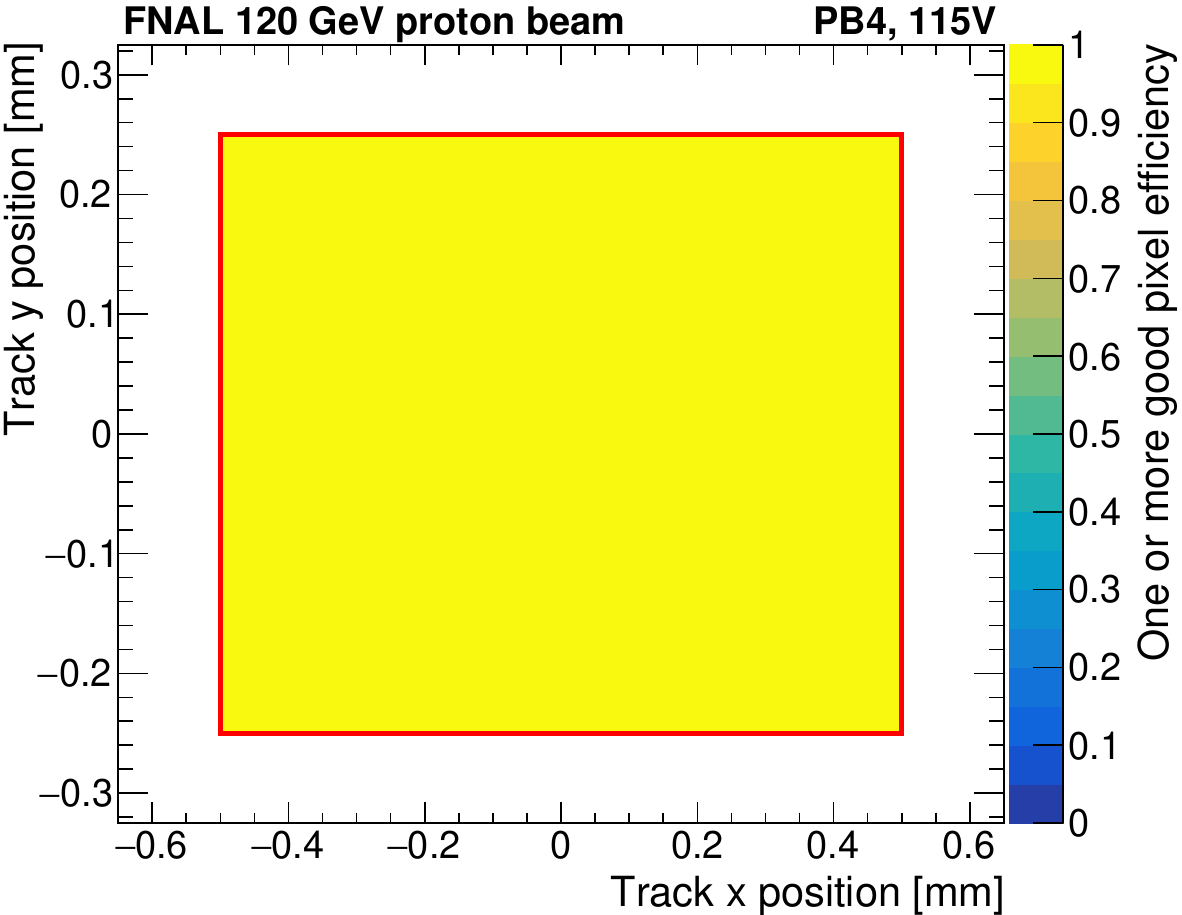}
    \caption{One or more-channel hit efficiency as a function of the $x$ and $y$ position coordinates for the BNL 4x4 pixel sensors with \SI{30}{\um} active thickness. The metal electrode geometry variations include small squares (leftmost), large squares (second from the left), squared circles (third from the left), and crosses (rightmost). The area within the red lines represents the region of interest.}
    \label{fig:pixelBNL-EffOneOrMore}
\end{figure}

\begin{figure}[H]
    \centering
    \includegraphics[width=0.24\textwidth]{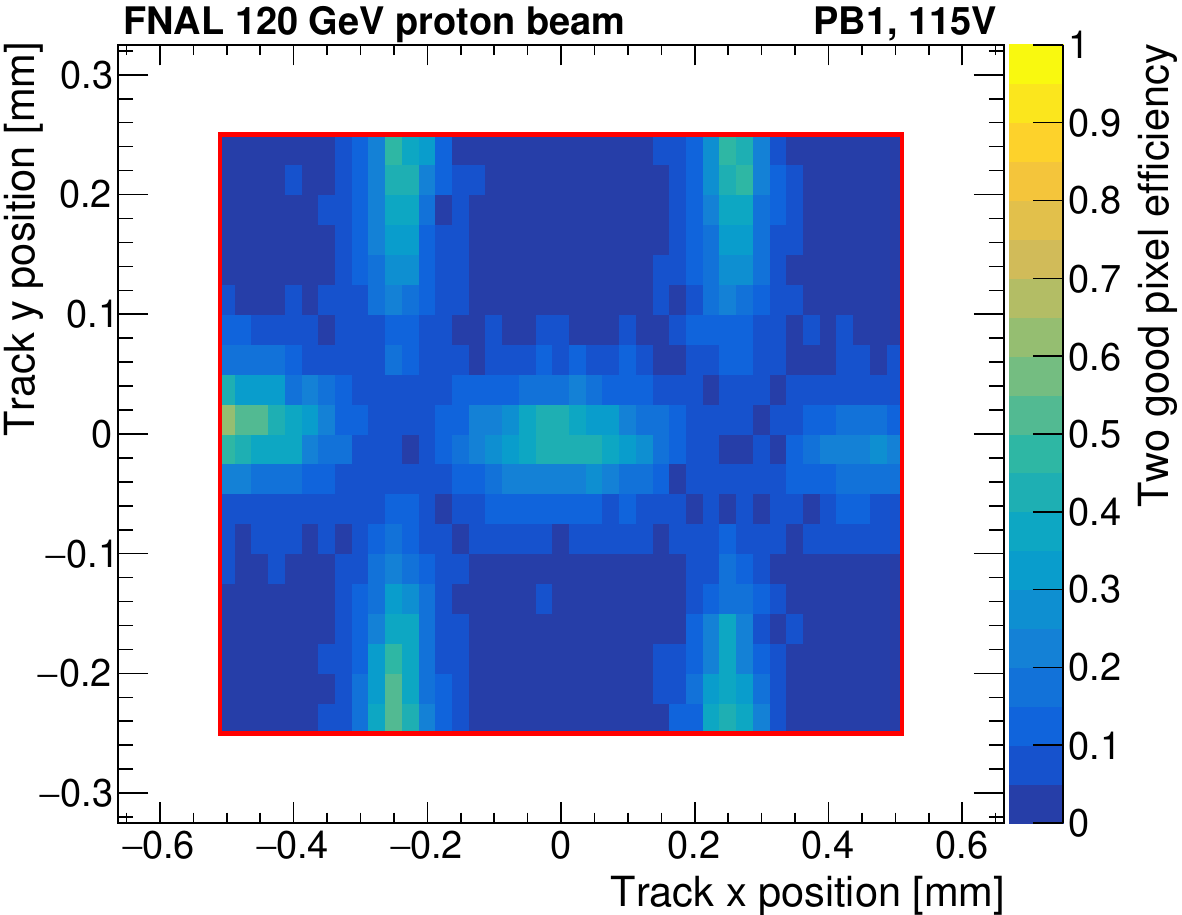}
    \includegraphics[width=0.24\textwidth]{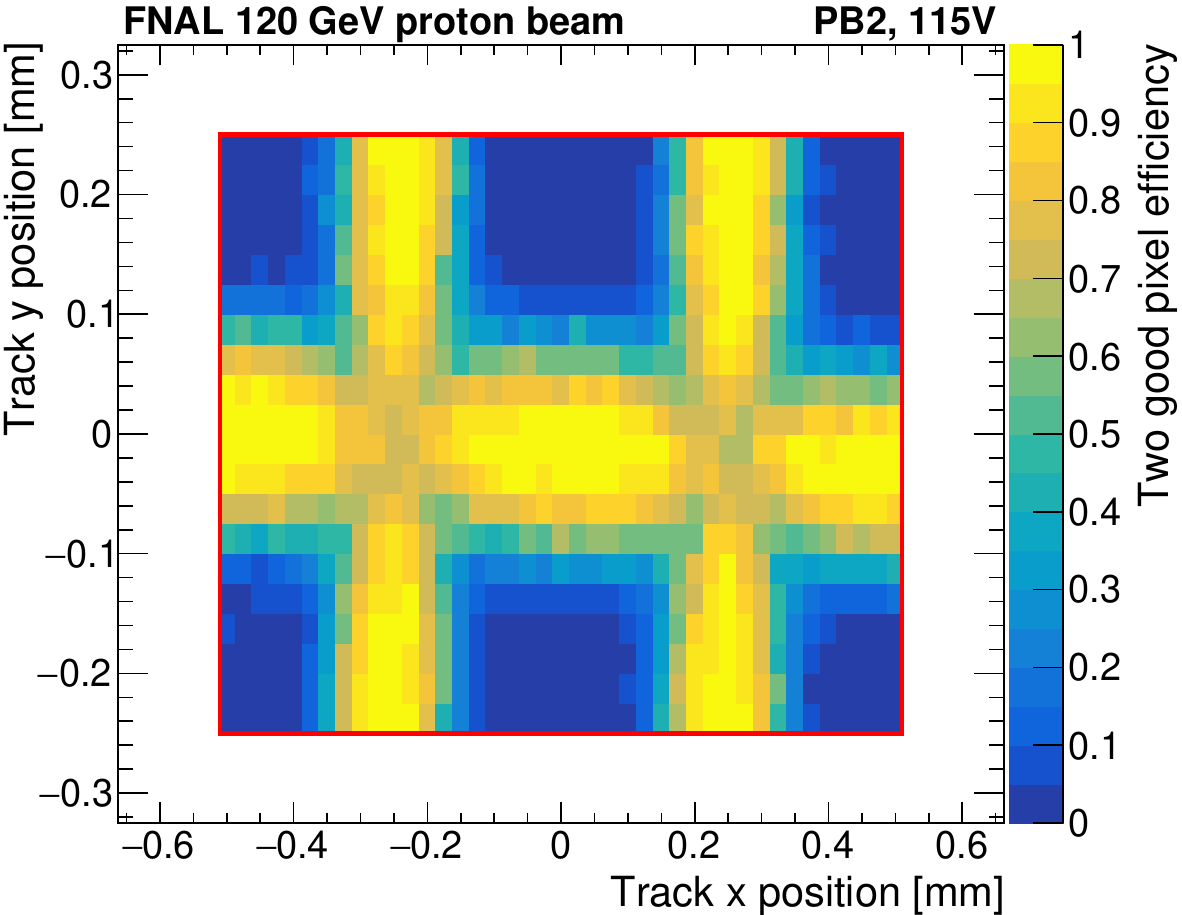}
    \includegraphics[width=0.24\textwidth]{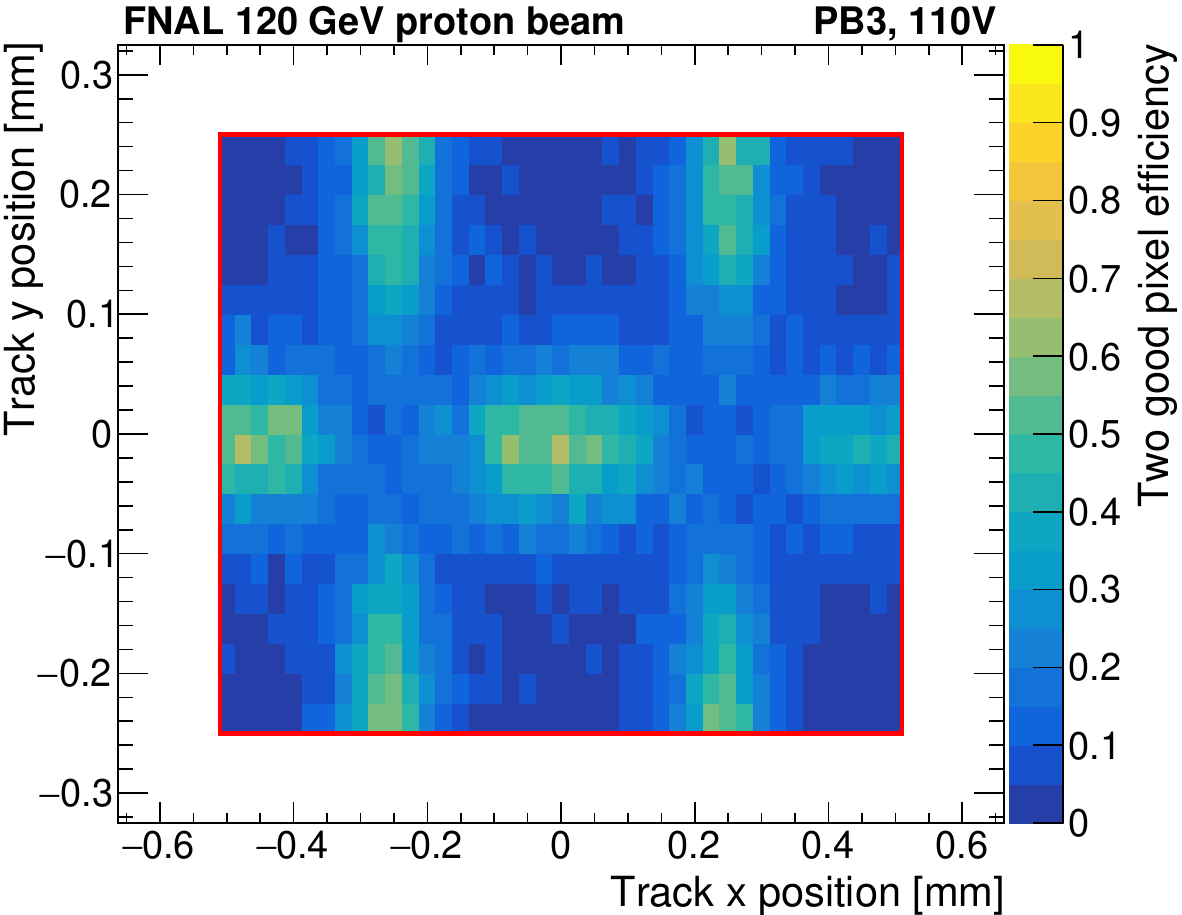}
    \includegraphics[width=0.24\textwidth]{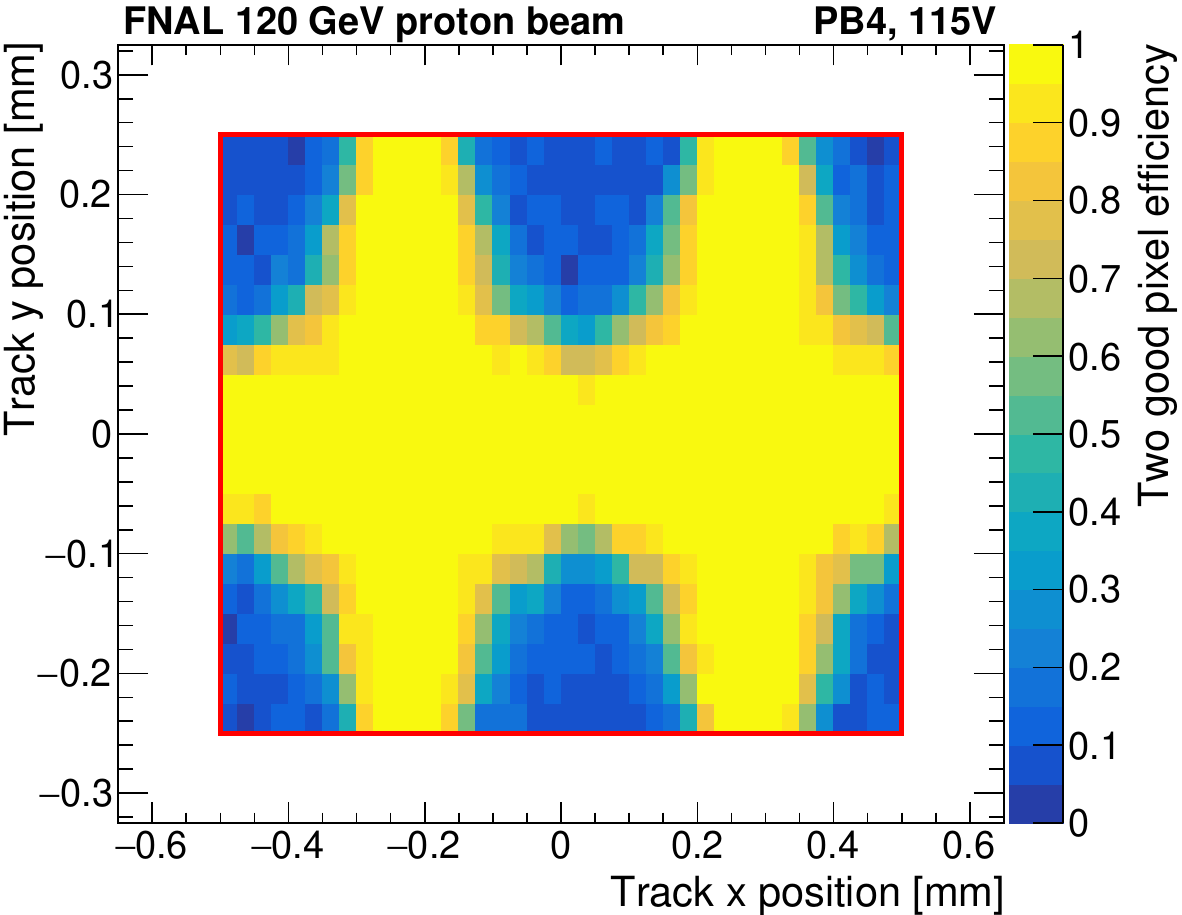}
    \caption{Two-channel hit efficiency as a function of the $x$ and $y$ position coordinates for the BNL 4x4 pixel sensors with \SI{30}{\um} active thickness. The metal electrode geometry variations include small squares (leftmost), large squares (second from the left), squared circles (third from the left), and crosses (rightmost). The area within the red lines represents the region of interest.}
    \label{fig:pixelBNL-Eff}
\end{figure}

The BNL-produced 4x4 pixel sensors are studied to understand the overall impact of different metal electrode geometries on the sensor performance. 
All sensors have an active thickness of 30 \si{\um} and the metal geometries considered include small squares (PB1), large squares (PB2), squared circles (PB3), and crosses (PB4), as shown in Figure~\ref{fig:bnl_sensor_pics}. 
Except for PB3, all other sensors were operated at the same bias voltage of 115 V. 
The PB3 sensor was instead operated at a slightly lower bias voltage of 110 V due to an earlier observed breakdown. 
Figures~\ref{fig:pixelBNL-Amp}, \ref{fig:pixelBNL-Risetime}, \ref{fig:pixelBNL-EffOneOrMore}, and~\ref{fig:pixelBNL-Eff} show the 2-dimensional maps of the mean maximum amplitudes, the signal risetimes, the one-or-more-channel hit efficiencies and the two-channel hit efficiency across the sensor as a function the track $x,y$ position, respectively. 

Figure~\ref{fig:pixelBNL-Amp} reveals that a larger signal size in the primary channel can be achieved with larger metalization, and both the cross-pixel and large squared-pixel sensors achieve fairly large signal amplitudes. 
However, the amplitude variations are more drastic in the large squared pixel sensors outside of the metalized regions as compared to the cross pixels, indicating better charge-sharing features for the cross-geometry pixels. 
The cross-pixels also have a fairly uniform spatial distribution of signal risetimes across the entire sensor, as shown in Figure~\ref{fig:pixelBNL-Risetime}. 
Although, the signal rise times in the metalized regions for the other three sensors are comparable, their gap regions primarily experience slower signals. 
It can be inferred that the faster signals and larger signal sizes with the cross-pixels should contribute to a lower jitter and the overall sensor time resolution.

We also assessed the two-channel hits efficiency for these sensors, which is a critical indicator for signal sharing and spatial reconstruction capabilities. 
From Figure~\ref{fig:pixelBNL-Eff}, the two-channel hits efficiency is notably low for the small squared-pixel sensors and the squared-circle-pixel sensors, which can be attributed to their smaller signal sizes. 
The large squares and cross-shaped pixel sensors show very high efficiencies in the gap regions, with the cross-shaped pixel sensor having a larger total area with efficiencies close to one. 
All sensors exhibit significantly lower two-channel hit efficiencies in the metalized regions, indicating poor spatial resolution in these regions. 
The one-or-more channel hit efficiencies are very close to unity everywhere on the large-squared and cross-pixel sensors, as shown in Figure~\ref{fig:pixelBNL-EffOneOrMore}. 
However, for the small-square and squared-circle pixel sensors, the gap regions suffer from very little signal and hence exhibit very low one-or-more channel hit efficiencies. 
From these observations it is inferred that the cross-shaped pixels have the best spatial resolution capabilities among all sensor geometries considered here.

The primary focus of this study was to compare various metal electrode geometries and perform a qualitative analysis of their 4D tracking capacity. 
The spatial variation in signal properties sufficiently illustrates the distinctions among these sensors, and consequently, allows us to conclude that the cross-pixel sensor demonstrates favorable characteristics for both temporal and spatial reconstruction.

\subsection{\textbf{Summary of Results}}
This section summarizes the experimental results that were presented in Sec~\ref{sec:results-long-strips},~\ref{sec:results-pixels-HPK} and ~\ref{sec:results-pixels-BNL}. 
In the previous sections, spatial resolutions were presented separately for the one-strip (column) and two-strip (column) populations in each sensor. To provide a single resolution value that fairly summarizes the performance, we define a quantity called the \textit{combined position resolution}, $\sigma_c$, which takes into account the resolution and relative proportion of both categories, as follows:
\begin{equation}\label{eq:combined-pos-res}
    \sigma_c = \sqrt{\frac{e_1\sigma_1^2 + e_2\sigma_2^2}{e_1 + e_2}}
\end{equation}
where $e_{1(2)}$ is the one (two) strip/column efficiency, and $\sigma_{1(2)}$ is the one (two) strip/column position resolution. The parameter $\sigma_{2}$ is calculated in bins of width \SI{50}{\um} as a function of the track $x$ position, and $\sigma_{1}$ is calculated in bins widths equaling the sensor's electrode pitch.

\begin{figure}[H]
    \centering
    \includegraphics[width=0.49\textwidth]{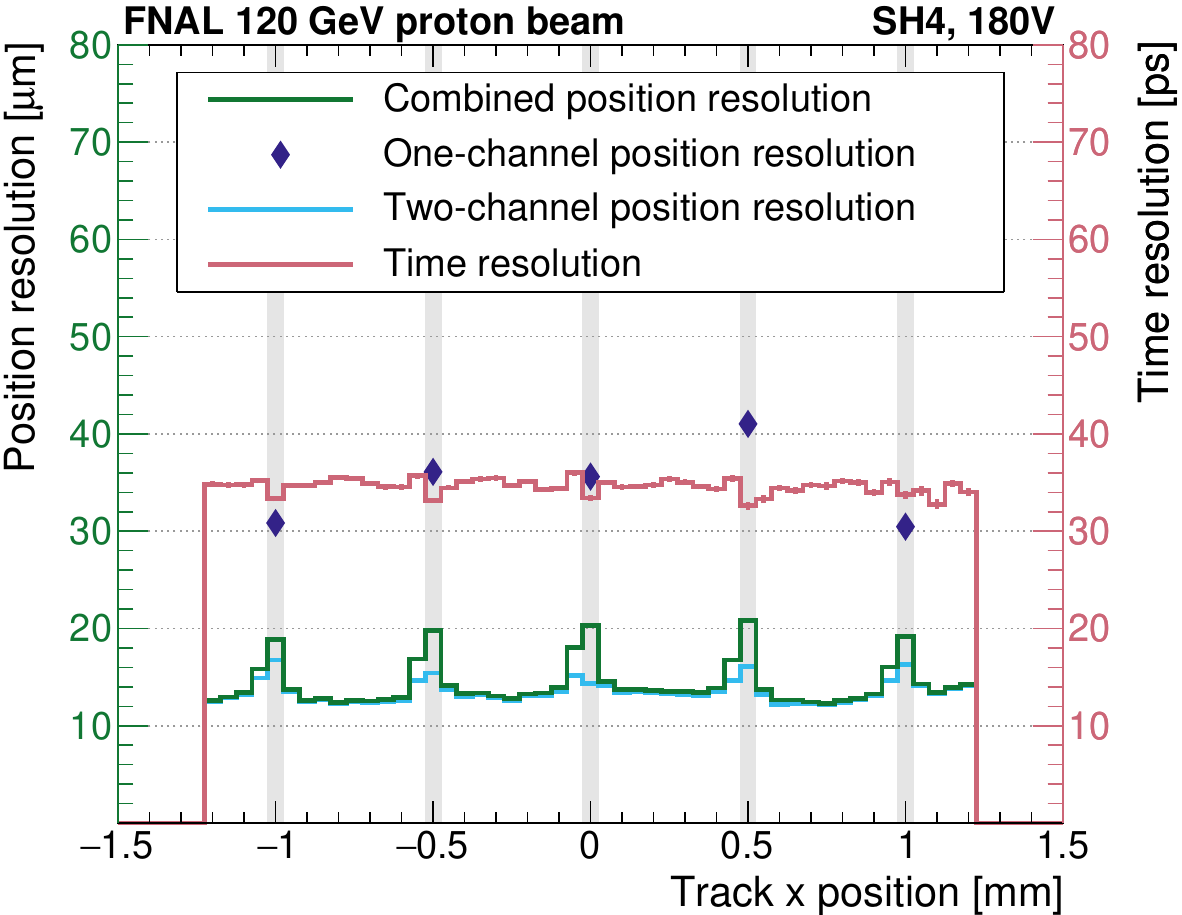}
    \includegraphics[width=0.49\textwidth]{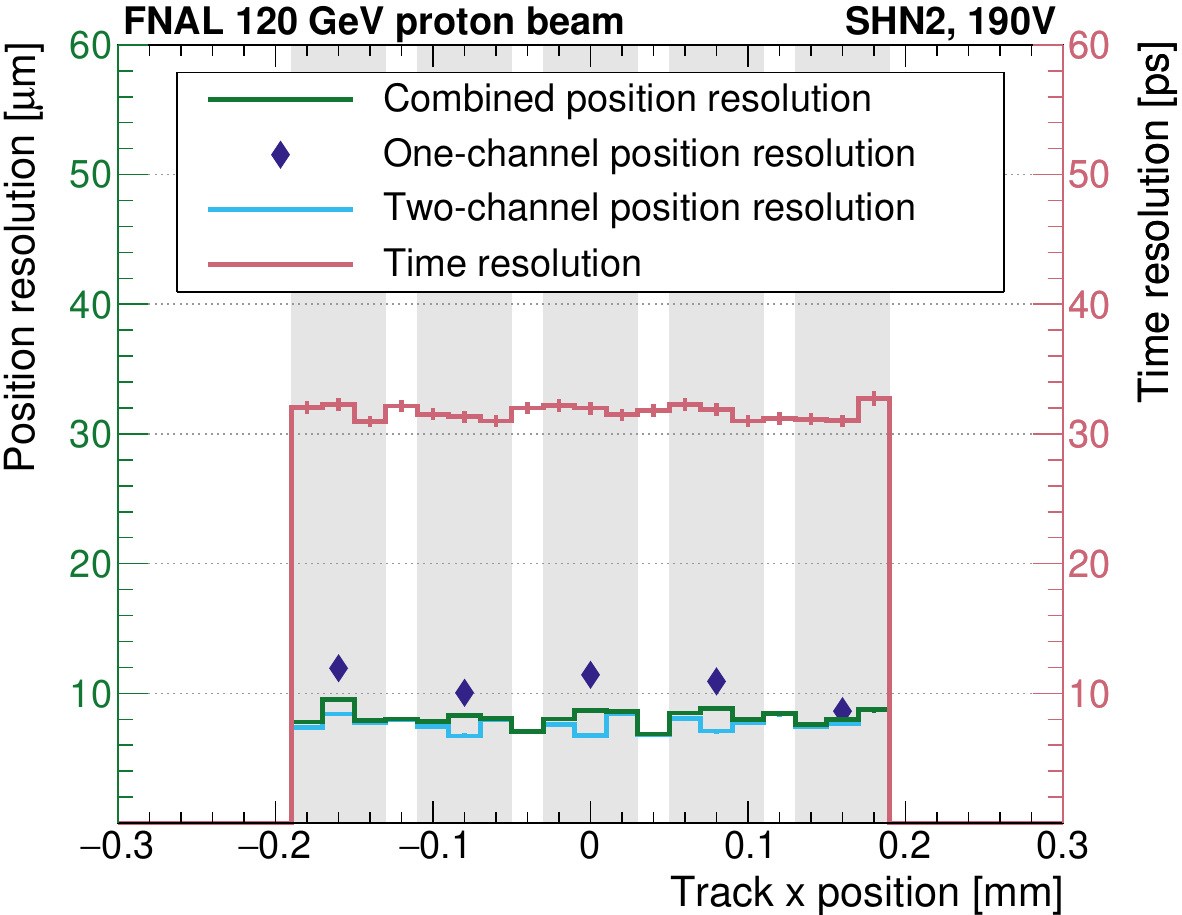}
    \caption{Summary of spatial and timing resolution values for the best performing HPK strip sensors with \SI{500}{\um} pitch with \SI{50}{\um} active thickness (left) and \SI{80}{\um} pitch with \SI{50}{\um} active thickness (right). Spatial resolution values have a reference tracker contribution of \SI{5}{\um} subtracted in quadrature. Time resolution values have an MCP reference contribution of \SI{10}{\ps} removed in quadrature.}
    \label{fig:HPK_strip_Summary}
\end{figure}

Figure~\ref{fig:HPK_strip_Summary} shows the two best-performing strip sensors from this survey. 
Figure~\ref{fig:HPK_strip_Summary} (left) shows the combined timing and position resolution performance of an HPK 500-\si{\um}-pitch strip sensor (SH4). 
This sensor (Figure~\ref{fig:HPK_strip_Summary} left) has simultaneous temporal and spatial resolutions of about \SI{35}{\pico\second} and \si{15}-\SI{20}{\um} respectively, across its entire surface, and is ideal for cases that require large area sensors with economic channel counts and power density constraints. 
In situations with a very high particle occupancy, the SHN2 sensor (Figure~\ref{fig:HPK_strip_Summary} right) with \SI{80}{\um} pitch can deliver a slightly better time resolution of $\sim$32 ps and \si{8}-\SI{10}{\um} spatial resolution.

\begin{figure}[H]
    \centering
    \includegraphics[width=0.49\textwidth]{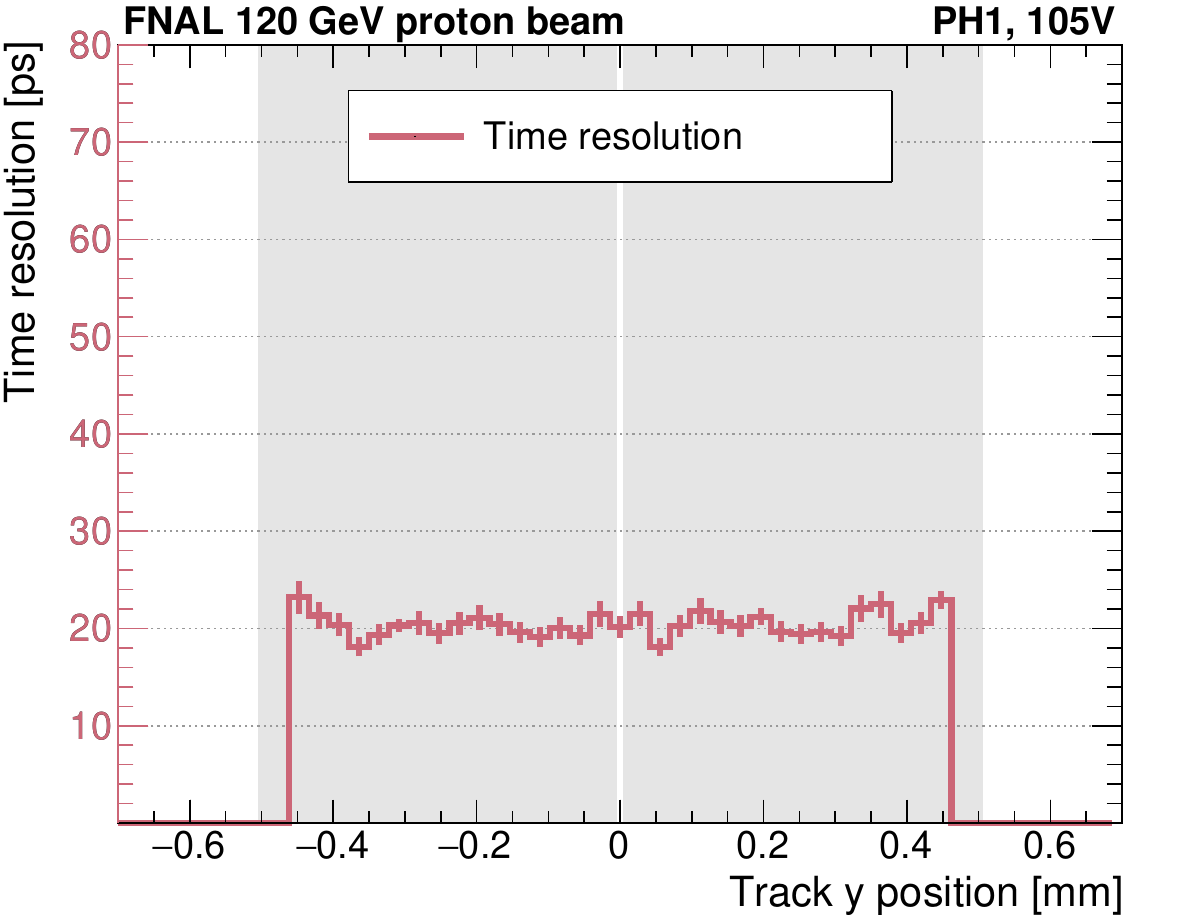}
    \includegraphics[width=0.49\textwidth]{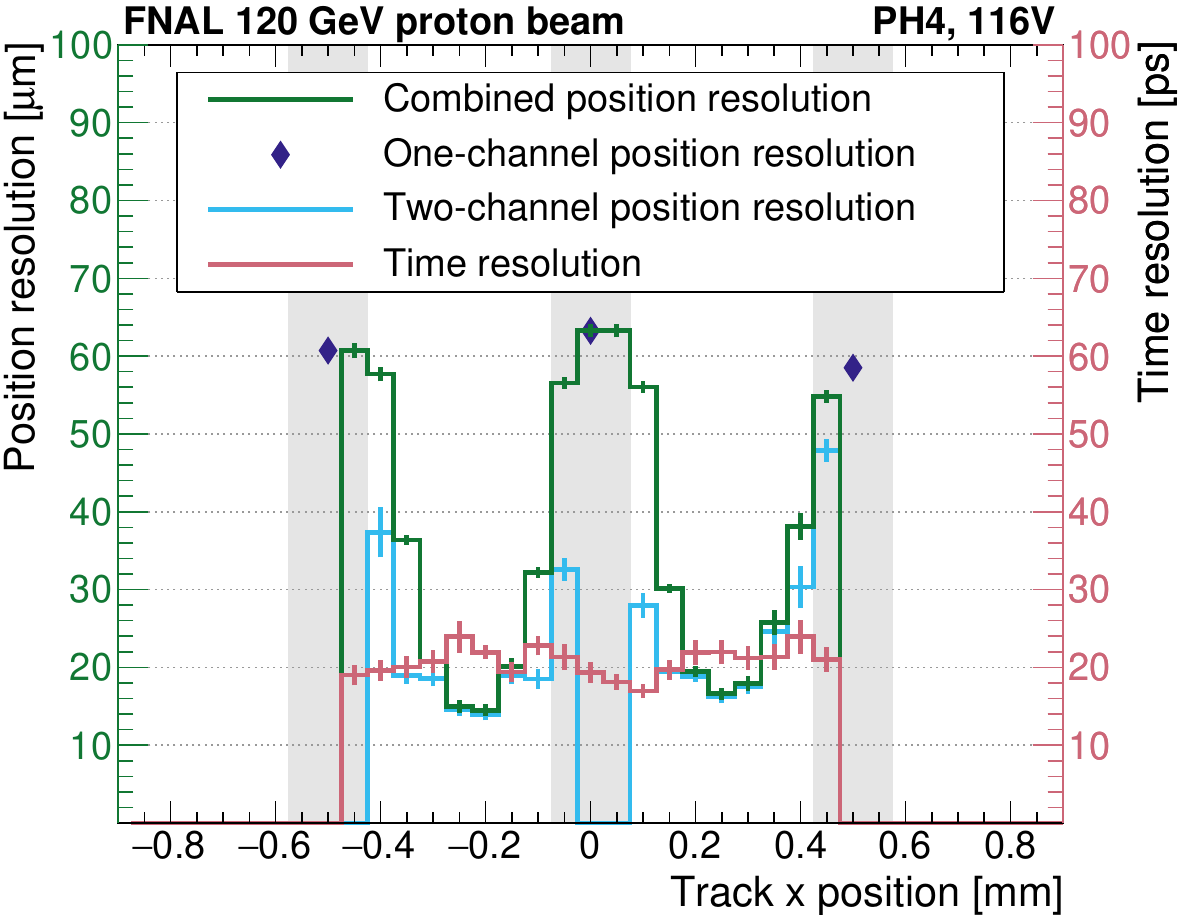}
    \caption{Summary of spatial and timing resolutions values for the best performing HPK pixel sensors with 2$\times$2 pads of \SI{500}{\um} metal width with \SI{20}{\um} active thickness (left) and 4$\times$4 pads of \SI{150}{\um} metal width with \SI{20}{\um} active thickness (right). Spatial resolution values have a reference tracker contribution of \SI{5}{\um} subtracted in quadrature. Two-channel reconstruction value is presented only if its efficiency is greater than 5\%. Time resolution values have an MCP reference contribution of \SI{10}{\ps} removed in quadrature.}
    \label{fig:HPK_pad_Summary}
\end{figure}

Figure~\ref{fig:HPK_pad_Summary} shows the two best-performing pixel sensors from this survey. 
The 2$\times$2 HPK pixel sensor (PH1) with \SI{20}{\um} active thickness can achieve a uniform time resolution of $\sim$20 ps (Figure~\ref{fig:HPK_pad_Summary} left) and is suitable for applications that require state-of-art timing resolutions with large amplitudes but can tolerate a position resolution of $O(\SI{150}{\um})$. 
The 4$\times$4 HPK pixel sensor with \SI{20}{\um} active thickness (PH4) can deliver $\sim$21 ps timing resolution and 20-70 \SI{}{\um} spatial resolution across the sensor surface (based on the particle hit location) and is better for applications that require the best timing performance with $\leq \SI{70}{\um}$ position resolution. 
Tables~\ref{tab:summary_strips} and~\ref{tab:summary_pix} summarizes the performance of all HPK strip sensors and pixel sensors studied in this campaign, respectively.

The metal electrode geometry survey studies using BNL pixel sensors also indicated that a cross-electrode geometry is more favorable as compared to square or circle configurations.

\begin{table}[H]
\centering
\begin{tabular}{l ccc |c| c c c | c c}
\multicolumn{1}{l|}{\multirow{3}{*}{Name}}  & \multirow{3}{*}{\begin{tabular}[c]{@{}c@{}} Act. \\Thick.\\ $[\si{\um}]$ \end{tabular}} & \multirow{3}{*}{\begin{tabular}[c]{@{}c@{}} Resis.\\$[ \si{\Omega / \sq}]$\end{tabular}} & \multicolumn{1}{c|}{\multirow{3}{*}{\begin{tabular}[c]{@{}c@{}} Cap.\\ $[\si{\pico F / \mm^2}]$\end{tabular}}} &\multicolumn{1}{c|}{\multirow{3}{*}{\begin{tabular}[c]{@{}c@{}} Time res.\\ $[\si{\ps}]$\end{tabular}}} & \multicolumn{3}{c|}{Spatial resolution [\si{\um}]} & \multicolumn{2}{c}{\multirow{2}{*}{\begin{tabular}[c]{@{}c@{}} Amplitude \\ $[\si{\mV}]$\end{tabular}}} \\ \cline{6-8}

\multicolumn{1}{c|}{} & \multicolumn{3}{c|}{} & \multicolumn{1}{c|}{} & \multicolumn{1}{c|}{One ch.} & \multicolumn{1}{c|}{Two ch.} & \multicolumn{1}{c|}{Combined}& &\\ \cline{6-10}
\multicolumn{1}{c|}{} & \multicolumn{3}{c|}{} & \multicolumn{1}{c|}{}& \multicolumn{1}{c|}{Res./Eff.} & \multicolumn{1}{c|}{Res./Eff.} & \multicolumn{1}{c|}{Res./Eff.} & \multicolumn{1}{c|}{Gap} & Metal \\ \hline


\multicolumn{10}{c}{(a) HPK - length: \SI{1.0}{\cm} -- pitch: \SI{500}{\um} -- metal width: \SI{50}{\um}}\\ \hline
\hline
\multicolumn{1}{l|}{SH1} & 20 & 1600 & 600 & 53 $\pm$ 1 & 116 / 67\% & 23 / 31\% & 96 / 98\% & 24 & 50\\
\multicolumn{1}{l|}{SH2} & 50 & 400 & 240 & 37 $\pm$ 1 & 39 / 6\% & 28 / 93\% & 29 / 100\% & 49 & 66\\
\multicolumn{1}{l|}{SH3} & 50 & 400 & 600 & 39 $\pm$ 1 & 37 / 4\% & 27 / 96\%& 28 / 99\%  & 43 & 58\\
\multicolumn{1}{l|}{SH4} & 50 & 1600 & 240 & 35 $\pm$ 1 & 36 / 2\% & 15 / 97\%& 16 / 99\%  & 69 & 127\\
\multicolumn{1}{l|}{SH5} & 50 & 1600 & 600 & 35 $\pm$ 1 & 54 / 4\% & 14 / 95\%& 18 / 99\%  & 74 & 129\\\hline
\hline
\multicolumn{10}{c}{(b) HPK - length: \SI{1.0}{\cm} -- pitch: \SI{500}{\um} -- metal width: \SI{100}{\um}}\\ \hline
\hline
\multicolumn{1}{l|}{SH6} & 20 & 1600 & 600 & 48 $\pm$ 1 & 124 / 76\% & 24 / 18\% & 112 / 94\% & 23 & 44\\
\multicolumn{1}{l|}{SH7} & 50 & 400 & 600 & 41 $\pm$ 1 & 70 / 9\% & 33 / 91\% & 38 / 100\% & 56 & 66\\ \hline
\hline
\multicolumn{10}{c}{(c) HPK - length: \SI{1.0}{\cm} -- pitch: \SI{80}{\um } -- metal width: \SI{60}{\um}}\\ \hline
\hline
\multicolumn{1}{l|}{SHN1} & 20 & 1600 & 240 & 30 $\pm$ 1 & 27 / 22\% & 11 / 77\%& 16 / 99\%  & 35 & 39\\
\multicolumn{1}{l|}{SHN2} & 50 & 1600 & 240 & 32 $\pm$ 1 & 17 / 13\% & 10 / 86\%& 11 / 99\%  & 68 & 74\\\hline
    

\end{tabular}
\caption{Summary table of the overall performance for HPK strip sensors of \SI{1.0}{\cm} strip length with (a) \SI{500}{\um} pitch and \SI{50}{\um} metal width, (b) \SI{500}{\um} pitch and \SI{100}{\um} metal width, and (c) \SI{80}{\um} pitch and \SI{60}{\um} metal width. The second column contains the characteristics of the sensor. The second, third, and fourth columns show active thickness in \si{\um}, sheet resistance in \si{\Omega/\sq}, and capacitance in \si{\pF/\mm^2} respectively.
The spatial resolution column shows the one channel, two or more channels, and combined spatial resolution in \si{\um} and efficiency.
The last two columns present the most probable value for the amplitude for direct hits in the gap/metal region.}
\label{tab:summary_strips}
\end{table}

\begin{table}[H] 
\centering
\begin{tabular}{l |ccc |c| c c c | c c}
\multicolumn{1}{l|}{\multirow{3}{*}{Name}}  & \multirow{3}{*}{\begin{tabular}[c]{@{}c@{}} Act. \\Thick.\\ $[\si{\um}]$\end{tabular}} & \multirow{3}{*}{\begin{tabular}[c]{@{}c@{}} Resis.\\ $[\si{\Omega / \sq}]$\end{tabular}} & \multicolumn{1}{c|}{\multirow{3}{*}{\begin{tabular}[c]{@{}c@{}} Cap.\\ $[\si{\pico F / \mm^2}]$\end{tabular}}} &\multicolumn{1}{c|}{\multirow{3}{*}{\begin{tabular}[c]{@{}c@{}} Time res.\\ $[\si{\ps}]$\end{tabular}}} & \multicolumn{3}{c|}{Spatial resolution [\si{\um}]} & \multicolumn{2}{c}{\multirow{2}{*}{\begin{tabular}[c]{@{}c@{}} Amplitude \\ $[\si{\mV}]$\end{tabular}}} \\ \cline{6-8}

\multicolumn{1}{c|}{} & \multicolumn{3}{c|}{} & \multicolumn{1}{c|}{} & \multicolumn{1}{c|}{One ch.} & \multicolumn{1}{c|}{Two ch.} & \multicolumn{1}{c|}{Combined} & &\\ \cline{6-10}
\multicolumn{1}{c|}{} & \multicolumn{3}{c|}{} & \multicolumn{1}{c|}{}& \multicolumn{1}{c|}{Res./Eff.} & \multicolumn{1}{c|}{Res./Eff.} & \multicolumn{1}{c|}{Res./Eff.} & \multicolumn{1}{c|}{Gap} & Metal\\ \hline
\hline

\multicolumn{10}{c}{(a) HPK 2 x 2 Squared -- pitch: \SI{510}{\um} -- metal width: \SI{500}{\um}}\\ \hline
\hline
\multicolumn{1}{l|}{PH1} & 20 & 1600 & 600 & 20 $\pm$ 1 $^{**}$ & - / - & - / - & - / - & 39 & 59\\
\multicolumn{1}{l|}{PH2} & 30 & 1600 & 600 & 24 $\pm$ 1 $^{**}$& - / - & - / - & - / - & 74 & 100\\
\multicolumn{1}{l|}{PH3} & 50 & 1600 & 600 & 35 $\pm$ 1 $^{**}$& - / - & - / - & - / - & 119 & 153 \\\hline
\hline
\multicolumn{10}{c}{(b) HPK 4 x 4 Squared -- pitch: \SI{500}{\um} -- metal width: \SI{150}{\um} }\\ \hline
\hline
\multicolumn{1}{l|}{PH4} & 20 & 400 & 600 & 21 $\pm$ 1 & 61 / 37\% & 21 / 63\% & 41 / 100\% & 35 & 110\\
\multicolumn{1}{l|}{PH6} & 50 & 400& 600 & 30 $\pm$ 1 & 56 / 25\% & 23 / 75\% & 34 / 100\% & 65 & 188 \\
\multicolumn{1}{l|}{PH7} & 50 & 1600 & 600 & 32 $\pm$ 1 & 61 / 25\% & 31 / 75\% & 41 / 100\% & 67 & 211\\\hline
\hline
\multicolumn{10}{c}{(c) HPK 4 x 4 Squared -- pitch: \SI{500}{\um} -- metal width: \SI{300}{\um} }\\ \hline
\hline
\multicolumn{1}{l|}{PH8} & 20 & 1600 & 600 & 21 $\pm$ 1 & 96 / 64\% & 15 / 36\% & 78 / 100\% &  49 & 142 \\ \hline
\end{tabular}
\caption{Summary table of the overall performance for HPK Pixel sensors with (a) \SI{510}{\um} pitch and \SI{500}{\um} metal width, (b) \SI{500}{\um} pitch and \SI{150}{\um}, and (c) \SI{500}{\um} pitch and \SI{300}{\um} metal width. The second, third, and fourth columns show active thickness in \si{\um}, sheet resistance in \si{\Omega/\sq}, and capacitance in \si{\pF/\mm^2} respectively.
The spatial resolution column shows the one channel, two or more channels, and combined spatial resolution in \si{\um} and efficiency.
The last two columns present the most probable value for the amplitude for direct hits in the gap/metal region. Sensors PH1, PH2, and PH3 are mostly metalized and do not include an estimate of spatial resolution. ($^{**}$) Quoted numbers use results from the Fermilab 16-channel board.}
\label{tab:summary_pix}
\end{table}

\section{Conclusions and Outlook}\label{sec:discussion}

A comprehensive survey of a set of strip and pixel AC-LGAD sensors in a \SI{120}{\GeV} proton beam was carried out. 
The study of the impact of various design parameters on the time and spatial resolution of these sensors, identified the guiding principles for designing sensors optimized for future collider experiments, for example the EIC.

The results show that reducing the sensor thickness has the potential to drastically improve the time resolution of AC-LGAD pixelated sensors. 
For pixel sensors a time resolution of around \SI{20}{\ps} can be achieved for \SI{20}{\um} sensors, while \SI{50}{\um} achieve around \SI{30}{\ps}, and uniform performance across sensor surface can be achieved with an optimized choice of sheet resistance. 
These pixel sensors are identified as ideal devices for future timing detectors. 

For multi-channel pixel sensors, a larger metal pad size is required to achieve reasonable signal amplitudes and faster rise times uniformly throughout the sensor surface. However, large metal pad size leads to degraded two-channel hit efficiencies. The overall system design and optimization, including the readout electronics and ASIC design aspects should all be considered to select the optimal electrode geometry. For example, one can consider alternative metal electrode geometries to ensure uniform response across the sensor surface, such as achieved by cross-shaped pixels. 

Further studies of these sensors with channel counts comparable to a realistic detector is needed, and this would require the advancement of specialized readout electronics appropriate for these devices, such as those developed by Ref.~\cite{Xie:2023flv}.

However, unlike the pixelated sensors, we observe that the long, coarse-pitch 20-\si{\um}-thick strip sensors have higher jitter than their 50-\si{\um}-thick counterparts, and as result have worse net time resolution despite their reduced Landau fluctuations.
The strip signal amplitudes tend to decrease with increasing strip length and with propagation across large inter-strip gaps, as shown in Ref.~\cite{Kita:2023dia}.
Compounding these effects with the smaller initial ionization from 20-\si{\um} sensors results in small signals with jitter too large to see the benefit of the reduced Landau component.
Future productions of long strips should try intermediate active thicknesses between 20--50~\si{\um}, such as 30 or 40~\si{\um}, whereas future pixel productions could attempt to decrease the active thickness further until they reach a limit as in the case of the current strips.

It was also observed in these studies that the AC-LGAD sensor performance strongly depends on the $n+$ sheet resistance, which should be tuned to achieve the specific design goals. 
For HPK strip sensors with low resistivity (\SI{400}{\ohm/\sq}), the signal size in the main readout channel was observed to be a limiting factor to achieve good time resolution, while the high resistivity (\SI{1600}{\ohm/\sq}) sensors provided the best performance.

We also studied the impact of metal electrode width (50 vs 100 \si{\um}).
It was observed that the overall performance of sensors with decreasing metal electrode width, ranging from \SI{100}{\um} to \SI{50}{\um}, was not significantly impacted. However, the two-strip efficiency increased over the whole sensor area, indicating an improvement in the overall spatial resolution for the 500-\si{\um}-pitch strip sensors.
Strip sensors similar to SH4 are identified as an ideal candidate for the future barrel timing layer of EIC. If the goal of achieving the ultimate timing resolution with large signal amplitudes is critical in specific applications, then the wider metal strips are a more promising approach, albeit at the cost of reduced position resolution.
This result is best seen by the mostly metalized pixel sensors shown in Sec.~\ref{sec:2x2_hpk}.

Furthermore, we observed that narrower pitch results in naturally smaller spatial resolution. 
However, the smaller pitch does not compensate for the loss in signal for \SI{1}{\cm} long strip devices in the sensors with small active thickness (20~\si{\um} as opposed to 50~\si{\um}), therefore no improvement is observed in the time resolution for such thin strip sensors.

The coupling capacitance of the AC electrodes does not have a significant impact on any sensor reported in this paper.
Perhaps, the variation in the capacitance is not substantial enough to see a dramatic difference.
However, it is worth noting that there is a small correlation between increasing signal amplitude with increasing the coupling capacitance at the cost of decreased signal sharing between neighboring channels. 

In summary, the development of 4D trackers utilizing AC-LGADs is ready for a major project application and the development of complementary readout electronics. The results presented in this paper show a wide range of sensors that can deliver excellent timing and position information.
The development of such sensors has reached a maturity level that further improvement must be made with a particular application in mind, for example specific detectors at the EIC or other future colliders.
If cost of cooling and large number of channels are critical for deciding on the optimal detector design then the 500-\si{\um}-pitch strip sensor (SH4) is an ideal device.
Alternatively, if high occupancy is the primary concern, the SHN2 device is an appropriate choice. The PH1 sensor is well-suited for applications requiring optimal time resolution and maximum signal amplitude, with a trade-off in positional performance. Lastly, if both optimal time and position resolution are essential, and the higher channel density and smaller signal sizes are not an issue, PH4 is the ideal choice.

\section*{Acknowledgements}

We thank the Fermilab accelerator and FTBF personnel for the excellent performance of the accelerator and support of the test beam facility, in particular M.~Kiburg, E.~Niner, N.~Pastika, E.~Schmidt and T.~Nebel. 
We also thank the SiDet department, in particular M.~Jonas and H.~Gonzalez, for preparing the readout board by mounting and wirebonding the LGAD sensors. 
Finally, we thank L.~Uplegger for developing and maintaining the telescope tracking system.

This document was prepared using the resources of the Fermi National Accelerator Laboratory (Fermilab), a U.S. Department of Energy, Office of Science, HEP User Facility. 
Fermilab is managed by Fermi Research Alliance, LLC (FRA), acting under Contract No. DE-AC02-07CH11359.
This work was also supported by the U.S. Department of Energy (DOE), Office of Science, Office of High Energy Physics Early Career Research program award. 
This work was also supported by funding from University of Illinois Chicago under Contract DE-FG02-94ER40865 with the U.S. Department of Energy (DOE), and Contract 418643 with Brookhaven National Laboratory.
This work has also been supported by funding from the California Institute of Technology High Energy Physics under Contract DE-SC0011925 with the U.S. Department of Energy.    
This work was supported by the Science Committee of Republic of Armenia (Research projects No. 22AA-1C009 and 22rl-037).
This work was supported by the Chilean ANID PIA/APOYO AFB230003 and ANID - Millennium Science Foundation - ICN2019\_044.


\bibliographystyle{elsarticle-num} 
\bibliography{main}{}

\end{document}